\documentclass[twocolumn, tighten]{aastex7_preprint}
\usepackage{lmodern, fix-cm}
\usepackage{verbatim}
\usepackage{soul}
\usepackage{lineno}
\usepackage{newtxtext,newtxmath}

\usepackage[T1]{fontenc}
\usepackage{multirow}
\usepackage{graphicx}	% Including figure files
\usepackage{amsmath}	% Advanced maths commands
\usepackage{amssymb}	% Extra maths symbols
\usepackage[utf8]{inputenc}
\usepackage{mathtools}

\usepackage{color}      
\usepackage{placeins}
\usepackage{longtable, booktabs, threeparttablex,}

\usepackage[caption=false]{subfig}
\usepackage{blindtext}
\usepackage[normalem]{ulem}
\usepackage{subfiles}
\makeatletter
\let\oldthebibliography\thebibliography
\let\endoldthebibliography\endthebibliography
\renewenvironment{thebibliography}[1]{%
  \oldthebibliography{#1}%
  \footnotesize % change bibliography font size
}{%
  \endoldthebibliography
}
\makeatother

\newcommand{\mi}[1]{\texttt{m12i}}
\newcommand{\mf}[1]{\texttt{m12f}}
\newcommand{\mm}[1]{\texttt{m12m}}
\newcommand{\mb}[1]{\texttt{m12b}}
\newcommand{\GLOBAL}[1]{global disturbance}
\newcommand{\PEAK}[1]{peak disturbance}
\newcommand{\Dglobal}{D_{\rm global}}
\newcommand{\Dpeak}{D_{\rm peak}}

\newcommand{\Msol}[1]{M\textsubscript{\(\odot\)}}

\usepackage{xcolor}
\definecolor{darkgreen}{rgb}{0,0.7,0}
\definecolor{darkpurple}{rgb}{0.6,0,0.6}

\newcommand{\uw}{Department of Astronomy and DiRAC Institute, University of Washington, 3910 15th Ave NE, Seattle, WA, 98195, USA}

\shorttitle{Streams in evolving hosts}

\begin{document}
\thispagestyle{empty} % This line removes the first page header
\title{No Stream Left Unscathed: The imprint of a host galaxy}

\correspondingauthor{Arpit Arora}
\email[show]{arora125@sas.upenn.edu}

\author[0000-0002-8354-7356]{Arpit Arora}
\affiliation{\uw}
\email{arora125@sas.upenn.edu}

\author[0000-0001-6957-1627]{Peter S. Ferguson}
\affiliation{\uw}
\email{pferguso@uw.edu}

\author[0000-0001-8042-5794]{Jacob Nibauer}
\affiliation{Department of Astrophysical Sciences, Princeton University, 4 Ivy Ln, Princeton, NJ 08544, USA}
\email{jnibauer@princeton.edu}
 
\author[0000-0003-2497-091X]{Nora Shipp}
\affiliation{\uw}
\email{nshipp@uw.edu}

\author[0009-0005-9440-3269]{Videep Reddy}
\affiliation{\uw}
\email{videepr@uw.edu}

\author[0000-0002-5038-9267]{Eugene Vasiliev}
\affiliation{University of Surrey, Guildford GU2 7XH, UK}
\email{eugvas@protonmail.com}

\author[0009-0003-9069-3399]{Jack Kohm}
\affiliation{\uw}
\email{jackkohm@uw.edu}

\author[0009-0008-3389-9848]{Laurella C. Marin}
\affiliation{\uw}
\email{emarin4@uw.edu}

\author[0000-0003-0872-7098]{Adrian M. Price-Whelan}
\affiliation{Center for Computational Astrophysics, Flatiron Institute, 162 Fifth Ave, New York, NY 10010, USA}
\email{aprice-whelan@flatironinstitute.org}

\author[0000-0002-8448-5505]{Denis Erkal}
\affiliation{University of Surrey, Guildford GU2 7XH, UK}
\email{d.erkal@surrey.ac.uk}

\author[0000-0003-0256-5446]{Sarah Pearson}
\affiliation{DTU Space, Technical University of Denmark, Elektrovej 327, DK- 2800 Kgs. Lyngby, Denmark}
\affiliation{DARK, Niels Bohr Institute, University of Copenhagen, Jagtvej 155A, 2200 Copenhagen, Denmark}
\email{sapea@dtu.dk}

\author[0000-0003-0603-8942]{Andrew Wetzel}
\affiliation{Department of Physics \& Astronomy, University of California, Davis, CA 95616, USA}
\email{awetzel@ucdavis.edu}

\author[0000-0001-6380-010X]{Jeremy Bailin}
\affiliation{Department of Physics and Astronomy, University of Alabama, Box 870324, Tuscaloosa, AL, 35487-0324, USA}
\email{jbailin@ua.edu}

\author[0000-0002-1109-1919]{Robert Feldmann}
\affiliation{Department of Astrophysics, University of Zurich, Zurich CH-8057, Switzerland}
\email{robert.feldmann@uzh.ch}

\begin{abstract}
Stellar streams from disrupted globular clusters are excellent probes of dark matter (DM) subhalos. Observed Milky Way streams display a remarkable diversity of features: spurs, gaps, kinks, cocoons, and density variations, many attributed to subhalo encounters. But how much of this diversity arises from the host itself? We simulate $\sim$15,000 globular cluster streams across four Milky Way-mass halos from the FIRE-2 cosmological simulations, evolved in basis function expansion potentials capturing the evolving disk, halo, and large-scale structure while excluding small-scale perturbers such as DM subhalos and giant molecular clouds. We find that roughly three quarters of streams develop complex features from the host potential, such as spurs, kinks, and cocoon-like envelopes. Even the smoothest streams exhibit 10--25\% width variation along their track and host overdensities and gaps at scales of ${\sim}2^\circ$, squarely in the $1^\circ$--$5^\circ$ range predicted for subhalo-induced gaps. Pericentric distance is the primary predictor of stream morphology, with ${\sim}15$ kpc separating smooth from disturbed streams and circular orbits beyond $\sim$20 kpc producing the smoothest streams. Only $\sim$70 out of $\sim$15,000 streams are free of detectable wiggles in the track at any scale. Analogs to observed features, such as the GD-1 spur and the ATLAS--Aliqa Uma kink, emerge even without the presence of subhalos. As next-generation surveys (LSST, Euclid, and Roman) resolve stream structure across hundreds of streams, the baseline established here, streams evolved without small-scale perturbers, becomes essential for extracting DM substructure constraints.

\end{abstract}

\section{Introduction} \label{sec:intro}

Stellar streams are gravitationally disrupted remnants of dwarf galaxies and globular clusters (GCs) stretched along their orbits by the host potential. They are among the most powerful tools available for mapping dark matter (DM) on galactic scales \citep[e.g.,][]{johnston1999tidal, helmi1999building, belokurov2006field}. A stellar stream's morphology and kinematics are set by the gravitational field through which it orbits. Therefore, streams serve as sensitive probes of the host potential at both the global, galaxy-wide
scale \citep[e.g.,][]{koposov2010constraining, bovy2016shape, malhan2019constraining, shipp2021measuring, koposov2023s, ibata2024charting} and the small scale of individual DM subhalos, since encounters subhalos are predicted to leave observable gaps, density perturbations, and off-track features in otherwise cold,
thin tidal tails \citep[e.g.,][]{ibata2002uncovering, johnston2002lumpy, carlberg2013gaps, erkal2015forensics, erkal2016number, banik2019effects, bonaca2019spur}.

Over the past decade the census of known Milky Way (MW) stellar streams has grown from a handful to nearly 100 \citep{mateu2023galstreams, bonaca2025stellar}, driven by wide-field photometric surveys such as the Dark Energy Survey \citep[DES; ][]{des2016dark, shipp2018stellar, shipp2019proper} and SDSS \citep{belokurov2006field, grillmair2006detection, bonaca2012cold}, and astrometric searches with \textit{Gaia} \citep{gaia2018gaia, malhan2018ghostly, ibata2019streams}. 
The next generation of surveys will push this census further still: the Vera C. Rubin Observatory's Legacy Survey of Space and Time (LSST) is expected to reach a coadded depth of $r\sim27.5$ magnitude over $\sim$20,000 deg$^2$ \citep{ivezic2019lsst}, the ESA \textit{Euclid} mission will image $\sim$15,000 deg$^2$ in the optical and near-infrared \citep{mellier2025euclid}, and the Nancy Grace Roman Space Telescope will deliver deep, high-resolution photometry over $\sim$3,000 deg$^2$ \citep{spergel2015wide, pearson2019detecting}. Together, these facilities will provide unprecedented photometric detail resolving density fluctuations, width variations, and off-track features at sub-degree angular scales for dozens to hundreds of streams simultaneously. With this level of detail, detecting the morphological imprints of individual DM subhalo encounters on streams becomes observationally feasible. But interpreting those detections requires knowing what a stream should look like before any subhalo perturbation.

This begs a fundamental interpretive question: what is the expected baseline morphology of a GC stream in a realistic host potential, before any DM subhalo encounters? The standard approach to constraining DM substructure from streams assumes that GC streams are intrinsically thin, cold, and smooth, so that any observed gaps, spurs, or density variations can be attributed to subhalo perturbations \citep[e.g.,][]{erkal2015forensics, bovy2017linear, bonaca2019spur, banik2021novel, nibauer2025measurement, nguyen2025forecasting}. 
GD-1, the ``jackpot'' stream\footnote{As termed by T.~S.~Li (private communication).}, hosts a prominent spur \citep{price2018off} and gap that have been interpreted as evidence for a close encounter with a $\sim$$10^{6}$--$10^{8}$ $M_\odot$ DM subhalo \citep{bonaca2019spur, bonaca2020high}, as well as a cocoon-like envelope whose origin may trace back to the progenitor's pre-accretion environment \citep{malhan2019butterfly, valluri2025gd}. GD-1's morphology and kinematics have been used to place direct constraints on the subhalo mass function \citep[e.g.,][]{banik2021novel, nibauer2025measurement}.

Yet a growing body of evidence challenges the assumption that GC streams are intrinsically featureless. The stream associated with Ophiuchus and Pal~5 is significantly perturbed by the Galactic bar \citep{price2016spending, hattori2016shepherding, pearson2017gaps, erkal2017sharper}, producing density variations and fan-like morphology that are entirely host-driven. The Orphan--Chenab stream shows large-scale misalignments attributed to the time-dependent influence of the LMC \citep{erkal2019total, koposov2023s, lilleengen2023effect}, and similar LMC-induced perturbations have been inferred for the Atlas--Aliqa~Uma (AAU) stream, Sagittarius stream and other outer-halo structures \citep{shipp2019proper, shipp2021measuring, vasiliev2021tango, brooks2025lmccalls}. Beyond direct morphological distortions, the LMC's infall also reshapes the encounter geometry between streams and DM subhalos, anisotropically boosting interaction rates for streams on certain orbits \citep{nadler2021effects,barry2023dark,arora2024lmc}. Even the collisional internal evolution of GC progenitors such as two-body relaxation, mass segregation, episodic mass loss can imprint density and width variations on the resulting streams \citep{balbinot2018devil, webb2019searching, ibata2020detection, banik2021n, gieles2021supra, panithanpaisal2025breaking}. In short, the MW is not a smooth, static, symmetric potential: it is a dynamically evolving system with a bar, spiral arms, a massive disk, an ongoing LMC interaction, and a cosmological assembly history that has shaped its potential over billions of years \citep[see recent review by][]{hunt2025milky}.

Despite this, most theoretical studies of GC stream morphology have relied on idealized, static, or mildly perturbed host potentials, adding a single component such as the LMC \citep[e.g.,][]{erkal2019total, vasiliev2021tango, shipp2021measuring} or the bar \citep[e.g.,][]{price2016spending,hattori2016shepherding, pearson2017gaps, erkal2017sharper, banik2019effects}, or using constrained N-body simulations that assume some degree of symmetry for the host \citep[e.g.,][]{lilleengen2023effect, brooks2025lmccalls, weerasooriya2025dancing}. The MW's DM halo is neither spherical nor static: it is triaxial, tilted with respect to the disk \citep{vera2013constraints, erkal2019total, shao2021twisted, vasiliev2021tango, han2022tilt, koposov2023s, nibauer2025galactic, dillamore2026geometry}, and shaped by both filamentary accretion and satellite interactions on timescales comparable to orbital periods of halo streams \citep[e.g.,][]{vera2013constraints,arora2025shaping, darragh2025shaping}. 
A complete accounting of host-driven stream complexity requires time-evolving potential models that capture the secular evolution of the halo, disk, bar, and large-scale substructure simultaneously, while excluding small-scale perturbers such as DM subhalo and giant molecular clouds, to establish a controlled baseline against which subhalo-induced perturbations can be measured.

Zoomed-in cosmological-baryonic simulations of MW-mass galaxies provide exactly this setting. While these simulations resolve the
host galaxy and the streams of its massive dwarf satellites \citep[e.g.,][]{panithanpaisal2021galaxy, sanderson2017modeling, riley2025auriga, shipp2025auriga, guerra2026probing} in detail, individual GC streams are not resolved at the baryonic particle masses available
($m_{\rm b} \sim 7{,}000\,M_\odot$), so we inject and evolve them
as test-particle populations in the simulated potential. In this work, we simulate a statistical population of $\sim$15,000 GC streams across four MW-mass halos from the FIRE-2 \textit{Latte} suite \citep{wetzel2023public}, spanning a range of assembly histories from isolated hosts to systems with recent LMC-mass and Sagittarius-mass mergers. Streams are evolved in time-dependent potentials modeled with basis function expansions (BFE) that capture the evolving bar, spiral structure, disk asymmetries, and halo response to satellite interactions \citep{arora2022stability, arora2024efficient}, while smoothing over small-scale perturbers by construction. The progenitor orbits span pericenters of 10--30 kpc and eccentricities from nearly circular to highly radial, broadly covering the orbital range of observed MW streams. 

\citet{panithanpaisal2025breaking} recently demonstrated a complementary approach with \textsf{CosmoGEMS}, modeling individual GC streams with self-consistent collisional dynamics and cosmologically motivated initial conditions from the \citet{grudic2023greatballs} GC formation model in the same FIRE-2 framework. \citet{holm2025catalog} similarly constructed stream catalogs using cosmologically motivated progenitor phase-space coordinates from the \citet{chen2024catalogue} GC populations. Here we trade that internal fidelity and cosmological realism in the initial conditions for population-level statistics, producing a controlled ensemble of $\sim$15,000 streams evolved without dark matter subhalo perturbations in a realistic cosmological potential.

In this work, we focus exclusively on the photometric morphology of the stream population, the projected on-sky structure accessible to imaging surveys such as LSST and Roman. We develop a suite of quantitative morphology metrics and apply them at a population level to ask: \textbf{how much morphological complexity do GC streams acquire from the host potential alone?} The answer establishes the \emph{complexity floor}: the baseline level of structure present even in the absence of any subhalo population. In subsequent work, we will introduce explicit DM subhalo populations to generate perturbed-stream catalogs and explore whether kinematic information (proper motions, line-of-sight velocities) can distinguish host-driven from subhalo-driven perturbations.

The paper is organized as follows. In Section~\ref{sec:methods} we summarize the methodology: the host simulations, the time-evolving BFE potential model, the GC stream generation pipeline, and methodological caveats. In Section~\ref{sec:morphology_metrics} we define the quantitative morphology metrics. In Section~\ref{sec:results} we present the metrics at a population level. We discuss implications for observed Milky Way streams in Section~\ref{sec:MW_context}. In Section~\ref{sec:drivers} we examine the physical origins of the complexity floor through a controlled comparison with a static, axisymmetric host potential. We summarize our findings and conclude in Section~\ref{sec:disc_conc}.

\section{Simulations and Methods} \label{sec:methods}

In this section we outline the methods. We first describe the galaxy simulations used in the analysis (Section~\ref{sec:sims}), then summarize the time-evolving, BFE-based potential model (Section~\ref{sec:pot_model}). We next briefly describe our GC stream simulation setup (Section~\ref{sec:GC_streams}), including progenitor initial conditions (Section~\ref{sec:GC_ics}) and the particle-spray formation model (Section~\ref{sec:particle_spray}). Finally, we discuss the caveats and limitations of the approach (Section~\ref{sec:caveats}).

\subsection{Milky Way analogs} \label{sec:sims}

We select four halos---\mi{}, \mf{}, \mm{}, and \mb{}---from the \textit{Latte} suite of zoomed-in cosmological–baryonic simulations of MW-mass galaxies, part of the Feedback In Realistic Environments (FIRE-2) project \citep{wetzel2023public}.\footnote{These simulations are publicly available \citep{wetzel2023public, wetzel2025second} at \url{http://flathub.flatironinstitute.org/fire}.} The simulations use the FIRE-2 physics model implemented in \texttt{GIZMO} \citep{hopkins2015code} and assume a \(\Lambda\)CDM cosmology consistent with Planck results \citep{collaboration2015planck}. For further details, see \citet{hopkins2018fire} for the FIRE-2 model and \citet{wetzel2023public, wetzel2025second} for the specific implementation. The \textit{Latte} suite resembles the MW in stellar mass, gas content, DM mass, and density profile \citep[e.g.,][]{wetzel2016reconciling, garrison2018origin, sanderson2020synthetic}.

The selected halos have total masses \({\sim}1\text{--}1.5\times10^{12}\ M_\odot\) and are resolved with initial baryonic particle masses \(m_{\rm b}=7{,}100\ M_\odot\) (stars and gas) and DM mass \(m_{\rm DM}=35{,}000\ M_\odot\). Plummer-equivalent gravitational softenings (typical FIRE-2 choices) are fixed for collisionless particles and adaptive for gas: \(\epsilon_{\rm DM}\approx40\ \textrm{pc}\), \(\epsilon_{\rm star}\approx4\ \textrm{pc}\), and adaptive gas softening with a minimum \(\epsilon_{\rm gas,min}\approx1\ \textrm{pc}\). Snapshots are saved at \(\sim\)25\,Myr cadence, and we focus exclusively on the last 5\,Gyr when integrating our GC stream population. The four halos sample a range of assembly histories and disk properties: \mb{} is the best MW–LMC analog (satellite-to-host mass ratio \(\approx\)1:8; most-massive satellite pericenter \(\approx\)40\,kpc) with its major satellite infall and pericentric passages occurring within the last 5\,Gyr \citep{arora2024lmc,garavito2024corotation}, \mm{} hosts the oldest and most massive galactic disk \citep{sanderson2020synthetic}, \mf{} undergoes two late-time, Sagittarius-mass mergers within the last 5\,Gyr \citep{arora2022stability} and, \mi{} remains the most isolated, with only minor mergers in the past 5\,Gyr. None of the four Latte halos develop a strong, long-lived bar \citep{ansar2025bar}, which would serve as an additional source of stream perturbations beyond what is captured here. We refer to \mb{} and \mf{} collectively as the ``merger hosts'' throughout. Key merger properties are compiled in \citet{garavito2024corotation} and the halo/galaxy properties in \citet{wetzel2023public, wetzel2025second}.

Before fitting a potential model to these simulations, we must define a consistent host-centered reference frame. These zoomed-in cosmological simulations are run in a periodic box frame on an expanding background. Since the zoom-in region does not have zero total momentum, we recenter the simulation onto the host galaxy frame using the shrinking spheres method \citep{power2003inner} to find the galactic center at each snapshot applied to star particles. To suppress center-of-mass jitter due to bursty star-formation and mergers, we spline-smooth the halo center trajectory (following the cubic-spline on each coordinate approach of \citealt{sanders2020models, vasiliev2021tango, arora2022stability}). This is absolutely crucial to construct a stable time-evolving potential model. Subsequently, we compute a single, fixed rotation matrix that aligns the galactic disk with the $XY$ plane at the present-day snapshot and apply this rotation to all snapshots. The resulting frame, host-centered with a fixed disk orientation, defines the galactocentric frame used throughout.

\subsection{Time-evolving potential} \label{sec:pot_model}

We use the time-evolving low-order multipole potential (TEMP) of \citet{arora2022stability}, built from a combination of basis function expansions (BFE) fitted to the host density at each snapshot. BFE have been shown to effectively reproduce orbits in both constrained \citep{garavito2021quantifying,lilleengen2023effect} and cosmological simulations \citep{lowing2011halo, sanders2020models, arora2024efficient}, including cases with massive satellites \citep[e.g.,][]{vasiliev2021tango, vasiliev2024dear, arora2024lmc}

We briefly describe the implementation. We represent the DM halo and hot gas (\(T_{\rm gas}\ge10^{4.5}\) K) with a spherical harmonic expansion evaluated on a logarithmic radial grid. The stellar bulge, flattened stellar disk, and cold gas (\(T_{\rm gas}\le10^{4.5}\) K) are represented by a Fourier expansion on a 2D \((R,Z)\) meridional grid. Expansions are fitted to all particles within 600\,kpc in the galactocentric frame (host-centered disk aligned with the \(XY\) plane at \(z=0\))\footnote{The Fourier-harmonic expansion assumes the disk is aligned with the \(XY\) plane; tests in \citet{arora2024efficient} show high-fidelity orbit recovery for moderate disk tilts and slow reorientation compared to orbital timescales.}. Expansion coefficients are computed with \texttt{Agama} \citep{vasiliev2019agama}. We restrict the expansions to a maximum order \(l_{\rm max}=4\) and \(m_{\rm max}=4\)  for both the spherical- and Fourier-harmonic components, which captures the dominant halo \citep[e.g.,][]{arora2022stability, weinberg2023new, arora2025shaping, darragh2025shaping} and large-scale structures such as the bar, spiral arms, and disk deformations \citep{wang2020basis, weinberg2021using, petersen2021using, hunt2025dark} driven by secular evolution and satellite interactions, while minimizing high-frequency noise and smoothing over small-scale substructure such as giant molecular clouds or dark subhalos.

Forces from the coefficients are interpolated in time to produce a continuous, time-dependent potential for orbit integration.\footnote{The correction force from spline-smooth halo center trajectory and scale-factor correction of \citet{arora2024efficient} are included as well.} Appendix~\ref{app:bfe_shot_noise} calibrates the shot-noise component of the BFE force field and shows that it contributes negligibly to stream morphology across the orbital range of our sample. Appendix~\ref{app:bfe_noise} verifies that the linear interpolation scheme introduces no detectable spurious structure into integrated orbits and bounds the cumulative velocity bias to below 0.1\% of the orbital velocity over the full 5 Gyr integration. Our typical snapshot cadence of \(\sim\)25\,Myr reproduces ensembles of orbits over multiple dynamical times. See Appendix~\ref{app:force_err_Nbody} for a direct comparison of BFE accelerations to tree-based evaluations of the FIRE-2 particle distribution. Full validation and interpolation/error details for these models in FIRE-2 simulations are given in \citet{arora2022stability,arora2024efficient}. 

\subsection{Simulating globular cluster streams}\label{sec:GC_streams}

\begin{figure*}[ht]
    \centering
    \includegraphics[width=\linewidth]{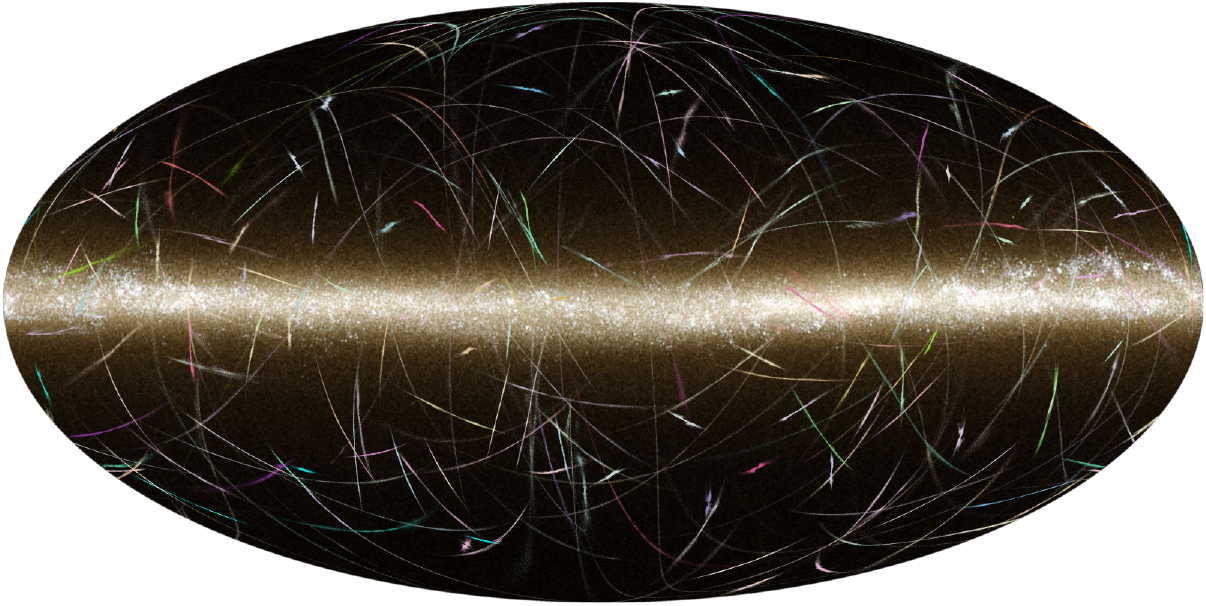}
    \caption{A random subset of 250 simulated GC streams (colored) shown in galactocentric Mollweide projection for the \mi{} halo at present day. The background shows a mock photometric map constructed from stellar populations in the host, with \textit{gri}-band luminosities assigned per star particle using \textsc{fsps} \citep{conroy2009propagation} and projected onto a \textsc{HEALPix} grid ($N_\mathrm{side} = 512$). Each stream is represented by 10,000 test particles. The ensemble exhibits a broad diversity of morphologies, from thin, dynamically cold streams to more diffuse and complex structures featuring gaps, spurs, and cocoon-like envelopes.}\label{fig:streams_mollweide_example}
\end{figure*}

We generate a statistical population of GC streams in each of the four FIRE-2 hosts described in Section~\ref{sec:sims}. Figure~\ref{fig:streams_mollweide_example} presents an all-sky view of 250 randomly selected GC streams in the \mi{} halo at present day, overlaid on a mock photometric map of the host constructed by assigning \textit{gri}-band luminosities to each star particle using the Flexible Stellar Population Synthesis code (\textsc{fsps}; \citealt{conroy2009propagation}). Each stream is represented by 10,000 test particles evolved in the time-dependent \mi{} potential for 5 Gyr. The population exhibits a striking variety of morphologies -- ranging from thin, dynamically cold streams to broader, more irregular structures with visible spurs, gaps, and cocoon-like components. Streams on orbits perpendicular to the disk plane tend to appear longer and thinner, while those interacting more strongly with the disk or bar show increased complexity along their tracks. We describe the progenitor initial conditions (Section~\ref{sec:GC_ics}) and stream formation model (Section~\ref{sec:particle_spray}) below.

\subsubsection{Progenitor profiles \& present-day distribution}\label{sec:GC_ics}

\begin{figure*}[ht]
    \centering
    \includegraphics[width=\linewidth]{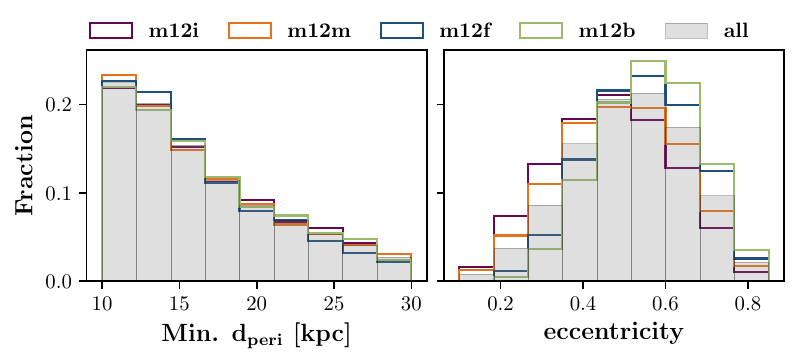}
    \caption{Distributions of progenitor minimum pericentric distance (left) and orbital eccentricity (right) for each halo (colored lines) and for the full sample (solid gray). Histograms are normalized to show the fraction of streams. Pericentric distance is taken as the minimum radius reached by the progenitor over the 5\,Gyr integration. Eccentricity is computed from the maximum apocenter and minimum pericenter as \(e=(r_{\rm apo}-r_{\rm peri})/(r_{\rm apo}+r_{\rm peri})\). The pericenter distributions are similar across halos, while the two merger hosts (\mb{}, \mf{}) show a higher fraction of radial (high-eccentricity) orbits.}
\label{fig:streams_peri_eccen}
\end{figure*}

We generate an initial sample of progenitors for each host by drawing present day phase-space coordinates from an isotropic, quasi-spherical distribution function based on the \citet{dehnen1993family} tracer density profile with scale radius $r_s = 6$ kpc and inner slope $\gamma = 1$, evaluated in the present-day host potential with enforced axisymmetry in each galaxy. The total mass of the tracer population is arbitrary as it serves only to define the phase-space sampling. This choice provides a smooth, physically motivated phase-space distribution that approximately follows the stellar halo.

We integrate each progenitor orbit backward for 5 Gyr in the time-evolving host potential and compute the minimum pericenter and maximum apocenter over that interval. Over this integration window, inner-halo streams ($d_{\rm peri} \sim 10$ kpc) complete $\sim$10 orbital periods, while outer-halo streams ($d_{\rm peri} \sim 30$ kpc) complete only $\sim$4, accumulating proportionally fewer pericentric passages. We apply orbital cuts requiring $10 \leq r_{\rm peri}/{\rm kpc} \leq 30$ and $20 \leq r_{\rm apo}/{\rm kpc} \leq 110$ and select roughly 4,200 present-day ICs per halo that meet the cuts. These limits broadly cover the radial range of observed MW streams and ensure both inner- and outer-halo orbits are sampled. The lower pericenter bound preferentially selects progenitors that undergo sufficient tidal stripping to form extended, observable streams while avoiding rapid disruption deep in the inner galaxy. For context, the Jacobi radius at the minimum pericentric distance sets the tidal boundary of each progenitor. Across the population, the Jacobi radius for orbits with with $r_{\rm peri} \geq20$~kpc ranges from $\sim$45 to $\sim$480 pc with a median of $\sim$78 pc. For a typical massive GC with a scale radius of $\sim$4--10 pc, this implies that these progenitors sit well within their tidal boundary and strip comparatively little material at each passage.

Figure~\ref{fig:streams_peri_eccen} shows the normalized (fractional) distributions of minimum pericenter and eccentricity for each halo (colored) and for the combined sample (solid gray). Eccentricity is computed from the orbit-wise maximum apocenter and minimum pericenter as \(e=(r_{\rm apo}-r_{\rm peri})/(r_{\rm apo}+r_{\rm peri})\). The pericenter distributions are statistically indistinguishable across halos (pairwise KS tests give p-value $\gtrsim 0.2$). \mb{} (green) and \mf{} (blue) with massive mergers show a larger fraction of highly radial/eccentric orbits. This radial bias arises because recent massive mergers violate the equilibrium assumed in the distribution function, shifting the sampled orbital mix toward more radial trajectories.

Each progenitor cluster is modeled with a time-static density profile following a King model characterized by central potential depth $W_0 = 5$ and a fixed scale radius of 4 pc\footnote{The particle-spray model is insensitive to the detailed progenitor profile. The scale radius primarily sets the Lagrange-point offset at ejection. We adopt 4 pc to match the calibration setup of \citet{chen2025improved}.}. Progenitor masses are drawn from a log-normal mass distribution consistent with present-day GC systems \citep[e.g.,][]{gnedin2014globular, brooks2025lmccalls}, spanning a range $10^4 - 10^7 \ \mathrm{M}_\odot$. Across this mass range, the tidal radius at the pericenters sampled in our orbital cuts remains much larger than the progenitor scale radius, so the fixed 4 pc choice does not affect the stripping dynamics or the resulting stream morphology. While in principle one could track where globular clusters form and evolve in a cosmological context \citep[e.g.,][]{li2017star, pfeffer2018mosaics, grudic2023greatballs, chen2024catalogue, pearson2024forecasting, holm2025catalog}, our goal here is to construct a statistical ensemble of progenitor orbits that provide representative phase-space coverage and a realistic mass spectrum.

\subsubsection{Particle-spray models for GC streams} \label{sec:particle_spray}

We model stream formation with a particle-spray (Lagrange cloud stripping) approach \citep[e.g.,][]{kupper2012more, fardal2015generation, chen2025improved}: 10,000 tracer particles per stream are released near the progenitor's Lagrange points and evolved as massless test particles in the host and progenitor potential. We adopt the improved particle-spray algorithm of \citet{chen2025improved}, which samples escape velocities from a calibrated multivariate Gaussian distribution measured in N-body experiments in a static-symmetric MW host and reproduces the action-space distribution of escaped stars accurately compared to full N-body in the collisionless regime. 

From each progenitor's present-day phase-space position we rewind the orbit for 5\,Gyr in the chosen time-evolving host potential, then forward-model stream formation by spawning two tracer particles per integration step (one for the leading arm and one for the trailing arm). Particles are ejected uniformly in time rather than weighted toward pericentric passages where stripping is strongest. This is a deliberate simplification: real mass loss is concentrated near pericenter, particularly for eccentric orbits, but uniform ejection provides a controlled baseline that isolates the effect of the host potential on stream morphology. Velocities and positional offsets at ejection are drawn from the empirical distributions described in \citet{chen2025improved}. We include the progenitor's local tidal radius when placing Lagrange-point offsets but do not evolve the progenitor's internal structure (i.e., the progenitor potential is static during the spray). Appendix~\ref{app:force_err_Nbody} validates this prescription against direct collisionless N-body simulations in our time-evolving potentials.

This simple prescription produces a computationally efficient baseline that captures the large-scale tidal morphology and kinematics formed by secular stripping (including leading/trailing asymmetries and epicyclic overdensities). It is intended as a controlled baseline for later experiments that add localized perturbations (e.g., subhalo impacts) \citep[e.g.,][]{nibauer2025streamsculptor} or more detailed collisional cluster evolution \citep{panithanpaisal2025breaking}. For correctness of the time-dependent integration we ensure orbit integration is performed in the appropriate host-centered inertial frame \citep{brooks2025lmccalls}.

\subsection{Caveats on the method} \label{sec:caveats}

The methods above define a controlled baseline of unperturbed by DM subhalos and other small-scale structures streams: secular stripping in a smooth, time-evolving host potential with no explicit subhalo population or collisional cluster evolution. This is by design, the baseline isolates the morphological complexity produced by the host potential alone and provides a reference for future experiments that add localized perturbers. Several simplifications follow from this choice.

BFE host potentials are approximate. With our choice \(l_{\rm max}, m_{\rm max}=4\) the expansion produces a relatively smooth field that captures large-scale features such as the bar, spiral arms, disk asymmetries \citep[e.g.,][]{weinberg1992detection, puerari1992fourier, weinberg2001noise, petersen2021using} and the lowest-order halo response to massive satellites \citep[e.g.,][]{garavito2021quantifying, weinberg2023new, arora2025shaping, darragh2025shaping} while suppressing high-frequency noise. This truncation necessarily smooths over small-scale structure that can also perturb streams, e.g., encounters with giant molecular clouds \citep{amorisco2016gaps} and low-mass DM subhalos \citep[e.g.,][]{ibata2002uncovering, johnston2002lumpy, erkal2015forensics, bonaca2019spur} are both expected to produce gaps, clumps and kinematic disturbances in cold streams \citep[e.g.,][]{banik2019effects}. Likewise, a snapshot cadence of \(\sim\)25\,Myr is sufficient for resolving orbits with periods \(\gtrsim200\) Myr outside \(\sim\)10 kpc (Appendix~\ref{app:bfe_noise}), but may fail to capture rapid, high-frequency phenomena. For example, resonant structure and phase-space features driven by a fast-rotating bar can evolve on substantially shorter timescales and require finer temporal sampling to resolve fully \citep[e.g.,][]{dehnen2000effect, monari2019signatures, chiba2021resonance}.

The particle-spray model (Lagrange cloud stripping) is intentionally simple and secular: we eject tracers uniformly in time and do not evolve the progenitor's internal structure. The model reproduces the large-scale morphology from collisionless stripping but neglects episodic, peri-centric-weighted mass loss and the progenitor’s internal, collisional evolution (mass loss, two-body relaxation, etc.), processes known to influence stream density, width and kinematics \citep[e.g.,][]{erkal2017sharper, balbinot2018devil, banik2021n}. A physically motivated alternative is to bias ejection toward peri-center passages or to sample an orbital-phase-dependent stripping kernel modeling internal cluster evolution (mass loss, two-body relaxation) would require N-body or hybrid \citep[e.g.,][]{vasiliev2021tango, limberg2025s} approaches and is outside the present scope. See, e.g., \citet{li2017star, pfeffer2018mosaics, weatherford2024stellar,Phillips_DirectNbody} for cluster-formation/evolution context.

Our progenitor initial conditions are drawn from a present-day equilibrium distribution function rather than from cosmologically motivated GC formation models \citep[e.g.,][]{grudic2023greatballs, chen2024catalogue, pearson2024forecasting, holm2025catalog}. Cosmologically formed GCs occupy a different region of phase-space, shaped by their formation environment and subsequent orbital evolution within the host. However, our analysis is structured to minimize sensitivity to this choice: the key morphological trends are presented as functions of orbital parameters (pericenter, eccentricity) rather than as population averages, so the results at a given orbit are insensitive to how that orbit was sampled. What could change with cosmologically motivated ICs is the \emph{relative weighting} of different orbital bins and therefore the population-level fractions (e.g., Table~\ref{tab:xi_classification}), but not the per-orbit morphological trends themselves. Moreover, the orbital distribution of MW GC streams will itself evolve as LSST and Roman uncover new streams across a wider range of orbits, making an agnostic sampling of orbital parameter space a practical advantage.

The \citet{chen2025improved} particle-spray algorithm we adopt was calibrated against collisionless N-body experiments in a single reference (static) MW potential. We have applied the same \citet{chen2025improved} ejection prescription within our TEMP framework and compared a small set of streams to collisionless N-body runs where we resolve progenitor self-gravity with a direct N-body code\footnote{Available at \url{https://github.com/appy2806/Nbody_streams/}.} (see Appendix~\ref{app:force_err_Nbody} for one such example). Those limited comparisons show reasonable agreement in bulk phase-space structure for the cases tested, but a full, statistical N-body validation across our entire ensemble is computationally expensive and beyond the present work. 

\section{Stream morphology metrics} \label{sec:morphology_metrics}
Stellar streams are among the most promising tools for detecting DM subhalos below the threshold of star formation \citep[e.g.,][]{ibata2002uncovering, johnston2002lumpy, carlberg2012dark, erkal2015forensics}. The basic idea is compelling: a close encounter between a cold stream and a DM subhalo should produce observable gaps, density perturbations, and off-track features in the stream's morphology \citep[e.g.,][]{erkal2016number, banik2019effects}, and these signatures have been used to place direct constraints on the subhalo mass function \citep[e.g.,][]{bonaca2019spur, banik2021novel, nibauer2025measurement}. 

Observed MW streams display a remarkable diversity of morphologies \citep[see][]{mateu2023galstreams, bonaca2025stellar}: large-scale asymmetries and broad structures \citep[e.g.,][]{sales2008genealogy, price2016spending, hattori2016shepherding,erkal2017sharper, pearson2017gaps, shipp2021measuring}, localized off-track features such as cocoons, spurs, or kinks \citep[e.g.,][]{malhan2019butterfly, bonaca2019spur,li2021broken, valluri2025gd}, and narrow gaps or density fluctuations along their length \citep{carlberg2013gaps, li2021broken, ferguson2021delve, tavangar2022fire}. But not all of these features necessarily come from subhalo encounters---the time-evolving host potential itself, including the bar, spiral arms, and massive satellite interactions, can produce morphologically similar signatures \citep[e.g.,][]{erkal2017sharper, erkal2018modelling, qian2022structure, dillamore2022impact, lilleengen2023effect, nibauer2024slant, brooks2025lmccalls}. Even in the absence of
time-dependence, chaotic orbits \citep{price2016chaotic} and orbital resonances in non-spherical potentials can sculpt streams into fans, shells, and other complex structures \citep{yavetz2021separatrix, yavetz2023stream}. Disentangling these origins requires quantitative morphological diagnostics that can be applied uniformly across a large stream population.

A useful starting point is a set of baseline, or \emph{null}, assumptions: that stream stars are distributed uniformly along the main track, that their cross-track distributions are approximately Gaussian in small segments as expected for collisionless debris in a smooth potential \citep[e.g.,][]{helmi1999building, eyre2011mechanics}, and that the overall structure can roughly be described by a single coherent orbit. Departures from these assumptions --- disturbed cross-track profiles, asymmetry, or excess small-scale density variations --- correspond to the diverse morphologies seen in observations and are exactly what one would search for as evidence of subhalo encounters or other perturbations.

In this section, we develop a suite of stream morphology metrics to quantify these departures. We group them into three categories: (1) stream-wide properties describing overall length and width (Section~\ref{sec:stream_length}), (2) local tests for off-track structure (Section~\ref{sec:local_NG}), and (3) measures of along-track structure that capture departures from smoothness, such as overdensities, gaps or excess small-scale power (Section~\ref{sec:power_spectrum}). While these baseline assumptions are reasonable in static, symmetric hosts, we ask: how do they hold up in a realistic, time-dependent galactic potential? Together, these metrics allow us to measure the \emph{complexity floor}: the baseline level of structure the host potential imprints on streams even in the absence of any subhalo population.

\subsection{Stream Length and Width} \label{sec:stream_length}

\begin{figure*}[ht]
  \centering
  \includegraphics[width=0.85\linewidth]{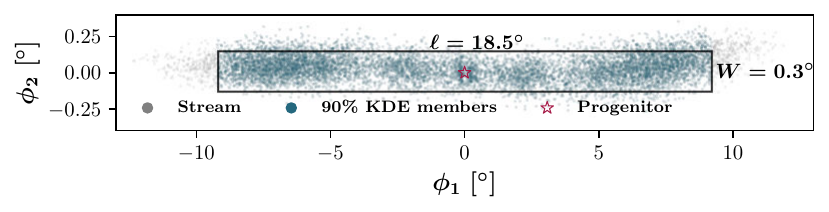}
  \caption{Example stream (gray points) in stream coordinates. Selected member stars within the 90\%-KDE mask are shown in blue. The progenitor is marked with a red star at \((\phi_1,\phi_2)=(0^\circ,0^\circ)\). The box indicates the straight segment length \(\ell\) and the quantile based width \(W\) derived from the 90\%-KDE mask. In this example the stream length is \(18.5^\circ\) and the width is \(0.30^\circ\). Note that x, y-axis aspects are unequal.}
  \label{fig:sample_stream_length_width}
\end{figure*}

\begin{figure*}[ht]
    \centering
    \includegraphics[width=\linewidth]{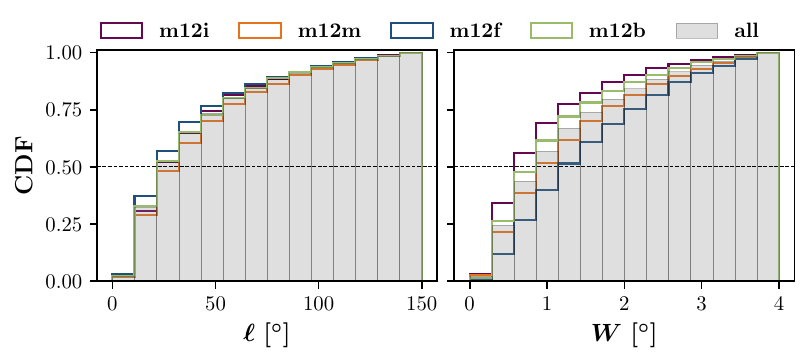}
    \caption{Cumulative distributions (CDFs) of stream length \(\ell\) (left) and width \(W\) (right) for each halo (colored lines) and for the full sample (solid gray). Curves are normalized to unity. \mf{} (blue) with its recent mergers produces, on average, shorter and thicker streams than the other halos.}
\label{fig:streams_len_wid}
\end{figure*}

Before testing for detailed morphological features, we first characterize each stream's basic spatial extent in the galacocentric coordinates. We quantify each stream with a single length \(\ell\) and a single width \(W\) measured in stream coordinates defined below with the progenitor fixed at \((\phi_1,\phi_2)=(0^\circ,0^\circ)\). For each stream, we compute a one-dimensional kernel density estimate (KDE) in \(\phi_1\) and retain the contiguous region enclosing 90\% of the KDE mass. This 90\%-KDE mask defines the coherent main track used for all metrics. Length \(\ell\) is the straight-segment span in \(\phi_1\) across that mask (reported in degrees). The straight-segment measure is sufficient for the comparisons presented here and for selecting along-track bins for local tests (more complex path reconstructions exist \citep{starkman2023fast} but are unnecessary for the diagnostics we present).

Width \(W\) is the 90\% quantile range of \(\phi_2\) for particles inside the \(\phi_1\) mask. The 90\%-KDE mask excludes poorly resolved low-density tails that become incoherent after \(\sim\)5\,Gyr and are sensitive to numerical resolution. The quantile width is robust to outliers and low-density wings but does not capture along-track width variation (see \(C_w\) in \S\ref{sec:local_NG}). Figure~\ref{fig:sample_stream_length_width} shows an example stream (gray points) with its 90\%-KDE members marked in blue and the progenitor as a red star at \((0^\circ,0^\circ)\). The rectangle indicates the reported \(\ell\) and \(W\) (example: \(\ell\approx18^\circ\), \(W\approx0.3^\circ\)). All subsequent analyses use the 90\%-KDE mask. 

The stream coordinate frame is defined by fitting a great circle to the whole stream track in the galactocentric position and velocity frame and then minimizing the $L_2$ spread in $\phi_2$ through arbitrary rotations around the stream spine. This produces the projection in which the stream appears narrowest and most coherent. Because our morphology metrics are computed in this best-case frame, they represent a lower bound on the apparent complexity: real structure that exists in the line-of-sight or radial direction can be projected away by this choice, but no spurious structure is introduced. Photometric surveys do not have the freedom to choose their viewing angle, so the morphological complexity measured here is conservative relative to what would be observed from any fixed vantage point such as the Sun.

Figure~\ref{fig:streams_len_wid} shows the cumulative distributions of stream length \(\ell\) (left) and width \(W\) (right) for each halo (colored) and for the combined sample (solid gray). The \(\ell\) distributions are broadly similar across halos, reflecting the range set by the number of pericenter passages. \mf{} (blue) contains a larger fraction of short streams: \({\sim}30\%\) of m12f streams have \(\ell\leq10^\circ\) compared with \({\sim}15\%{-}20\%\) in the other halos. The LMC-analog host m12b (green) follows the combined distribution as well. Widths show larger halo-to-halo variation: the isolated \mi{} (purple) produces predominantly thin streams ( \({\sim}75\%\) with \(W\leq1.5^\circ\) ), while \mf{} is thicker (only \({\sim}50\%\) of streams have \(W\leq1.5^\circ\) ) and \mm{} (orange), with its massive disk, also has a higher fraction of thicker streams. These differences track recent assembly and disk properties across the sample.

\subsection{Off-Track Structure: Local Disturbance Metrics} \label{sec:local_NG}

\begin{figure*}[ht]
    \centering
    \includegraphics[width=\linewidth]{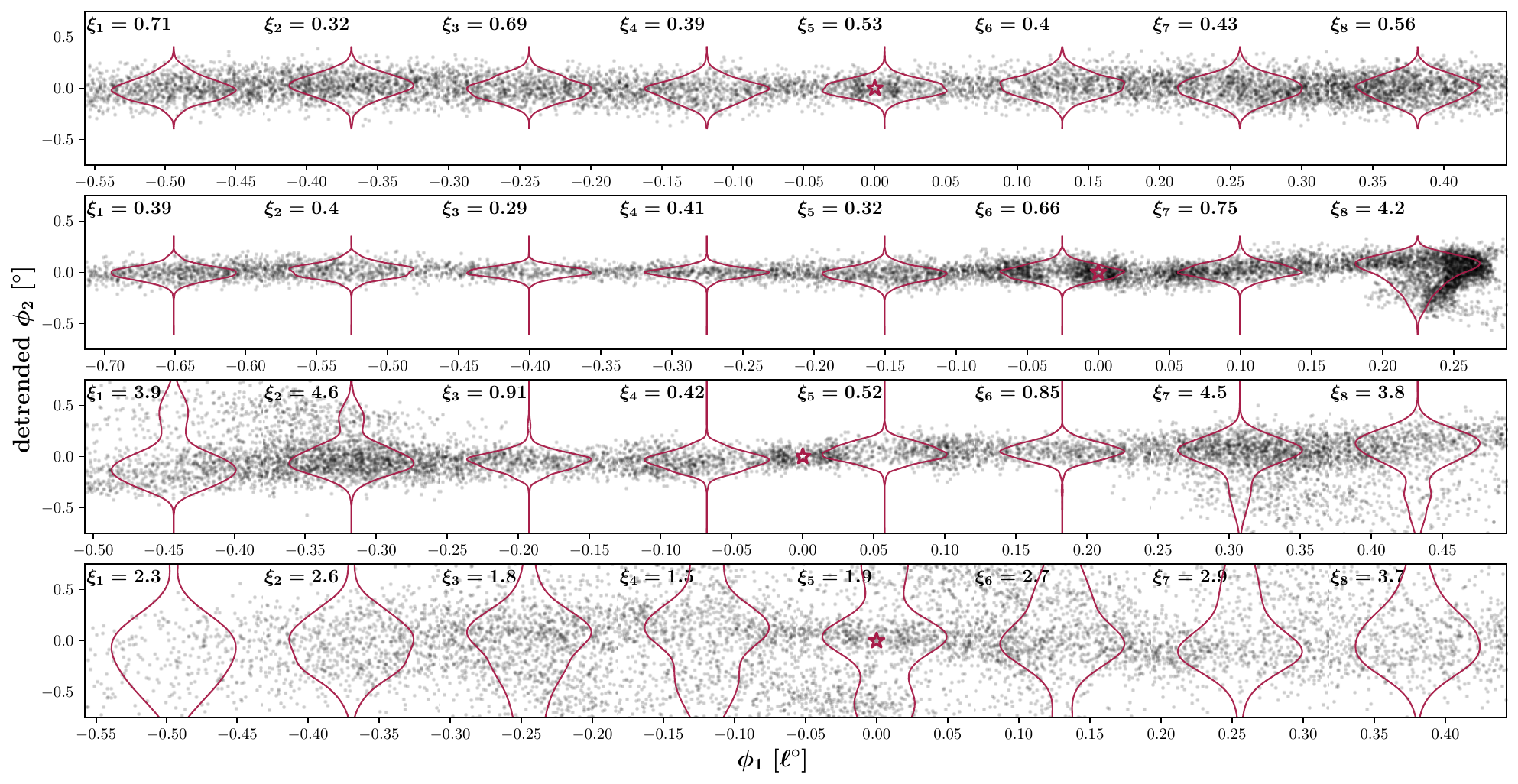}
    \caption{Representative examples of local off-track structure for four GC streams in stream coordinates (four rows). Streams are divided in eight equal-length (\(\ell/8\)) \(\phi_1\) bins after detrending \(\phi_2(\phi_1)\) by a cubic polynomial. The red violin curves display the local detrended \(\phi_2\) particle distribution in each region, the annotated number is the disturbance score \(\xi_i\) for that bin (Appendix~\ref{app:wass_dis_Gauss}). The progenitor is marked with a red star at \((\phi_1,\phi_2)=(0^\circ,0^\circ)\). Streams are ordered by increasing overall disturbance: the top row is locally Gaussian in all bins, the second row contains a single strong localized deviation, the third row shows mixed behavior (Gaussian in the middle regions with energetic tails), and the bottom row is disturbed across all bins. Streams lengths range from \(20^\circ\)–\(50^\circ\). The progenitor need not occupy the same bin in every row.} \label{fig:sample_streams_NonGauss}

\end{figure*}

A smooth stream stripped in a static potential should produce a narrow, Gaussian off-track ($\phi_2$) distribution profile in any small segment along its length. A spur, kink, cocoon, or any other off-track feature will break that local Gaussianity --- broadening the profile, adding secondary peaks, or producing asymmetric tails. We exploit this expectation to build a segment-by-segment test for off-track structure along each stream.

We test for local departures from Gaussianity along each stream track using the off-track coordinate \(\phi_2\) (the same procedure can be applied to kinematic observables such as \(v_{\rm los}\) or proper-motion components). The method proceeds as follows.  First, we detrend the stream track by fitting a low-order polynomial to \(\phi_2(\phi_1)\) and subtracting the fit to remove large-scale curvature and median trends. Second, we divide the coherent 90\%-KDE track (Sec.~\ref{sec:stream_length}) into eight equal-length (\(\ell/8\)) bins in \(\phi_1\) and collect the \(\phi_2\) samples within each bin. Third, for each bin \(i\) we compute the disturbance metric \(\xi_i\) using the Wasserstein distance between a reference standardized normal and the \(\phi_2\) distribution in the bin, with significance tested at the 5\% level by bootstrap resampling matched to the bin sample size. Appendix~\ref{app:wass_dis_Gauss} describes the metric and bootstrap procedure in detail. 

Figure~\ref{fig:sample_streams_NonGauss} shows the region-wise, detrended \(\phi_2\) distributions (red violin curves) and the computed disturbance score \(\xi_i\) (annotated on the top-left corner for each region) for four example streams (rows). By construction, \(\xi_i<1\) indicates consistency with the local Gaussian cross-track profile at the 5\% significance level. Larger \(\xi_i\) values signal increasingly strong off-track structure such as secondary peaks, asymmetric tails, or broad wings. The example streams range from \({\sim}10^\circ\) to \({\sim} 50^\circ\) in length. The progenitor (red star) is placed at \(\phi_1=0^\circ\) but need not lie in the same bin for every stream.

The top-row stream is locally Gaussian in all 8 regions (\(\xi_i\lesssim0.7\)) and would be classified as smooth. The second-row stream is Gaussian in most regions but shows a single, strong localized departure in the rightmost bin, the kind of feature that could be mistaken for a subhalo-induced spur. The third-row stream has a Gaussian core (middle four bins) flanked by energetic, off-track components in outer bins (several bins with \(\xi_i\gtrsim4\)). The bottom-row stream is disturbed across all bins. Visual inspection confirms that the \(\xi_i\) score reliably tracks the presence and strength of localized structure.

We compute the \(\xi_i\) sequence along the track for each stream across all halos and summarize a stream with a small set of diagnostics:

\begin{itemize}
  \item {\boldmath\({f_{\rm Gauss}} = N(\xi_i<1)/N_{\rm bins}\)}: fraction of bins consistent with the Gaussian null (e.g., \(f_{\rm Gauss}\ge0.5\) means at least half of the stream track is locally Gaussian).
  \item {\bf Global disturbance} ({\boldmath\(\Dglobal{})\)} \(\equiv \mathrm{median}(\xi_i)\): overall level of off-track structure along the stream. Low values indicate a stream that is broadly consistent with a narrow, Gaussian track. High values indicate global broadening or disruption.
  \item {\bf Peak disturbance \boldmath\(\Dpeak{}\)} \(\equiv \max(\xi_i)\): strength of the single strongest localized departure anywhere along the track. Sensitive to isolated features such as a spur, kink, or cocoon, even if the rest of the stream appears smooth.
  \item {\bf Width variation} {\boldmath$(C_w)$} $\equiv \dfrac{\mathrm{std}(\sigma_i) }{\mathrm{mean}(\sigma_i)}\times 100 \% $, where \(\sigma_i\) is the cross-track standard deviation in bin \(i\). A high \(C_w\) indicates indicates that the stream pinches and flares significantly along its length.
  
\end{itemize}

\begin{table}[ht]
\caption{Off-track diagnostics for the four example streams shown in Figure~\ref{fig:sample_streams_NonGauss}. Columns list the fraction of bins consistent with the Gaussian null ($f_{\rm Gauss}$), the global disturbance $\Dglobal$, the peak disturbance $\Dpeak$, and the width variation $C_w$.}
\label{tab:sample_nonG}
\begin{tabular}{ccccc}
% \toprule
\textbf{Stream} & \boldmath$f_{\rm Gauss}$ & \boldmath$\Dglobal{}$ & \boldmath$\Dpeak{}$ & $C_w (\%)$\\
\midrule
\midrule
row 1 & 1.00 & 0.48 & 0.71 & 10\\
row 2 & 0.88 & 0.41 & 4.20 & 42\\
row 3 & 0.50 & 2.37 & 4.60 & 48\\
row 4 & 0.00 & 2.48 & 3.70 & 13\\
\bottomrule
\end{tabular}
\end{table}

These diagnostics act as complementary filters: \(\Dglobal{}\) and \(f_{\rm Gauss}\) classify streams by overall smoothness, \(\Dpeak{}\) flags strongly localized structure, and \(C_w\) quantifies relative width variability. In practice, $\Dglobal{}$ supersedes $f_{\rm Gauss}$ as the primary classification metric because it is continuous and captures the degree of departure rather than a binary count. $f_{\rm Gauss}$ is retained as an intuitive cross-check. Table~\ref{tab:sample_nonG} lists these diagnostics for the four example streams shown in Figure~\ref{fig:sample_streams_NonGauss}. 

Stream~4 has a low $C_w$ despite a large $\Dglobal{}$ because the track is broadly heated in $\phi_2$: the stream is uniformly thick, so $\sigma_i$ is large but varies little between bins. By contrast, streams with a single strong localized perturbation (e.g., stream~2 and stream~3) show elevated $C_w$ because one or two bins have much larger $\sigma_i$, driving up the relative width variation.

The connection to observations is direct. Low $\Dglobal{}$ selects streams that would appear overall smooth in photometric \citep[e.g.,][]{des2016dark, ivezic2019lsst} and astrometric \citep{gaia2018gaia} surveys, although they can still exhibit significant width variation (high \(C_w\)) along the track. Conversely, a large \(\Dpeak{}\) identifies streams with localized, observable features such as cocoons, spurs, or kinks. The disturbance metric provides a quantitative measure of off-track structure that simple moment-based or KS-style tests do not robustly capture for small, unevenly sampled bins.

The choice of eight bins is a practical hyperparameter: it balances spatial resolution with sufficient sample counts in low-density streams. We tested alternative binning schemes (10 and 12 equal-length bins, and equal-star-count bins) and find no statistically significant differences in the population-level diagnostics. Using 8 bins preserves power in the lowest-density regions while remaining stable across tests.

\subsection{Along-Track Density Structure} \label{sec:power_spectrum}

Gaps and overdensities along a stream's track are the primary morphological signatures used to search for DM subhalo encounters \citep[e.g.,][]{carlberg2012dark, erkal2015forensics, erkal2016number}. Detecting such features requires distinguishing real density variations from the statistical noise inherent in any finite stellar sample. We approach this by measuring the 1D power spectrum of fractional density fluctuations along the stream track \citep{bovy2017linear} and comparing it against a Monte Carlo Poisson null. This allows us to ask, for each stream: how strong are the density fluctuations, at what angular scales do they appear, and are they statistically significant?

\begin{figure*}[ht]
    \centering
    \includegraphics[width=\linewidth]{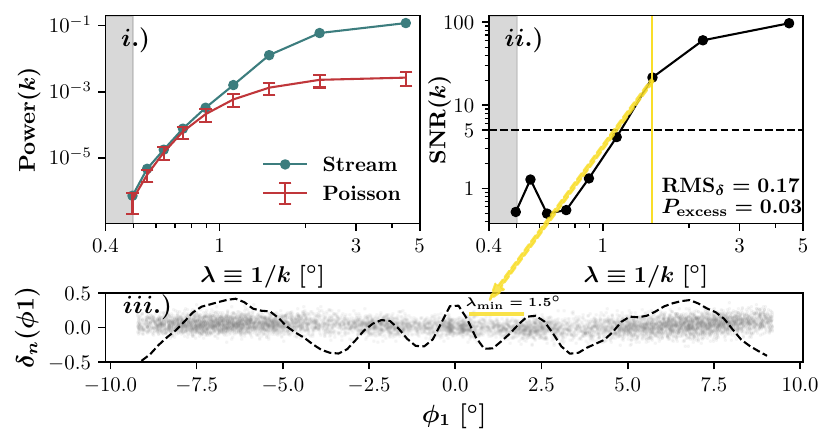}
    \caption{1D power-spectrum analysis for an example, visually smooth stream. \textit{i.)} Top left: Power spectrum \(P(k)\) of the fractional density residual \(\delta_n(\phi_1)\) (blue) plotted against angular separation \(\lambda=1/k\). The red curve show the median and \(\pm1\sigma\) from 2,000 null realizations sampling a uniform along-track distribution with the same particle counts and window. \textit{ii.)} Top right: Signal-to-noise \( \mathrm{SNR}(k) = \big[P(k)-\langle P^{\rm null}(k)\rangle\big]/\sigma_{P^{\rm null}}(k)\) relative to the Poisson null. The horizontal dashed line marks \(\mathrm{SNR}=5\) and the vertical yellow line marks \(\lambda_{\rm min}\), the minimum angular scale exceeding this threshold. Annotated values show $\mathrm{RMS}_\delta = 0.17$ and $P_{\rm excess} = 0.03$ for this stream. The gray shaded region marks the Nyquist scale $\lambda_{\rm nyq}\approx0.5^\circ$. \textit{iii.)} Bottom row: Fractional density residual \(\delta_n(\phi_1)\) plotted along the stream track with the stream rendered in the background. The yellow bar shows \(\lambda_{\rm min}\) for reference. Over- and under-densities at scales \(\gtrsim\lambda_{\rm min}\) are visually apparent.}
    \label{fig:example_power_spec_1d}
\end{figure*}

The procedure is as follows. We estimate the one-dimensional number density \(n(\phi_1)\) along the coherent 90\%-KDE track using a KDE with bandwidth $0.25^\circ$, comparable to the angular resolution of matched-filter stream searches in wide-field photometric surveys \citep[e.g.,][]{shipp2018stellar, li2021broken, ferguson2021delve}. We then remove large-scale trends by fitting and subtracting a cubic polynomial $\hat{n}(\phi_1)$ to form the fractional density residual $\delta_n(\phi_1) = (n-\hat{n})/\hat{n}$. Finally, we compute the 1D power spectrum of $\delta_n$ using the Welch method \citep{welch1967use} as implemented in \texttt{scipy.signal.welch}:

\begin{equation}
    P(k) = \big|\mathcal{F}[\delta_n(\phi_1)]\big|^2,
\end{equation}

\noindent where \(\mathcal{F}\) is the Fourier transform. We convert spatial frequency \(k\) to angular separation \(\lambda=1/k\) for interpretability.

To assess significance, we construct a Poisson null by drawing 2,000 Monte Carlo realizations that uniformly sample the same along-track window and particle count as the stream. For each realization we compute \(P^{\rm null}(k)\) and record the mean \(\langle P^{\rm null}(k)\rangle\) and dispersion \(\sigma_{P^{\rm null}}(k)\). The per-scale signal-to-noise is

\begin{equation}
    \mathrm{SNR}(k)=\frac{P(k)-\langle P^{\rm null}(k)\rangle}{\sigma_{P^{\rm null}}(k)}.
    \label{eq:SNR_k}
\end{equation}

The Monte Carlo null captures sampling- and window-dependent features introduced by the Welch procedure and is therefore preferred over an analytic white-noise floor. We use a Hann window for Welch and set \texttt{nperseg = max(8, number of bins/8)} with \texttt{noverlap = floor(0.5 * nperseg)}. The KDE bandwidth of $0.25^\circ$ sets the smallest resolvable angular structure and implies a conservative Nyquist-like scale $\lambda_{\rm nyq} \approx 0.5^\circ$ (shaded region in Figure~\ref{fig:example_power_spec_1d}). We restrict interpretation to scales above this limit. We tested alternative KDE bandwidths, segment lengths, overlap fractions, and binning grids (including equal-star-count sampling). Single-stream measurements can shift with parameter choices, but the population-level distributions and orbital trends presented in Section~\ref{sec:results_power} are robust to all variations tested.

We report three per-stream diagnostics throughout the analysis:

\begin{itemize}
  \item {\bf Minimum detectable scale} ({\boldmath \(\lambda_{\min}\)}): the smallest angular separation at which \(\mathrm{SNR}\ge5\). This is the finest scale at which along-track density structure is statistically confirmed above the noise.
  \item {\bf RMS density fluctuation} ({\boldmath $\mathrm{RMS}_\delta$}) $=\sqrt{\langle\delta_n^2\rangle}$: the root-mean-square fractional density fluctuation along the track. A stream with $\mathrm{RMS}_\delta = 0.10$ shows roughly 10\% peak-to-trough density variation.
  \item {\bf Excess power} ({\boldmath $P_{\rm excess}$}) $=\int_{k_1}^{k_2}[P(k)-\langle P^{\rm null}(k)\rangle]\,dk$: integrated power above the Poisson null, computed only over scales where $P(k)>\langle P^{\rm null}(k)\rangle$. This captures the total amount of statistically significant density structure across all detectable scales.
\end{itemize}

Figure~\ref{fig:example_power_spec_1d} illustrates the method for an example stream. The top-left panel shows \(P(k)\) (blue) and the MC null band (red). The top-right panel shows \(\mathrm{SNR}(k)\) with \(\lambda_{\min}\) marked in yellow. The bottom panel shows \(\delta_n(\phi_1)\) with the stream rendered in the background. This stream has $\mathrm{RMS}_\delta = 0.17$ (roughly 17\% peak-to-trough density variation), $P_{\rm excess} = 0.03$, and significant power at angular scales $\gtrsim1^\circ$ with $\lambda_{\min}\approx1.5^\circ$.

Together, these three metrics answer complementary questions. $\mathrm{RMS}_\delta$ measures how strong the density variations are. $P_{\rm excess}$ measures how much total structure exceeds the noise floor. And $\lambda_{\min}$ identifies the finest angular scale at which that structure is detectable, providing a direct point of comparison with the gap scales predicted from DM subhalo encounters \citep[e.g.,][]{erkal2015forensics, erkal2016number, banik2019effects}. As a point of reference, we also apply the same metrics to 1,000 streams evolved in a static, axisymmetric MW potential as a null comparison for the time-evolving FIRE-2 hosts, presented in Section~\ref{sec:drivers}.

\section{Stream Morphology Across the Population}\label{sec:results}

How \emph{morphologically complex} are GC streams that form in a realistic, time-evolving host potential? In this section we answer that question at a population level, applying the morphology metrics described in Section~\ref{sec:morphology_metrics} to the full stream population across all four halos. We organize the analysis around three complementary views of stream structure: off-track features visible in the cross-track features such as spurs, cocoons, bifurcations, and kinks, quantified by $\Dglobal$ and $\Dpeak$ (\S\ref{sec:results_nonGauss}); width modulation along the stream track by $C_w$ (\S\ref{sec:results_width}); and along-track density structure such as overdensities and gaps, quantified by $\mathrm{RMS}_\delta$, $P_{\rm excess}$, and $\lambda_{\min}$ (\S\ref{sec:results_power}). Large-scale track distortions such as wiggles and misalignments are partially captured by the cross-track metrics, as our detrending preserves structure on scales comparable to the bin size. For each category, we characterize the population distributions, identify trends with orbital parameters, and connect the results to observable stream morphologies. We then synthesize these views into the complexity floor (\S\ref{sec:complexity_floor}). 

Throughout, we restrict the analysis to streams with $\ell \geq 10^\circ$, corresponding to the typical angular extent of MW streams accessible to photometric surveys \citep{ivezic2019lsst, mateu2023galstreams} and excluding poorly resolved streams with insufficient track length for the binned diagnostics. After this cut, the combined sample contains 14,787 streams (3,909 in \mi{}, 4,093 in \mm{}, 3,493 in \mb{}, and 3,292 in \mf{}), with the merger hosts retaining fewer streams consistent with stronger tidal disruption producing shorter, less coherent debris. All metrics are computed on the 90\%-KDE track described in Section~\ref{sec:stream_length}.

\subsection{Off-Track Features Along the Stream}
\label{sec:results_nonGauss}

Streams are usually detected as coherent overdensities on the sky. A smooth, thin stream appears as a single narrow ridgeline, while one with cocoons, spurs, or off-track extensions present as broader, structured features. This distinction is the most immediate
visual diagnostic of stream complexity and the most directly connected to what imaging surveys observe. 

We quantify it with two complementary metrics defined in Section~\ref{sec:local_NG}. Global disturbance ($\Dglobal{}$), which measures the typical level of off-track structure averaged along the full stream. Low values indicate a narrow, well-behaved track, while high values indicate global broadening or warping. Peak disturbance ($\Dpeak{}$) isolates the single strongest localized departure anywhere along the track. A stream can score low on \GLOBAL{} yet high on \PEAK{} if it has an otherwise clean track with one prominent spur or kink, exactly the kind of feature commonly attributed to a DM subhalo encounter. Because our stream coordinate frame is defined to minimize cross-track spread (Section~\ref{sec:stream_length}), these metrics capture off-track structure in the most favorable projection. Any features detected here would appear at least as prominent from a fixed observational vantage point.

\subsubsection{A Visual Atlas of Off-Track Features}
\label{sec:results_xi_atlas}

\begin{figure*}[ht]
    \centering
    \includegraphics[width=\linewidth]{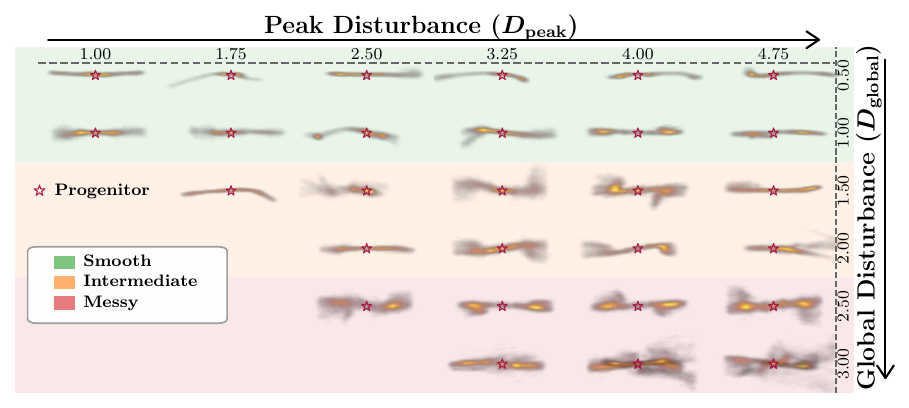}
    \caption{Representative streams arranged on a 2D grid with columns showing increasing \PEAK{} ($\Dpeak{}$) (left to right: 1.0 to 4.75 in steps of 0.75) and rows showing increasing \GLOBAL{} ($\Dglobal{}$) (top to bottom: 0.25 to 3.0 in steps of 0.5) at fixed $\Dpeak{}$ per row. The red star marks the progenitor at $(0^\circ, 0^\circ)$. Background shading indicates the three morphological classes: green for smooth ($\Dglobal{} \leq 1$), yellow for intermediate ($1 < \Dglobal{} \leq 2$), and red for messy ($\Dglobal{} > 2$). Smooth streams in the top-left are thin and largely featureless and, moving right along any row reveals progressively stronger localized features, while moving down any column reveals overall broadening and structural complexity.}\label{fig:streams_phi2_nonGauss_2d}
\end{figure*}

Photometric surveys observe streams as projected on-sky structures, and the most immediate diagnostic of stream complexity is its visual morphology: does it appear as a thin line, or does it show broader, structured features? To build intuition for what our metrics capture across different levels of complexity, Figure~\ref{fig:streams_phi2_nonGauss_2d} shows a representative sample of streams on a 2D grid, with columns ordered by increasing \PEAK{} ($\Dpeak{}$) (left to right, from 1.0 to 4.75 in steps of 0.75) and rows ordered by increasing \GLOBAL{} ($\Dglobal{}$) (top to bottom, from 0.25 to 3.0 in steps of 0.5). The red star marks the progenitor position at $(0^\circ, 0^\circ)$ in each panel. Background shading divides the grid into three broad regions: green ($\Dglobal{}\leq 1$), yellow ($1 < \Dglobal{} \leq 2$), and red ($\Dglobal{}> 2$), corresponding to smooth, intermediate, and messy streams.

Moving from left to right across any row, localized off-track features become progressively stronger: the leftmost streams appear thin and featureless, while streams toward the right show isolated spurs or off-track extensions. A notable example is the stream at $(\Dpeak{}, \Dglobal{}) \approx (3.25, 1.0)$, which appears globally smooth yet has a clear extended spur in one bin. Moving top to bottom down any column, the track itself becomes globally broader and more complex, with some intermediate streams showing a combination of a smooth core track flanked by energetic tails in outer bins. At larger $\Dpeak{}$, the intermediate streams ($1 < \Dglobal{} \leq 2$) can develop bimodal cross-track profiles. For example, the stream near $(\Dpeak{}, \Dglobal{}) \approx (4.0, 1.5)$ shows a clear double-peaked distribution in at least one bin, and in some cases the progenitor is visually displaced from the main dense track, as in the stream near $(4.75, 2.0)$.

The messy regime ($\Dglobal{} > 2$, red background) hosts the most morphologically complex structures. Some streams display self-similar wraps or bifurcations, where the leading and trailing arms overlap or cross back onto themselves. The stream near $(4.0, 3.0)$ is a clear example, with the progenitor completely displaced from the densest region of the track. Others, such as the stream near $(2.5, 2.5)$, show a broad cocoon-like envelope with no obvious thin ridgeline. If detected in full by photometric surveys, these complex morphologies would not be consistent with a simple tidal tail from a single compact progenitor, and they would be poor candidates for constraining DM subhalo populations from stream perturbations alone.

Visual inspection and the metric definitions together motivate the three-tier qualitative morphological classification used throughout the remainder of the analysis. A stream is \emph{smooth} if $\Dglobal{} \leq 1$. By construction, this means at least half of its track bins show no
significant off-track structure. A stream is \emph{intermediate} if $1 <\Dglobal{} \leq 2$, indicating that significant fractions of the track are disturbed but the stream retains some coherent structure. A stream is \emph{messy} if $\Dglobal{}> 2$, where the cross-track distribution is disturbed across most of the track.

\subsubsection{Orbital Trends in Off-Track Structure}
\label{sec:results_xi_peri}

\begin{figure*}[ht]
    \centering
    \includegraphics[width=\linewidth]{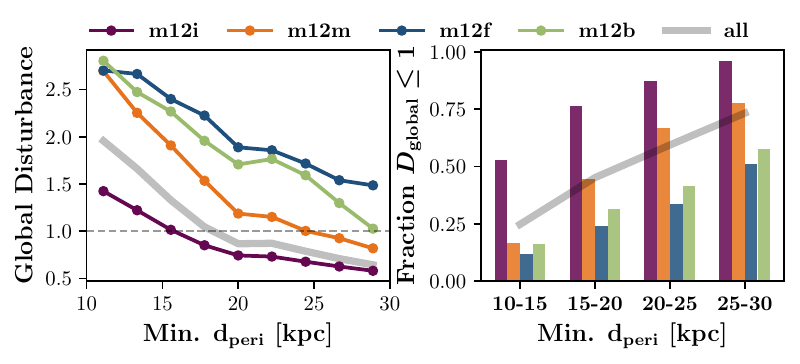}
    \caption{Left: Median \GLOBAL{} ($\Dglobal{}$) as a function of progenitor minimum pericenter for each halo (colored) and the combined sample (gray). Right: Fraction of smooth streams ($\Dglobal{} \leq 1$) in pericenter bins. The scatter at fixed pericenter is large (not shown).\mi{} (purple) reaches $\Dglobal{} \leq 1$ by $\sim$15 kpc and shows $\sim$80\% smooth streams beyond 20 kpc. Halo \mm{} reaches this threshold near $\sim$20 kpc. Halos \mb{} and \mf{} decline more slowly with larger scatter. Across all halos, visible off-track structure decreases and the smooth fraction increases at larger pericenters.}\label{fig:med_xi_peri_frac}
\end{figure*}

Streams whose progenitors orbit closer to the galactic center plunge deeper into the complex inner potential of the host, where the bar, spiral arms, and disk all contribute to perturbing the stream track. These streams also complete more orbital periods over the 5 Gyr integration, accumulating perturbations over more pericentric passages. In the merger hosts, recent satellite passages add further
perturbations across a range of radii. Progenitors on orbits with larger pericenters spend more time in the comparatively quiescent outer halo. This predicts a correlation between a stream's minimum pericentric distance and its off-track complexity, which we test here.

Figure~\ref{fig:med_xi_peri_frac} (left) shows the bin averaged \GLOBAL{} ($\Dglobal{}$) as a function of minimum pericentric distance for each halo (colored) and the combined sample (gray). The scatter at fixed pericenter is large (not shown). The trend is clear: progenitors with smaller pericenters show systematically higher $\Dglobal{}$, as they are more strongly perturbed by the complex inner potential of FIRE-2 halos \citep[e.g.,][]{sanderson2020synthetic, arora2024efficient, mccluskey2024disc, ansar2025bar}. The right panel shows the corresponding smooth fraction ($\Dglobal{} \leq 1$), which increases with pericenter for all halos. The orbital geometry driving this trend
is discussed further in Figure~\ref{fig:xi_peri_eccen_all}. 

The pericenter threshold at which most streams become smooth differs across halos, reflecting their different assembly histories and
potential structures. \mi{}(purple), the most isolated host, reaches $\Dglobal{} \leq 1$ by $\sim$15 kpc and has roughly 80\% smooth streams beyond 20 kpc. \mm{} (orange), with its massive disk, reaches this threshold near 20 kpc. \mb{} (green) and \mf{} (blue), whose
recent major mergers perturb streams even at intermediate radii, decline more slowly and show larger scatter at fixed pericenter. Despite these halo-to-halo differences, the trend is universal: orbits at larger pericenters produce smoother streams. For outer-halo progenitors with $r_{\rm peri} \geq 20$ kpc, the typical Jacobi radius lies in the range $\sim$40--480 pc (median $\sim$78 pc) in our host halos, substantially larger than the characteristic scale radius of GC progenitors. This places them in a weak stripping regime where only the outermost, weakly bound stars are removed. Progenitors that do strip sufficiently to form extended tidal tails produce thin, dynamically cold, and morphologically smooth streams, but some may retain enough mass that they remain observationally indistinguishable from intact clusters.   

\begin{figure*}[ht]
    \centering
    \includegraphics[width=\linewidth]{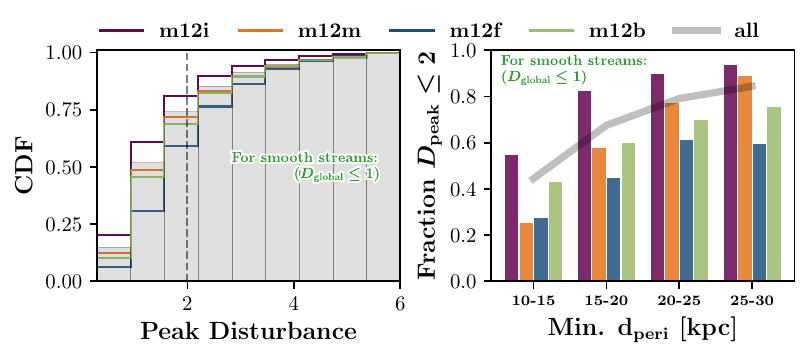}
    \caption{Left: CDF of \PEAK{} $\Dpeak{}$ for smooth streams (\GLOBAL{} $\Dglobal{} \leq 1$) in each halo (colored) and the combined sample (gray). Right: Fraction of smooth streams with $\Dpeak{} \leq 2$ in pericenter bins. A \PEAK{} value of 2 corresponds to a localized
    off-track departure twice as strong as expected from statistical noise alone (see Appendix~\ref{app:wass_dis_Gauss}), above which
    off-track features are visually identifiable. \mi{} has $\sim$75\% of smooth streams with \PEAK{} $\leq 2$, \mf{} (blue) has roughly 55\%. Even globally smooth streams regularly contain at least one localized detectable off-track feature.}\label{fig:max_smooth_xi_peri_frac}

\end{figure*}

Even among globally smooth streams, a substantial fraction of streams shows strong localized off-track feature in at least one segment of the track. Figure~\ref{fig:max_smooth_xi_peri_frac} (left) shows the CDF of \PEAK{} $\Dpeak{}$ for smooth streams (\GLOBAL{} $\Dglobal{} \leq 1$) in all halos (colored) and in the combined sample (gray). Across the combined sample, roughly 50\% of smooth streams have {} $\Dpeak{} > 1$, meaning they contain at least one segment with a detectable off-track feature such as a spur, kink, or cocoon. \mi{} (purple) fares best at $\sim$60\% with $\Dpeak{} \leq 1$, while \mf{} (blue) falls to roughly 30\%. 

A \PEAK{} value of 2 provides a natural reference: by construction, it corresponds to a localized off-track departure twice the normalized distance expected from a smooth cross-track profile at the same sample size (Appendix~\ref{app:wass_dis_Gauss}). Visual inspection of streams in Figure~\ref{fig:streams_phi2_nonGauss_2d} confirms that streams at or above this value consistently show identifiable off-track features, while those below it do not. Roughly 75\% of smooth \mi{} streams fall below this threshold, compared with $\sim$55\% for \mf{}. The right panel shows the corresponding pericenter trend. These results indicate that even globally smooth streams frequently carry at least one localized off-track feature, consistent with features observed in MW streams such as GD-1 \citep[e.g.,][]{price2018off, bonaca2019spur} and the stream associated with Pal~5 ($r_{\rm peri} \sim 7$--$8$ kpc; \citealt{pearson2017gaps, erkal2017sharper}), whose low pericenter places it well within the disturbed regime of our classification.

\subsubsection{Morphological Classification Across Orbital Space}
\label{sec:results_xi_orbital}

\begin{figure*}[ht]
    \centering
    \includegraphics[width=\linewidth]{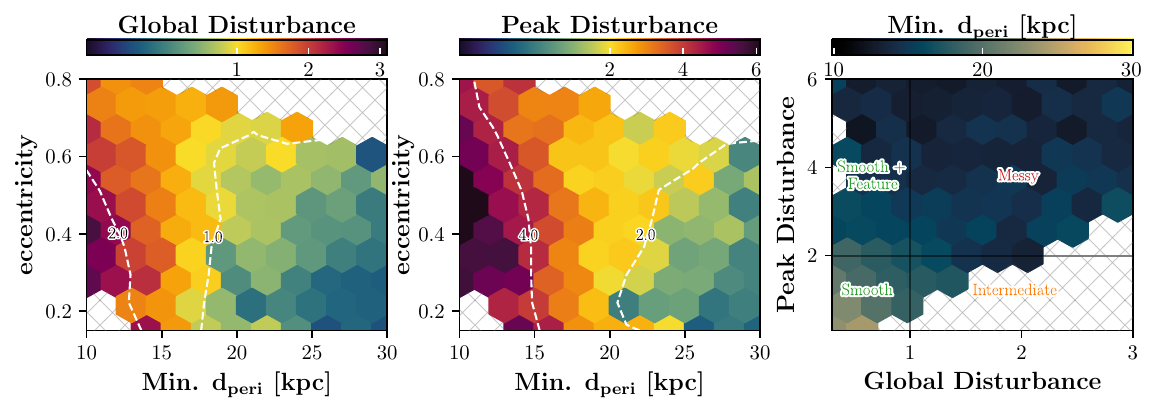}
     \caption{Average \GLOBAL{} $\Dglobal{}$ (left) and \PEAK{} $\Dpeak{}$ (middle) across all streams in pericenter--eccentricity space, color-coded by the average value in each bin (bluer indicates lower, smoother values and, redder indicates higher, messier values). Contours mark $\Dglobal{}= 1, 2$ (left) and $\Dpeak{}= 2, 4$ (middle). Right: \PEAK{} versus \GLOBAL{} for individual streams, color-coded by minimum pericenter. Dashed lines at $\Dglobal{}= 1$ and $\Dpeak{}= 2$ mark the four morphological tiers. Large-pericenter, low-eccentricity orbits produce the smoothest streams and, at small pericenters, nearly circular orbits are systematically messier than more radial ones, consistent with longer time spent near the complex inner potential.}\label{fig:xi_peri_eccen_all}
\end{figure*}  

The pericenter trend in the previous section treats each orbital parameter in isolation. In practice, a stream's morphology depends on the full orbital geometry and not just on how close it approaches the galactic center, but how much time it spends there, which is set by the interplay of pericenter and eccentricity. Figure~\ref{fig:xi_peri_eccen_all} maps \GLOBAL{} (left) and \PEAK{} (middle) across the pericenter--eccentricity plane for the full combined sample.

The dominant trend is with pericenter: the smooth region ($\Dglobal{}\leq 1$) is confined to pericenters $\gtrsim$15--20 kpc, and the messy region ($\Dglobal{}> 2$) dominates at $\lesssim 10$ kpc across all eccentricities, consistent with Figure~\ref{fig:med_xi_peri_frac}. The eccentricity dependence is more subtle and changes sign with radius. At small pericenters, nearly circular orbits are systematically more disturbed than radial ones: progenitors on circular orbits at $\lesssim 15$ kpc are confined to the inner regions and continuously accumulate perturbations from the inner disk structure. More radial orbits at the same pericenter spend a significant fraction of their period at apocenter, far from these structures, effectively diluting the integrated perturbation.

At larger pericenters the pattern reverses: circular orbits remain at a roughly constant galactocentric distance in a quiescent region, while more eccentric orbits at the same pericenter spend significant time at large apocenter before returning for repeated pericentric passages, accumulating perturbations at each return. The transition scale of $\sim$15 kpc is suggestive --- it coincides with the extent of the massive disk in \mm{} \citep{sanderson2020synthetic, mccluskey2024disc} --- but we lack the controlled experiments needed to attribute the threshold to any single potential component. We defer a causal decomposition to future work. The \PEAK{} (middle) panel in Figure~\ref{fig:xi_peri_eccen_all} shows no comparably strong eccentricity trend, though circular orbits at large pericenters still tend to favor lower values.

The right panel of Figure~\ref{fig:xi_peri_eccen_all} motivates a four-tier refinement of the three-tier classification introduced above. Within the smooth population ($\Dglobal{}\leq 1$), we distinguish \emph{Smooth} streams, which satisfy $\Dglobal{}\leq 1$ and $\Dpeak{}\leq 2$ (no significant off-track features anywhere along the track), from \emph{Smooth + Feature} streams, which satisfy $\Dglobal{}\leq 1$ but $\Dpeak{}> 2$ (an otherwise clean track with at least one detectable off-track departure such as a spur or kink). These four tiers map directly onto what imaging surveys would see: smooth streams appear as featureless thin ridgelines, smooth + feature streams show a clean track with an isolated spur or kink, intermediate streams have patchy, partially disrupted structure and, messy streams are globally disturbed. As expected the Smooth tier is concentrated at larger pericenters and slightly more circular orbits (Figure~\ref{fig:streams_peri_eccen_tier4} in Appendix~\ref{app:supp}). Table~\ref{tab:xi_classification} summarizes the thresholds and population fractions.

\setlength{\tabcolsep}{4pt}  % default is 6pt
\begin{table}[ht]
\caption{Four-tier morphological classification based on the \GLOBAL{} and \PEAK{} off-track metrics. Per-halo fractions vary with
host assembly history. See Section~\ref{sec:results_xi_peri} for halo-by-halo trends.}
\label{tab:xi_classification}
\begin{tabular}{lccc}
\textbf{Class} & \boldmath$\Dglobal{}$ & \boldmath$\Dpeak{}$ & \textbf{Fraction (all halos)} \\
\midrule
\midrule
Smooth       & $\leq 1$             & $\leq 2$      & 26\%\\
Smooth + Feature    & $\leq 1$             & $> 2$   & 13\% \\
Intermediate       & $1 < \cdot \leq 2$ & ---        & 27\% \\
Messy              & $> 2$          & ---        & 34\% \\
\bottomrule
\end{tabular}
\end{table}

In summary, stream morphological complexity is primarily set by the progenitor's pericentric distance, with a secondary dependence on the host's assembly history. The $\sim$15 kpc pericenter threshold separating smooth from messy streams is robust across halos, though its precise value is potential-dependent: the most isolated host (\mi{}) produces the highest smooth fractions, while the merger hosts (\mf{}, \mb{}) show elevated off-track structure at all pericenters. Across the combined sample, the Smooth tier with no detectable off-track features anywhere along their track, comprises only 26\% of the population, while the Messy tier alone accounts for 34\%. Roughly three quarters of all streams show some departure from a clean, narrow track, whether localized or global. This is the first piece of the complexity floor: most streams carry off-track structure. The remaining diagnostics,
width variation and along-track density, will further narrow the population of genuinely featureless streams. For the remainder of the analysis, we focus primarily on the smooth populations as the most promising targets for constraining DM substructure from stream morphology.

\subsection{Width Variation Along the Stream}\label{sec:results_width}

The off-track metrics in Section~\ref{sec:results_nonGauss} identify whether a stream's cross-track profile is disturbed, but they do not capture how much the stream's width itself varies along its length. A stream can appear well-behaved in every segment yet still pinch and flare significantly from one part of the track to the next. The width variation $C_w$ (Section~\ref{sec:local_NG}) quantifies this: it measures the relative standard deviation of the per-segment width along the track, and is sensitive to any perturbation that modulates the stream's thickness without displacing stars off the main ridgeline. Figure~\ref{fig:streams_grid_cw_2d} in Appendix~\ref{app:supp} shows a random selection of smooth streams across $C_w$ bins with unequal aspect along $\phi_1$--$\phi_2$ to give the reader a visual sense of what different levels of width variation look like.

\begin{figure}[ht]
    \centering
    \includegraphics[width=\linewidth]{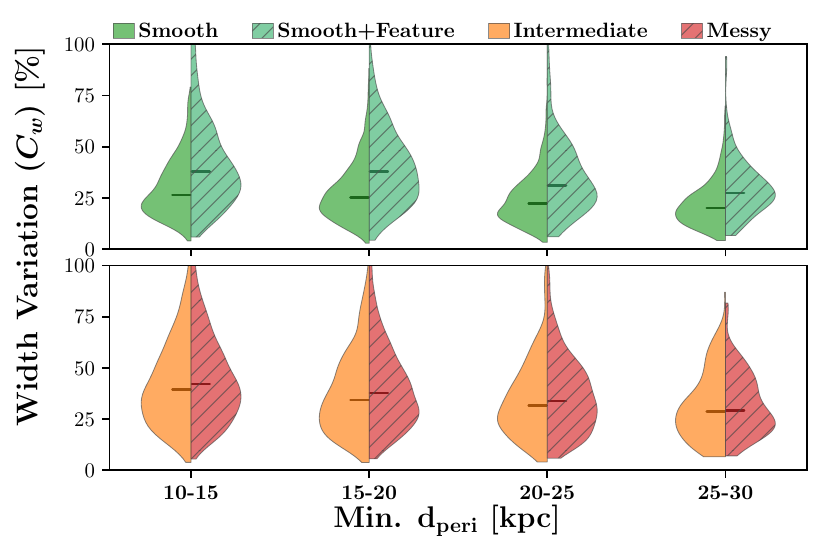}
    \caption{Violin PDFs of width variation $C_w$ in pericenter bins for Smooth (green) and Smooth + Feature (green, hatched) streams (top row), and Intermediate (yellow) and Messy (red, hatched) streams (bottom row), included for completeness. Black lines mark the median of each distribution. Smooth streams show a modest declining trend with pericenter (medians $\sim$26\% to $\sim$20\%). The remaining three categories show no trend and broadly match each other with medians near $\sim$35\%.}
    \label{fig:cw_violin}
\end{figure}

As with off-track structure, we expect streams whose progenitors orbit closer to the galactic center to experience stronger tidal forcing at each pericentric passage, leaving a detectable imprint in their width profiles. Figure~\ref{fig:cw_violin} shows violin PDFs of width variation $C_w$ in pericenter bins, split by morphological tier (Table~\ref{tab:xi_classification}). For Smooth streams (green, top row), $C_w$ shows a modest but statistically significant declining trend with pericenter: the median decreases from $\sim$26\% at $d_{\rm peri} \in [10, 15]$ kpc to $\sim$20\% at $d_{\rm peri} \in [25, 30]$ kpc, and the distributions tighten at larger pericenters. This trend is not driven by the larger absolute stream widths at smaller galactocentric radii. $C_w$ measures the relative variation in width from segment to segment along the track, not the width itself, so the declining trend reflects genuinely less uniform width profiles at smaller pericenters. We interpret this as a residual imprint of tidal shocking: streams on orbits with smaller pericenter experience more impulsive tidal forcing at each passage, leaving a detectable signature in their width profiles even when the cross-track distribution remains well-behaved. The eccentricity trend (not shown) is weak and largely absent. The mild signal present is connected to the higher fraction of radial orbits in \mf{} and \mb{} (Figure~\ref{fig:streams_peri_eccen}) rather than eccentricity itself driving $C_w$.

The Smooth + Feature, Intermediate, and Messy tiers (bottom row and hatched distributions) show only minor trend with pericenter and broadly match each other with medians near $\sim$35\%. The similar medians are somewhat misleading: a Smooth + Feature stream has a higher $C_w$ than a Smooth stream not because it is globally broader, but because the localized off-track feature in one or two segments inflates the width variance. This mirrors what we see in the 2D atlas (Figure~\ref{fig:streams_phi2_nonGauss_2d}): streams at $(\Dglobal, \Dpeak) \approx (1.5, 1.75)$ and $(2.0, 2.25)$ and $(2.0, 2.25)$ can appear broadly thick without any dominant localized feature, while Smooth + Feature streams have an otherwise narrow track with one off-track segment driving the width variance (see stream 2 in Table~\ref{tab:sample_nonG} and Figure~\ref{fig:sample_streams_NonGauss}). The similar medians across these three classes reflect different physical origins of width variation rather than equivalent morphologies.

\begin{figure}[ht]
    \centering
    \includegraphics[width=\linewidth]{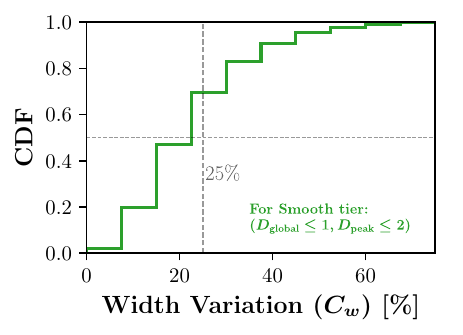}
    \caption{CDF of width variation $C_w$ for Smooth streams. All Smooth streams have $C_w \geq 10\%$, meaning no Smooth stream is perfectly uniform in width. The median is $\sim$22\% with $\sim$50\% of streams below this value.}
    \label{fig:cw_cdf}
\end{figure}

Figure~\ref{fig:cw_cdf} shows the CDF of $C_w$ for Smooth streams. All Smooth streams have $C_w \geq 10\%$, and $\sim$50\% fall below $\sim$22\%. No stream in our sample is perfectly uniform in width. This sets a direct observational expectation: even the cleanest streams in a time-evolving host show $\gtrsim 10\%$ relative width variation along their length. Width variation at this level is in principle detectable in deep photometric surveys with sufficient stellar counts per resolution element, and should be accounted for as a baseline when interpreting width modulations as evidence for subhalo impacts or other localized perturbations \citep[e.g.,][]{erkal2017sharper, balbinot2018devil, banik2019effects,nibauer2024slant}. 

$C_w$ captures modulations in the stream's transverse extent but is by construction insensitive to density variations along the ridgeline itself. We turn to that complementary axis next.

\subsection{Along-Track Density Structure}
\label{sec:results_power}

Gaps, clumps, and density variations along a stream's track are among the most sought-after signatures of DM subhalo encounters \citep[e.g.,][]{carlberg2013gaps, erkal2015forensics, erkal2016number, banik2019effects}. The standard expectation is that a GC stream stripped in a smooth potential should produce a relatively smooth density profile along its length, so that any detected density variations can be attributed to an external perturber. But is this expectation justified in a realistic, time-evolving host? 
Here we test this directly, asking whether Smooth streams (already selected to have clean cross-track profiles) evolved without subhalos perturbers still carry detectable density structure along their track, and at what angular scales. We quantify this with three diagnostics (Section~\ref{sec:power_spectrum}): the RMS fractional density fluctuation along the track ($\mathrm{RMS}_\delta$), the integrated excess power above the Poisson null ($P_{\rm excess}$), and the minimum angular scale with $\mathrm{SNR} \geq 5$ ($\lambda_{\min}$). The analysis in this section is restricted to the Smooth tier ($\Dglobal \leq 1$, $\Dpeak \leq 2$).

\subsubsection{Amplitude of Density Fluctuations}
\label{sec:results_rms_excess}

\begin{figure*}[ht]
    \centering
    \includegraphics[width=\linewidth]{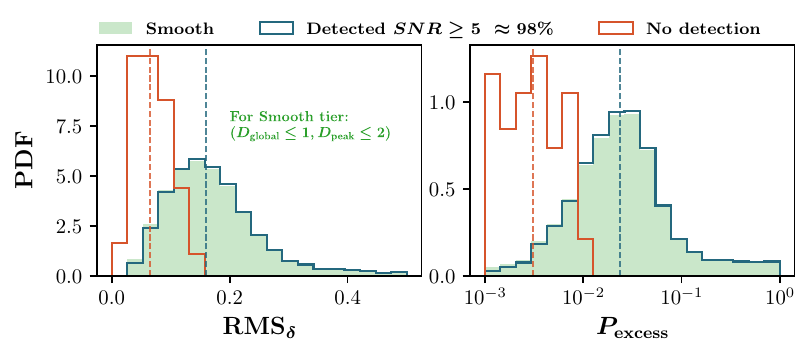}
    \caption{PDF of $\mathrm{RMS}_\delta$ (left) and $P_{\rm excess}$ (right) for Smooth streams (green), split into streams with a detected along-track feature ($\mathrm{SNR} \geq 5$ at some $\lambda$; blue) and those with no detection (orange). Vertical dashed lines mark the median of each subsample. Detected streams have a median $\mathrm{RMS} = 0.17$ and median $P_{\rm excess} = 0.026$, compared to $0.07$ and $0.0035$ for non-detected streams. The two populations are well separated, with detected streams showing $\sim$2.5$\times$ higher RMS amplitude and $\sim$7$\times$ higher integrated excess power.}
    \label{fig:smooth_rms_excess_pdf}
\end{figure*}

How strong are the density variations along the track of a Smooth stream? A stream with large fractional density fluctuations would show visible overdensities and underdensities --- clumps and gaps --- in sufficiently deep photometric data, while one with small fluctuations would appear uniform. We quantify the amplitude with $\mathrm{RMS}_\delta$, the root-mean-square fractional density variation along the track, and $P_{\rm excess}$, the integrated power above the Poisson noise floor (Section~\ref{sec:power_spectrum}). Our particle-spray model ejects stars uniformly in time rather than preferentially at pericenter, likely underestimating the true along-track density variation relative to a full N-body simulation with realistic non-uniform mass loss. The density structure we report here is therefore a conservative lower bound.

Figure~\ref{fig:smooth_rms_excess_pdf} shows the PDF of $\mathrm{RMS}_\delta$ (left) and $P_{\rm excess}$ (right) for Smooth streams (green), split by whether a significant along-track feature was detected ($\mathrm{SNR} \geq 5$; blue) or not (orange). Vertical dashed lines mark the median of each subsample. The two populations are well separated. Streams in the detected sample have a median $\mathrm{RMS}_\delta = 0.17$, corresponding to roughly 17\% peak-to-trough fractional density variation along the track. Streams in the non-detected sample have a median of $0.07$, a factor of $\sim$2.5$\times$ lower. The separation is even more pronounced in $P_{\rm excess}$: streams in the detected sample have a median of $0.026$ compared to $0.0035$ for the non-detected sample, a factor of $\sim$7$\times$. 

These amplitudes show a statistically significant but modest dependence on orbital parameters: streams with smaller pericenters carry slightly stronger density fluctuations, though the trend is driven primarily by the innermost bin ($d_{\rm peri} \in [10, 15]$ kpc, median $\mathrm{RMS}_\delta = 0.192$), while streams beyond $\sim$20 kpc cluster around $0.157$--$0.165$ with no further decrease. A similar pattern holds with eccentricity, where more eccentric orbits show modestly higher amplitudes. $P_{\rm excess}$ follows the same trends. In practical terms, Smooth streams carry real along-track density structure across all orbital bins, with orbit setting only a secondary modulation on its amplitude.

The 17\% RMS in detected streams represents real, statistically confirmed structure. Whether it is photometrically recoverable depends on stellar counts per resolution element: at $\sim$30 stars per bin, Poisson noise alone contribute $\sim$18\% scatter, placing these streams right at the detection threshold of current wide-field surveys \citep[e.g., DES $10\sigma$ depth $r = 23.9$;][]{Bechtol:2026}. LSST-depth observations will be almost 3 magnitudes deeper \citep[$10\sigma$ depth r $\sim 26.75$,][]{ivezic2019lsst}, which will substantially increase stellar counts for nearby streams, pushing the Poisson floor well below $\sim$10\% and making the along-track structure in the majority of our Smooth sample photometrically
recoverable. The non-detected sample at 7\% RMS are not structureless streams in an absolute sense, but recovering their density fluctuations would require $\gtrsim$200 stars per resolution element, challenging even for LSST except for the nearest and richest streams.

\subsubsection{Minimum Detectable Angular Scale}
\label{sec:results_lambda}

\begin{figure}[ht]
    \centering
    \includegraphics[width=\linewidth]{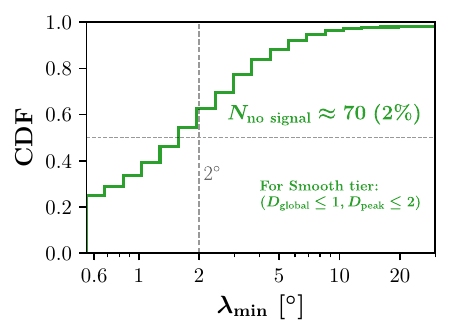}
    \caption{CDF of $\lambda_{\min}$, the minimum detectable angular feature for Smooth streams with a detection. Roughly 2\% of Smooth streams have no detected along-track feature at $\mathrm{SNR} \geq 5$. The median $\lambda_{\min}$ is $\sim$2$^\circ$, with the distribution spanning $0.5^\circ$--$10^\circ$. Most Smooth streams host detectable along-track structure at degree-scale separations.}
    \label{fig:lamba_cdf}
\end{figure}

\begin{figure}[ht]
    \centering
    \includegraphics[width=\linewidth]{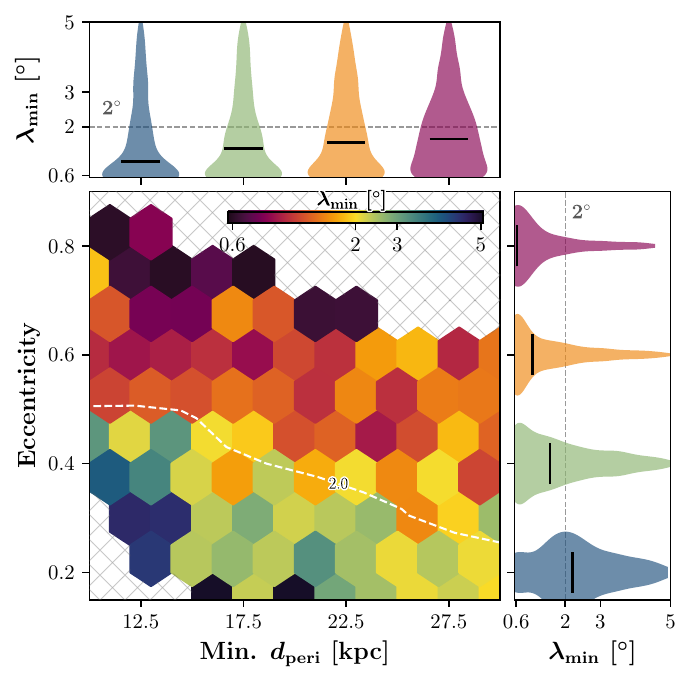}
    \caption{Main panel: Smooth streams with a detection in pericenter--eccentricity space, color-coded by median $\lambda_{\min}$ per bin (blue: larger angular scales and, red: smaller angular scales). The contour marks $\lambda_{\min} = 2^\circ$. Top panel: violin PDFs of $\lambda_{\min}$ in pericenter bins, larger pericenters show a subtle shift toward larger separations. Right panel:  violin PDFs in eccentricity bins. Circular orbits ($e \lesssim 0.3$) show systematically larger $\lambda_{\min}$, while more eccentric orbits concentrate structure at smaller angular scales. The eccentricity trend is more pronounced than the pericenter trend for this metric.}
    \label{fig:lambda_corner_violin}
\end{figure}

\begin{figure*}[ht]
    \centering
    \includegraphics[width=\linewidth]{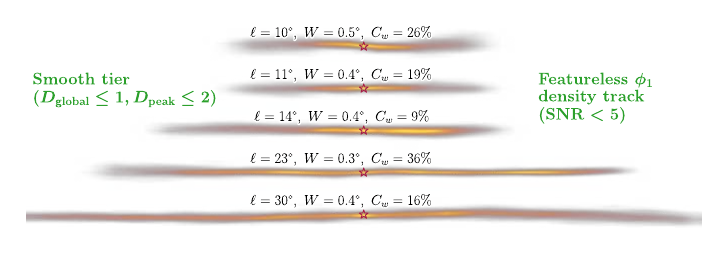}
    \caption{Five randomly selected Smooth streams with no detected along-track feature ($\mathrm{SNR} < 5$ at all $\lambda$), sorted by increasing length ($10^\circ$--$30^\circ$). Stream widths span $0.3^\circ$--$0.5^\circ$ with width variation $C_w \sim15$--$35\%$.}
    \label{fig:featureless_streams}
\end{figure*}

If Smooth streams carry real density structure, at what angular scales does it appear and could it be confused with the imprint of a DM subhalo encounter? The minimum detectable angular scale $\lambda_{\min}$ answers this directly: it is the smallest angular separation at which the along-track density power exceeds the Poisson noise floor at $\mathrm{SNR} \geq 5$ (Section~\ref{sec:power_spectrum}). A stream with $\lambda_{\min} \sim 2^\circ$ hosts structure at scales comparable to the gaps expected from subhalo flybys. A stream with no detection at any scale is the closest our sample comes to a genuinely featureless track.

Figure~\ref{fig:lamba_cdf} shows the CDF of $\lambda_{\min}$ for Smooth streams with a detection. Roughly 2\% of Smooth streams ($\sim$70 out of $\sim$3,800) show no detectable along-track feature at $\mathrm{SNR} \geq 5$, these are the closest to genuinely featureless streams in our sample. Among detected streams, the median $\lambda_{\min}$ is $\sim$2$^\circ$, with most streams spanning roughly $0.5^\circ$--$5^\circ$. This scale sits squarely in the range where DM subhalo flyby encounters are expected to leave imprints \citep[e.g.,][]{erkal2015forensics, erkal2016number}, which typically produce gaps of $1^\circ$--$5^\circ$ depending on subhalo mass and encounter geometry. On the basis of photometric morphology alone, degree-scale density features in a Smooth stream are indistinguishable from subhalo-induced perturbations, motivating the use of additional observables such as kinematics to disentangle their origins.

Figure~\ref{fig:lambda_corner_violin} maps $\lambda_{\min}$ across pericenter--eccentricity space. The pericenter trend is subtle: the distributions across pericenter bins overlap substantially and shift only modestly toward larger $\lambda_{\min}$ at larger pericenters. The eccentricity trend is more pronounced and is the dominant pattern in the main panel. The $\lambda_{\min}= 2^\circ$ contour roughly bisects the plot horizontally, with circular orbits ($e \lesssim 0.3$--$0.4$ in the outer regions, $e \lesssim 0.25$ in the inner regions) showing systematically larger $\lambda_{\min}$ and eccentric orbits concentrating power at smaller angular scales. This is consistent with the cross-track picture in Section~\ref{sec:results_xi_orbital}: more eccentric orbits experience repeated pericentric passages that deposit structure at shorter scales along the track, while circular orbits at large pericenters remain the least dynamically perturbed.

For context, restricting the along-track analysis to Smooth streams is a conservative choice but not the only informative one. Across all four morphological tiers, the detection fraction is similar ($\sim82-86\%$), and the median $\lambda_{\rm min}$ for intermediate and messy (1$^\circ$ and 1.3$^\circ$, respectively) is comparable to the Smooth value, indicating that along-track density structure is largely independent of cross-track morphological complexity. The exception is the Smooth + Feature tier, whose median $\lambda_{\rm min} = 0.61 ^\circ$ is significantly smaller: the same localized perturbation that drives $\Dpeak > 2$ in the cross-track direction also imprints power at shorter along-track scales, consistent with a single impulsive event modulating both axes simultaneously.

Figure~\ref{fig:featureless_streams} shows five randomly selected Smooth streams with no detected along-track feature, sorted by increasing length. These represent the closest analog in our simulated ensemble to the idealized thin, cold streams often assumed in subhalo perturbation analyses \citep[e.g.,][]{bonaca2019spur, nibauer2025measurement, nguyen2025forecasting}. Even these streams show non-negligible width variation along the track ($C_w \sim 15$--$35\%$), a reminder that width variation is a distinct axis of complexity from the cross-track and along-track density metrics. They are thin, coherent, and free of detectable density structure, but not perfectly uniform.

\subsection{Summary of the metrics: The Complexity Floor}\label{sec:complexity_floor}

The three morphological diagnostics developed in this work --- off-track structure, width variation, and along-track density --- each probe a distinct axis of stream complexity. Applied together, they reveal that genuinely featureless streams are extraordinarily rare. Most streams show detectable off-track features somewhere along their track, whether bimodal cross-track profiles, localized spurs, or broad cocoon-like envelopes. Those that pass this first filter still carry 10--25\% width variation from segment to segment. And among the smoothest, narrowest streams, nearly all host overdensities and gaps at angular scales of $\lambda_{\min} \sim 2^\circ$, squarely in the range where DM subhalo encounters are expected to leave imprints \citep{erkal2015forensics, erkal2016number}. Only $\sim$70 streams out of $\sim$15,000 off-track features and no detectable along-track density structure at any scale. Even these carry non-negligible width variation along their length ($C_w \sim 15$--$35\%$).

This establishes a complexity floor: the baseline morphological structure that the host potential imprints on GC streams in the complete absence of DM subhalos or other small-scale perturbers. The floor is conservative in two respects. Our stream coordinate frame minimizes cross-track spread (Section~\ref{sec:stream_length}), so the features we detect appear in the most favorable projection, and our particle-spray model ejects stars uniformly in time, likely underestimating the true along-track density variation. Streams observed from a fixed vantage point, with realistic non-uniform mass loss, would generically appear more complex. 

The floor also varies across the population, depending strongly on pericentric distance and orbital shape, and is modulated by the host potential structure itself. On the basis of photometric morphology alone, host-driven and subhalo-driven perturbations at these scales are indistinguishable. Kinematic information may offer a path forward for separating the two, which we will explore in future work. In Section~\ref{sec:drivers} we examine the physical origins of this complexity through a controlled comparison with a static, axisymmetric host potential, and outline preliminary evidence for the role of the time-evolving disk in the inner halo.

\section{MW Streams in Context}\label{sec:MW_context}

\begin{figure*}[ht]
    \centering
    \includegraphics[width=\linewidth]{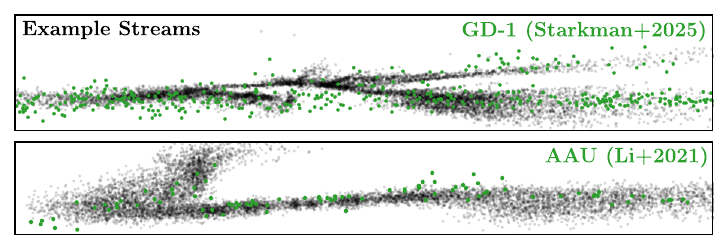}
    \caption{Two simulated GC streams (black) formed in time-evolving host potentials without any DM subhalo encounters, each shown at a viewing angle chosen to reveal a off-track features extending from the main track. Top row: a spur-like feature compared with observed GD-1 members (green) near the spur from \citet{starkman2025stream}, shifted along $\phi_1$ to align the features. Bottom row: a kink-like feature compared with observed AAU members (green) from \citet{li2021broken}, similarly shifted. The morphological similarity in angular scale and structure is notable in both cases, despite neither simulated stream containing any subhalo-induce perturbations.}\label{fig:GD_1_example}
\end{figure*}

The morphological complexity floor established in Section~\ref{sec:results} raises a natural question: do the features in our simulated streams resemble those observed in real MW streams? We explore this with a qualitative comparison, focusing on two iconic examples.

GD-1's spur and density fluctuations have been widely interpreted as evidence for a population of low-mass dark matter subhalos \citep{bonaca2019spur, banik2021novel, nibauer2025measurement}. Our stream catalog contains morphological analogs to this feature. By visually inspecting our stream catalog in galactocentric coordinates, we identified streams with spur-like extensions along their track. We then rotate the viewing angle around the stream spine to reveal the feature in $\phi_1$--$\phi_2$ space. The selected stream orbits with a minimum pericenter of $\sim$10 kpc and maximum apocenter of $\sim$24 kpc. The pericenter of GD-1 is $14~\rm{kpc}$ with an apocenter of $22~\rm{kpc}$ \citep{bonaca2020high}.

Figure~\ref{fig:GD_1_example} (top row) shows the selected stream, with observed GD-1 members from \citet{starkman2025stream} overplotted for comparison (shifted along $\phi_1$ to align the spur regions). The morphological similarity in angular scale and structure is notable. This stream was evolved in a time-evolving host potential with no subhalo encounters of any kind. 

The ATLAS--Aliqa Uma (AAU) stream provides another compelling case. AAU exhibits a prominent kink-like feature along its track \citep{li2021broken, nguyen2025forecasting} as well as large-scale asymmetries that have been attributed to the influence of the LMC \citep{shipp2021measuring}. Figure~\ref{fig:GD_1_example} (bottom row) shows a simulated stream from our catalog, orbiting with a minimum pericenter of $\sim$14 and maximum apocenter of $\sim$25 kpc, with a qualitatively similar kink. 

These are not isolated cases. The bifurcated structure observed in Jhelum \citep{shipp2019proper, bonaca2019multiple} resembles the bimodal cross-track profiles seen in our Intermediate tier (Figure~\ref{fig:streams_phi2_nonGauss_2d}). Large-scale misalignments reported in the Orphan--Chenab stream \citep{erkal2019total, shipp2019proper, koposov2023s}, often attributed to the LMC's influence, find natural analogs in the merger hosts \mf{} and \mb{}, where massive satellite passages leave a detectable imprint on stream morphology across a wide range of orbital radii. Broad, cocoon-like envelopes similar to those around Jhelum \citep{shipp2018stellar} and GD-1 \citep{malhan2019butterfly} appear throughout our Messy tier. Notably, none of the four FIRE-2 hosts develop a strong, long-lived bar \citep{ansar2025bar}, yet the features we find already overlap with the resonant structures predicted from stream-bar interactions \citep{price2016spending, hattori2016shepherding, pearson2017gaps}.

In each case, the host potential produces morphological structure at the angular scales and amplitudes observed in the MW, without invoking any subhalo population.

The spurs, kinks, bifurcations, and density variations that define the observed morphological landscape of MW streams all have analogs in our host-driven catalog, produced without invoking small-scale substructure. In future work, we will add explicit DM subhalo populations to the same host potentials and ask whether the two sources of morphological complexity can be separated with kinematic information along the stream track. Our results demonstrate that streams across a broad range of orbital parameters and host environments are needed to fully characterize the interplay between host-driven and subhalo-driven structure. Next-generation surveys will deliver exactly this, and the host-driven complexity floor we establish here will be essential for interpreting what they find.

\section{What Drives the Complexity Floor?}\label{sec:drivers}

\begin{figure*}[ht]
    \centering
    \includegraphics[width=\linewidth]{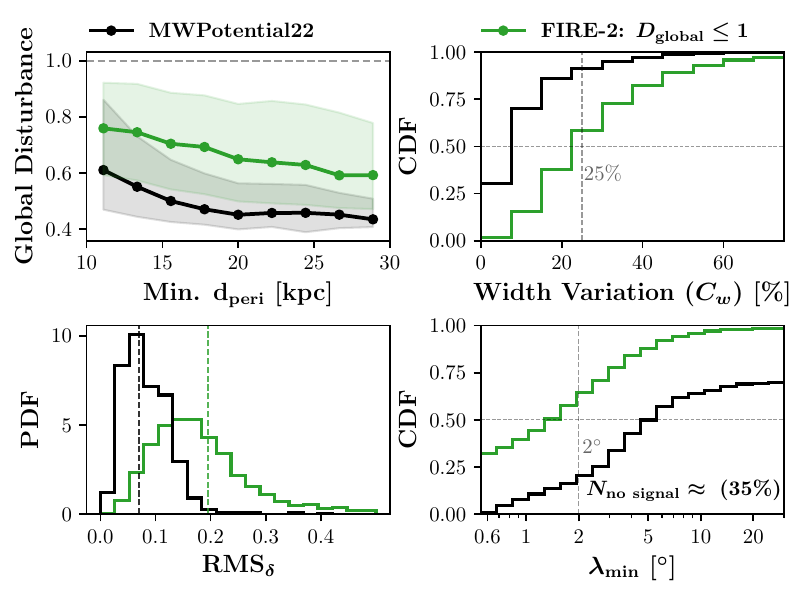}
    \caption{Key morphological metrics for 1,000 GC streams evolved in a static, axisymmetric \texttt{MilkyWayPotential2022} \citep{price2017gala, darragh2023escargot} (black), compared to Smooth streams ($\Dglobal \leq 1$) from the FIRE-2 sample (green). The same orbital cuts and metric definitions are applied to both samples. \textit{Top left:} $\Dglobal$ versus minimum pericentric distance. \textit{Top right:} CDF of width variation $C_w$. \textit{Bottom left:} PDF of $\mathrm{RMS}_\delta$. \textit{Bottom right:} PDF of $\lambda_{\min}$. In the static-axisymmetric potential, $\Dglobal$ remains below 1 (within $1\sigma$) at all pericenters, roughly 75\% of streams have $C_w$ below 20\%, the median $\mathrm{RMS}_\delta$ is $\sim$0.07 for nearly all streams, and $\sim$35\% show no detectable $\lambda_{\min}$ at $\mathrm{SNR} \geq 5$. Only $\sim$15\% of streams show detectable structure around $\sim$2$^\circ$ angular scales. The comparison demonstrates that the morphological complexity documented in our FIRE-2 sample is driven by the time-evolving and/or non-axisymmetric nature of the host potential rather than by the stream formation process itself.}\label{fig:mwpot22_comparison}
\end{figure*}

The complexity floor established in Section~\ref{sec:results} is a population-level outcome: most GC streams develop morphological features even in the absence of small-scale perturbers. Which aspects of the host potential are responsible? Here we present a controlled comparison against a static, axisymmetric host and outline preliminary evidence pointing toward the time-evolving, non-axisymmetric disk as a contributor in the inner halo.

To isolate the role of the time-evolving, non-axisymmetric potential from the stream formation process itself, we apply the same morphology metrics to 1,000 GC streams evolved in a static, axisymmetric \texttt{MilkyWayPotential2022} \citep{price2017gala,darragh2023escargot}, using identical orbital cuts and particle-spray setup as our FIRE-2 runs. This potential represents a smooth, present-day MW-like host with no time evolution, no bar or spiral structure, and no satellite interactions.

 Across all four diagnostic axes, the full population of static-potential streams are systematically less complex than the FIRE-2 Smooth population as shown in Figure~\ref{fig:mwpot22_comparison}. $\Dglobal$ stays below 1 (within $1\sigma$) at all pericenters, $C_w$ is below 20\% for $\sim$75\% of streams, and the median $\mathrm{RMS}_\delta \sim 0.07$ is less than half the FIRE-2 Smooth median. $\sim$35\% of streams show no detectable along-track structure at $\mathrm{SNR} \geq 5$, compared to only $\sim$2\% of FIRE-2 Smooth streams, and only $\sim$15\% of the static-potential streams show detectable structure below $\sim$2$^\circ$, compared to roughly 60\% of FIRE-2 Smooth streams at SNR$\leq5$. The contrast indicates that the complexity we document in Sections~\ref{sec:results} and \ref{sec:MW_context} is shaped by the time-evolving, non-axisymmetric nature of the host potential, rather than by the spray algorithm or the progenitor setup.

Given that the time-evolving, non-axisymmetric host shapes the complexity floor, which components of the potential carry the most weight? The $\sim$15 kpc pericenter threshold separating smooth from disturbed streams (Figure~\ref{fig:med_xi_peri_frac}) coincides with the extent of the massive disk in \mm{} \citep{sanderson2020synthetic, mccluskey2024disc}, pointing to the time-evolving disk as a likely contributor in the inner halo \citep[see also][]{nibauer2024slant}. Our preliminary analysis supports this picture. Stream bifurcations seen in \mi{} \citep[e.g., GC-1 in][]{panithanpaisal2025breaking} coincide with the onset of strong non-axisymmetric ($m=2$) disk structure, and do not occur when the same streams are evolved with the disk restricted to its axisymmetric ($m=0$) component or held static in its full non-axisymmetric form, indicating that both the non-axisymmetry and its time evolution are required (Warren et al., in prep.). The basis function expansion framework we use \citep{arora2022stability, arora2024efficient} is well suited to address this question directly, enabling or disabling individual components to isolate their contribution \citep[as in][]{lilleengen2023effect, brooks2025lmccalls}. A systematic decomposition across the full stream population is the focus of Warren et al.\ (in prep.). A MW-like bar would likely push the inner-halo complexity floor higher still, given its observed influence on streams such as Pal~5 \citep{pearson2017gaps, erkal2017sharper} and Ophiuchus \citep{hattori2016shepherding, price2016spending}. The FIRE-2 hosts used here lack strong, long-lived bars \citep{ansar2025bar}.

For outer-halo streams, where the disk's influence weakens, the asymmetric and time-evolving shape of the DM halo itself \citep{arora2025shaping, darragh2025shaping} and the chaotic nature of orbits in a triaxial potential \citep{yavetz2021separatrix, yavetz2023stream, price2016chaotic} may play a larger role. Massive satellite interactions are a third contributor. The basis function expansion framework captures these passages in its higher-order coefficients, and we find that the merger hosts \mf{} and \mb{} produce more disturbed stream morphologies than the more isolated \mi{} across the full range of pericenters sampled (Figure~\ref{fig:med_xi_peri_frac}). The complexity floor we measure is the integrated signature of these drivers acting together. Disentangling their individual contributions, and separating them from a subhalo-driven baseline, is the next step toward a complete picture of stream complexity in MW-mass hosts.

\section{Summary and Conclusions} \label{sec:disc_conc}

Stellar streams from disrupted GCs are among the most promising probes of DM substructure at the low-mass end of the halo mass function, where subhalos are expected to be entirely dark \citep[e.g.,][]{ibata2002uncovering, johnston2002lumpy, carlberg2012dark}. The power of this approach rests on the assumption that GC streams are intrinsically thin, cold, and relatively smooth, so that any observed gaps, spurs, or density variations can be attributed to subhalo encounters. But streams also trace the gravitational potential of their host, and MW-mass galaxies are not smooth, static, or symmetric: they harbor bars, spiral arms, massive disks, and undergo satellite interactions that reshape the potential on timescales comparable to stream orbital periods \citep[e.g.,][]{price2016spending, hattori2016shepherding,amorisco2016gaps, erkal2017sharper, pearson2017gaps, banik2019effects, shipp2021measuring, lilleengen2023effect, brooks2025lmccalls, arora2025shaping, hunt2025milky}. Establishing the morphological baseline that the host potential itself produces is therefore essential before stream perturbations can be confidently attributed to DM subhalos.

In this work, we have simulated a population of ${\sim}15,000$ GC streams across four MW-mass halos from the FIRE-2 \textit{Latte} suite \citep{wetzel2023public, wetzel2025second}, spanning a range of assembly histories from isolated hosts to systems with recent LMC-mass and Sagittarius-mass mergers \citep{garavito2024corotation, arora2025shaping}. Streams are evolved in time-dependent potentials modeled with basis function expansions \citep{arora2022stability, arora2024efficient} that capture the secular evolution of the halo, disk, bar, and large-scale structure while deliberately excluding small-scale perturbers such as DM subhalos and giant molecular clouds. We developed a suite of quantitative morphology metrics and applied them at a population level to establish the \emph{complexity floor}: the baseline morphological structure present even in the absence of any subhalo population. Our key conclusions are:  

\begin{itemize}
    \item \textbf{\textit{GC streams are morphologically complex from the host potential alone.}} Across the combined sample, only 26\% of streams qualify as Smooth (no significant off-track features anywhere along the track), while 34\% are Messy with globally disrupted morphologies (Table~\ref{tab:xi_classification}). Roughly three quarters of all streams show some departure from a clean, narrow track, whether a localized spur or kink (Smooth + Feature, 13\%) or broad structural complexity (Intermediate, 27\%). Even the Smooth tier is not featureless: these streams carry a median width variation of $\sim$22\% (Figure~\ref{fig:cw_cdf}), with no stream below below 10\%, and median along-track density fluctuations of $\sim$17\% RMS (Figure~\ref{fig:smooth_rms_excess_pdf}). Only $\sim$2\% of Smooth streams show no detectable along-track density structure at our sensitivity threshold (Figure~\ref{fig:lamba_cdf}). These fractions are specific to our sampled orbital distribution and host potentials and should not be interpreted as direct predictions for the MW GC stream population.

    \item \textbf{\textit{Pericentric distance is the primary predictor of stream morphology.}} Streams whose progenitors orbit closer to the galactic center experience stronger perturbations from the complex inner galaxy structures and complete more orbital periods over the integration time, producing systematically more disturbed morphologies. Across all four halos, a pericenter threshold of $\sim$15 kpc robustly separates the smooth and disturbed population (Figure~\ref{fig:med_xi_peri_frac}). At fixed pericenter, the eccentricity dependence reverses with radius: nearly circular orbits in the inner halo ($\lesssim$15 kpc) are more disturbed than eccentric ones because they never escape to larger radii, while circular orbits in the outer halo remain the least perturbed (Figure~\ref{fig:xi_peri_eccen_all}). However, outer-halo progenitors ($r_{\rm peri} \geq 20$ kpc) have Jacobi radii ($\sim$40--480 pc, median $\sim$78 pc) far exceeding typical GC scales, placing them in a weak stripping regime where some may not produce observable streams at all.
    
    \item \textbf{\textit{Stream morphology varies across host assembly histories.}} All four halos produce both smooth and messy streams with broadly similar orbital trends. However, the most isolated host (\mi{}) produces the highest smooth fractions, while the merger hosts (\mf{}, \mb{}) show elevated disturbance at all pericenters (Figure~\ref{fig:med_xi_peri_frac}). These halo-to-halo differences reflect distinct assembly histories and potential structures rather than differences in the orbital sampling.

    \item \textbf{\textit{Along-track density structure appears at angular scales where DM subhalo encounters leave imprints.}} The median minimum detectable angular scale for Smooth streams is $\lambda_{\min} \sim 2^\circ$ (Figure~\ref{fig:lamba_cdf}), squarely in the $1^\circ$--$5^\circ$ range predicted for gaps from DM subhalo flybys \citep{erkal2015forensics, erkal2016number}. On the basis of photometric morphology alone, degree-scale density features in a Smooth stream due to the host potential are indistinguishable from subhalo-induced perturbations. More eccentric orbits concentrate structure at smaller angular scales, while circular orbits at large pericenters show systematically larger $\lambda_{\min}$ (Figure~\ref{fig:lambda_corner_violin}).

    \item \textbf{\textit{Observed MW stream features have analogs in our host-driven catalog.}} We identify spur-like features comparable in morphology to GD-1's spur, and kink-like features resembling that of the ATLAS--Aliqa Uma stream (Figure~\ref{fig:GD_1_example}). Bifurcations and cocoon-like envelopes similar to Jhelum and fanning similar to Pal~5 appear throughout the Intermediate and Messy tiers. These features arise naturally from the host potential, suggesting that some observed stream morphologies may not require invoking subhalo encounters.

    \item \textbf{\textit{The complexity floor is shaped by the time-evolving, non-axisymmetric host.}} Streams evolved in a static, axisymmetric \texttt{MilkyWayPotential2022} \citep{price2017gala, darragh2023escargot} show systematically less complexity across all diagnostic axes (Figure~\ref{fig:mwpot22_comparison}). The $\sim$15 kpc pericenter threshold separating smooth from disturbed streams coincides with the extent of the massive disk in \mm{}, and preliminary tests with the disk's non-axisymmetric structure selectively disabled point to the time-evolving, non-axisymmetric disk as a likely contributor in the inner halo (Warren et al., in prep.). Other potential drivers, such as the time-evolving asymmetric DM halo or chaotic orbits in a triaxial potential, may play a larger role for outer-halo streams.
\end{itemize}

These results have important implications for programs that use streams to constrain the DM subhalo mass function \citep[e.g.,][]{bonaca2019spur, banik2021novel, nibauer2025measurement, nguyen2025forecasting}. The features we document are not fine-tuned but emerge generically from the secular tidal field of the host across a broad range of orbits and assembly histories. The complexity floor we measure motivates incorporating a host-driven morphological baseline into stream perturbation analyses. The morphological complexity arising from the time-dependent host potential and from subhalo passages likely precludes a clean interpretation of features found in any single stream. Disentangling the two sources of structure may only be possible in a statistical sense, across large populations of streams spanning diverse orbits and host environments. 

Our Lagrange-stripping/particle-spray stream formation model is intentionally simple (Section~\ref{sec:caveats}). It captures the large-scale tidal morphology from collisionless stripping but releases particles uniformly in time rather than modeling if/when stars actually escape the progenitor. It also does not model the progenitor's internal collisional evolution. These processes are known to influence stream density, width, and kinematics \citep[e.g.,][]{erkal2017sharper, balbinot2018devil, banik2021n, weatherford2024stellar, panithanpaisal2025breaking, Phillips_DirectNbody} and could add further morphological structure beyond what we report here. Similarly, our progenitor initial conditions are drawn from a present-day distribution function rather than from cosmologically motivated GC formation models \citep[e.g.,][]{grudic2023greatballs, chen2024catalogue, holm2025catalog, panithanpaisal2025breaking}. While our key trends are presented as functions of orbital parameters and are therefore insensitive to the specific choice of progenitor sampling, the complexity floor we measure is conservative in this respect as well.

The next generation of photometric surveys \citep{ivezic2019lsst, spergel2015wide} will bring the complexity floor into sharp observational focus. LSST-depth observations will push the Poisson floor below $\sim$10\% RMS for nearby streams, making the along-track density structure we find in the vast majority of our Smooth sample photometrically recoverable. The morphological diversity documented here, from subtle width variations to prominent spurs and kinks, will be directly observable in next-generation stream catalogs. Interpreting this wealth of morphological detail will require the kind of host-driven baselines established in this work.

Kinematic information may offer a path forward for separating host-driven from subhalo-driven perturbations. Proper motions and line-of-sight velocities along the stream track are sensitive to the velocity coherence and phase-space structure of perturbations, which may differ between impulsive subhalo encounters and the secular tidal field of the host. In future work, we will introduce explicit DM subhalo populations into the same host potentials to generate perturbed stream catalogs and explore whether the full 6D phase-space information can disentangle subhalo-driven perturbations from the various host-driven sources of complexity, including the bar, disk, halo asymmetry, and satellite interactions, each of which may leave distinct signatures. The complexity floor we establish here is the foundation for that separation.

\begin{contribution}
AA led the project, ran the stream simulations, developed the morphology metrics, performed the analysis, and wrote the paper. PSF and NS contributed to the development and interpretation of the morphology metrics and editing of the manuscript. JN and APW conceived the project and provided guidance throughout. VR helped with the production pipelines for the stream simulations. EV performed detailed testing and validation of the integration methods and is the developer and maintainer of the \texttt{Agama} code used for potential modeling and orbit integration. JK identified the observed MW stream analogs presented in Section~\ref{sec:MW_context}. LCM helped run a test suite of simulations. All authors provided feedback and comments on the manuscript.
\end{contribution}

\begin{acknowledgments}
We thank Robyn E. Sanderson, Alex Drlica-Wagner, Ting S. Li, Tom Quinn, and Adriana Dropulic for valuable discussions that shaped this work. AA also acknowledges the FIRE collaboration for access to simulation data, and the Scientific Computing Team at the Flatiron Institute for computing resources. AA acknowledges support from Gordon and Betty Moore foundation. The DiRAC Institute is supported through generous gifts from the Charles and Lisa Simonyi Fund for Arts and Sciences, Janet and Lloyd Frink, and the Washington Research Foundation. JN is supported by a National Science Foundation Graduate Research Fellowship, Grant No. DGE-2039656. Any opinions, findings, and conclusions or recommendations expressed in this material are those of the author(s) and do not necessarily reflect the views of the National Science Foundation. SP was supported by a research grant (VIL53081) from VILLUM FONDEN and co-funded by the European Union (ERC, BeyondSTREAMS, 101115754). Views and opinions expressed are, however, those of the author(s) only and do not necessarily reflect those of the European Union or the European Research Council. Neither the European Union nor the granting authority can be held responsible for them. EV acknowledges support from an STFC Ernest Rutherford fellowship (ST/X0040066/1), and believes that STFC doesn't particularly care about his opinions and views regarding stellar streams. AW received support from via NSF CAREER award AST-2045928. JB acknowledges support from the National Aeronautics and Space Administration Grant No. 80NSSC24K1227.

This research made use of the Claude language model (Anthropic) for assistance with text editing and code debugging. All scientific conclusions are the authors’ own.
\end{acknowledgments}

\software{
This work made use of the following software packages: \texttt{python} \citep{python}, \texttt{Jupyter} \citep{kluyver2016jupyter}, \texttt{Agama} \citep{vasiliev2019agama}, \texttt{numpy} \citep{numpy}, \texttt{scipy} \citep{scipy},  \texttt{matplotlib} \citep{matplotlib}, \texttt{pandas} \citep{mckinney2011pandas}, \texttt{cmasher} \citep{cmasher}, and \texttt{healpy} \citep{healpy2019}. 
This research has made use of the Astrophysics Data System, funded by NASA under Cooperative Agreement 80NSSC21M00561.}

\appendix

\section{Stochastic Noise in the Reconstructed Force Field}\label{app:bfe_shot_noise}

In this section, we calibrate the stochastic noise inherent in the BFE reconstruction of each FIRE-2 snapshot. The basis-function coefficients are least-squares projections of a finite particle distribution onto a truncated basis, and finite particle count contributes a shot-noise component that adds to the smooth force field. We test whether this noise has amplitudes and spatial scales relevant for stream morphology.

\subsection{Bootstrap construction of noise realizations}\label{app:bootstrap}

We generate independent realizations of the BFE noise field by bootstrap-resampling the particle population at representative snapshots spanning the integration window. At each snapshot, we refit the BFE coefficients following the procedure described in Section~\ref{sec:pot_model} \citep{arora2022stability, arora2024efficient}, with identical $l_\mathrm{max} = m_\mathrm{max} = 4$, identical radial and meridional grids, and the same host-centric principal-axis frame. For each bootstrap realization $b = 1, \dots, N_\mathrm{boot}$, we draw $N$ particle indices with replacement from the snapshot and refit. We adopt $N_\mathrm{boot} = 50$.

We evaluate the total force $\mathbf{F}^{(b)}(\mathbf{x}, t_k)$ from each realization on a Fibonacci spiral grid of 50,000 directions across 15 logarithmically spaced radial shells from 10 to 30 kpc, spanning the orbital range of our stream sample. The Fibonacci spiral places points at uniform increments in $\cos\theta$ and golden-angle increments in azimuth, yielding a near-uniform angular distribution on the sphere without the pole crowding of a regular $(\theta, \phi)$ grid \citep[e.g.,][]{gonzalez2010measurement}. At each grid point $\mathbf{x}_i$ we compute the realization mean
\begin{equation}\label{eq:bfe_mean}
    \langle \mathbf{F}\rangle (\mathbf{x}_i, t_k) = \frac{1}{N_\mathrm{boot}}\sum_{b=1}^{N_\mathrm{boot}} \mathbf{F}^{(b)}(\mathbf{x}_i, t_k),
\end{equation}
the residual $\delta\mathbf{F}^{(b)} = \mathbf{F}^{(b)} - \langle\mathbf{F}\rangle$, and the pointwise noise amplitude
\begin{equation}\label{eq:bfe_sigma}
    \sigma_F^2(\mathbf{x}_i, t_k) = \frac{1}{N_\mathrm{boot} - 1}\sum_{b=1}^{N_\mathrm{boot}} |\delta\mathbf{F}^{(b)}(\mathbf{x}_i, t_k)|^2.
\end{equation}
The fractional noise $\sigma_F / |\langle\mathbf{F}\rangle|$ characterizes the amplitude. To characterize the spatial structure, we compute the correlation of the noise vector field,
\begin{equation}\label{eq:bfe_corr}
    C(\Delta r, t_k) = \frac{\langle \delta\mathbf{F}^{(b)}(\mathbf{x}_1) \cdot \delta\mathbf{F}^{(b)}(\mathbf{x}_2)\rangle_{b,\,|\mathbf{x}_1 - \mathbf{x}_2| = \Delta r}}{\sigma_F(\mathbf{x}_1, t_k)\,\sigma_F(\mathbf{x}_2, t_k)},
\end{equation}
where the bracket denotes an average over bootstrap realizations and over pairs of grid points with separation $\Delta r$. By construction $C(0, t_k) = 1$ and $C \to 0$ at separations beyond which the noise is uncorrelated. The differential force variance between two points separated by $\Delta r$ is $\langle |\delta\mathbf{F}_1 - \delta\mathbf{F}_2|^2\rangle = 2 \sigma_F^2 [1 - C(\Delta r)]$, the quantity that controls differential perturbations across a stream.

\subsection{Noise amplitude and spatial structure}\label{app:bfe_results}
 
\begin{figure*}[ht]
    \centering
    \includegraphics[width=\linewidth]{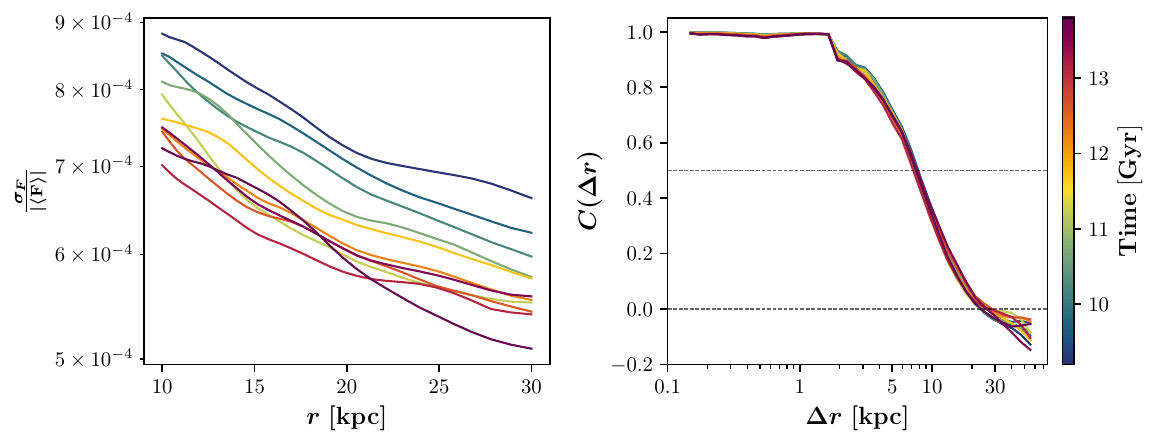}
    \caption{Bootstrap shot-noise calibration of the BFE potentials, computed by refitting each snapshot 50 times with bootstrap-resampled particles. \textit{Left:} median fractional force noise $\sigma_F / |\boldsymbol{\langle \bf F \rangle}|$ as a function of galactocentric radius, with one curve per snapshot (colored). The amplitude is sub-percent across the orbital range and decreases outward. \textit{Right:} Spatial correlation function $C(\Delta r)$ of the noise vector field. The noise is fully coherent across the stream cross-section and partially decorrelated along its length, scaling with the basis angular resolution $r/l_\mathrm{max}$.}
    \label{fig:bfe_noise}
\end{figure*}
 
Figure~\ref{fig:bfe_noise} shows $\sigma_F / |\langle\mathbf{F}\rangle|$ as a function of galactocentric radius (left) and $C(\Delta r)$ as a function of pair separation (right) across snapshots spanning the integration window. The fractional noise sits at $5$--$7 \times 10^{-4}$ across $10 < r < 30$ kpc, varies by less than a factor of two across the 5 Gyr window, and decreases monotonically outward as the basis becomes increasingly well-matched to the smoother density profile of the outer halo. The correlation function plateaus at $C \approx 1$ for $\Delta r \lesssim 1.5$ kpc, falls to $1/2$ at $\Delta r \sim 7$ kpc, and crosses zero beyond $\Delta r \sim 25$ kpc. The coherence scale is consistent with the angular resolution of the basis, $\Delta r_\mathrm{coh} \sim r/l_\mathrm{max}$.

We translate the per-snapshot noise into an integrated velocity drift over the 5 Gyr window. The bootstrap resamples particles independently at each snapshot, so realizations at $t_k$ and $t_{k+1}$ are uncorrelated by construction. We therefore measure the per-snapshot noise amplitude directly, but cannot quantify the temporal coherence of the noise in the underlying simulation, where the same particles persist between snapshots. We bound the integrated effect by treating each snapshot's noise as independent, which approximates the noise as a random walk. We note that the spatial structure of the noise (both amplitude and coherence length) is self-similar across snapshots and the fractional amplitude is $\sim 10^{-3}$ throughout, so even under fully coherent temporal accumulation the integrated kick budget remains well below the stream internal velocity dispersion. At $r = 20$ kpc, $|\langle\mathbf{F}\rangle| \sim V_\mathrm{circ}^2/r \sim 2400$ (km/s)$^2$/kpc and $\sigma_F \sim 1.4$ (km/s)$^2$/kpc. The per-snapshot kick is $\Delta v_k \sim \sigma_F \cdot \Delta t_\mathrm{snap} \sim 0.04$ km\,s$^{-1}$ at the cadence $\Delta t_\mathrm{snap} = 25$ Myr. Accumulated over $N_\mathrm{steps} \sim 200$ snapshots, the bulk drift is $\sigma_{\Delta v,\mathrm{bulk}} \sim \sqrt{N_\mathrm{steps}} \cdot \Delta v_k \sim 0.5$ km\,s$^{-1}$, with the differential drift across a 10 kpc separation enhanced by $\sqrt{2[1-C(L)]} \sim 1.2$ to $\sim 0.6$ km\,s$^{-1}$. Both remain below the stream internal velocity dispersion and accumulate as smooth secular drifts rather than the localized velocity offsets characteristic of subhalo encounters.
 
The spatial structure of the noise determines its effect on streams. Across the typical stream width of $\sim$0.1 kpc, the noise is fully coherent, producing no differential kick across the cross-section. Along the typical stream length of $\sim$10 kpc, it is partially decorrelated, with a differential variance factor of $2[1 - C(L)] \sim 1.4$. Even at full decorrelation, the fractional noise of $\sigma_F/|\langle\mathbf{F}\rangle| \sim 10^{-3}$ is three orders of magnitude below unity, and the dominant variation a stream particle experiences along its orbit is the smooth change in the tidal field between pericenter and apocenter, not the noise. The radial trends reported in Section~\ref{sec:results} therefore reflect the genuine evolution of the tidal field rather than accumulated noise: streams with smaller pericenters develop more morphological structure because the smooth force field changes more sharply near pericenter, not because they accumulate larger noise budgets. The BFE shot noise contributes negligibly to stream morphology.

\section{Integration Accuracy of the Time-Interpolated Potentials}\label{app:bfe_noise}

The time-evolving potentials used throughout this work are constructed from multipole and cylindrical-spline expansions fit to each FIRE-2 snapshot, with the basis-function coefficients linearly interpolated in time between consecutive snapshots (\texttt{Agama} \texttt{type=Evolving} potential model with \texttt{interpLinear=true}). This appendix calibrates two distinct sources of integration error introduced by the linear-in-time reconstruction. Throughout, $\mathbf{F}(\mathbf{x},t)$ denotes the total gravitational force per unit mass, $t_k$ the snapshot timestamps, $\Delta t_k = t_{k+1}-t_k$, $\tau = t-t_k$, and dots denote partial derivatives in time at fixed $\mathbf{x}$.

The first concern is the integrator's behavior. Linear interpolation yields a force field that is $C^0$ in $t$, with slope discontinuities at every $t_k$, while \texttt{Agama}'s adaptive 8th-order Runge-Kutta scheme  is formally derived under $C^2$ smoothness. The second concern is the fidelity of the force-field. Linear interpolation reproduces the snapshot values $\mathbf{F}_k = \mathbf{F}(\mathbf{x},t_k)$ and captures the average slope across each interval, but misses the local curvature of the underlying $\mathbf{F}_\mathrm{true}(t)$. We address each concern in turn.

\subsection{Integrator behavior at snapshot boundaries}\label{app:integrator}

We first verify empirically that the slope discontinuities in $\mathbf{F}_\mathrm{lin}(t)$ at $t_k$ do not produce localized integrator artifacts that could mimic the signature of a substructure encounter. We integrate
16 test progenitor orbits backward 5 Gyr from their present-day phase-space coordinates in the
linearly-interpolated potential using four schemes: \texttt{Agama}'s adaptive RK-8, scipy's DOP853 (an independent 8th-order RK implementation), scipy's RK45, and a 0.5 Myr fixed-step kick-drift-kick leapfrog. Leapfrog serves as the reference because it is symplectic, so its error does not rely on the smoothness of $\mathbf{F}$ that the RK schemes assume. If kinks were amplifying RK error terms involving high derivatives of the force, the RK trajectories would drift away from leapfrog with the residual concentrating at $t_k$ crossings.

Across the 16 orbits, the maximum position residual between any RK scheme and leapfrog over the integration window has median $2 \times 10^{-3}$ kpc and worst case $3 \times 10^{-2}$ kpc, well below the typical stream width of $\sim 0.1$ kpc. Maximum velocity residuals have median $1.4 \times 10^{-2}$ km\,s$^{-1}$ and worst case $0.23$ km\,s$^{-1}$. Binning residuals into $\pm 1$ Myr windows around each $t_k$ and comparing to inter-snapshot regions, we find on-kink to off-kink ratios of $0.95$--$0.96$ for both position and velocity across all three RK schemes. The residuals are uniform in time relative to $t_k$ which implies that the integrator does not concentrate error at the discontinuities. We conclude that the slope kinks introduce no detectable spurious structure into the integrated trajectories.

\subsection{Force-field residual from linear interpolation of forces}\label{app:curvature}

We now quantify how far the linearly-interpolated velocity field departs from a smoother reconstruction of the same FIRE snapshots. Taylor expanding the true force around $t_k$ at fixed $\mathbf{x}$,

\begin{equation}\label{eq:F_true_taylor}
    \mathbf{F}_\mathrm{true}(t) = \mathbf{F}_k + \dot{\mathbf{F}}_k \tau
    + \tfrac{1}{2} \ddot{\mathbf{F}}_k \tau^2 + \mathcal{O}(\tau^3),
\end{equation}
and constructing the linear chord through the snapshot endpoints,

\begin{equation}\label{eq:F_lin_chord}
    \mathbf{F}_\mathrm{lin}(t) = \mathbf{F}_k + \frac{\mathbf{F}_{k+1} - \mathbf{F}_k}{\Delta t_k} \tau,
\end{equation}
substitution of the Taylor expansion of $\mathbf{F}_{k+1}$ into the chord slope gives $(\mathbf{F}_{k+1} -\mathbf{F}_k)/\Delta t_k = \dot{\mathbf{F}}_k + \tfrac{1}{2} \ddot{\mathbf{F}}_k \Delta t_k +
\mathcal{O}(\Delta t_k^2)$. The constant and $\dot{\mathbf{F}}_k$ terms cancel against the truth. The chord absorbs the average slope across the interval exactly. Only the curvature is nonzero,

\begin{equation}\label{eq:F_residual}
    \mathbf{F}_\mathrm{lin}(t) - \mathbf{F}_\mathrm{true}(t)
    = \tfrac{1}{2} \ddot{\mathbf{F}}_k \tau (\Delta t_k - \tau)
    + \mathcal{O}(\Delta t_k^3).
\end{equation}

This is the Lagrange remainder for linear interpolation, vanishing at both endpoints and parabolic in between, with peak amplitude $\tfrac{1}{8} |\ddot{\mathbf{F}}_k| \Delta t_k^2$ at the interval midpoint. Integrating over one interval yields the per-knot velocity bias,
\begin{equation}\label{eq:dvk}
    \delta\mathbf{v}_k = \int_{t_k}^{t_{k+1}}
    \!\!\big[\mathbf{F}_\mathrm{lin}(t) - \mathbf{F}_\mathrm{true}(t)\big] \, dt
    = \tfrac{1}{12} \ddot{\mathbf{F}}_k \Delta t_k^{\,3} + \mathcal{O}(\Delta t_k^{\,4}).
\end{equation}

We accumulate $\delta\mathbf{v}_k$ over the integration window $[8.8, 13.8]$ Gyr, which contains $\sim200$ snapshots with median $\Delta t_k = 22$ Myr. We estimate $\ddot{\mathbf{F}}_k$ via the non-uniform centered second difference of the snapshot forces and report the cumulative coherent and random-walk bias,
\begin{equation}\label{eq:dv_cum}
    \Delta \mathbf{v}_\mathrm{coh}(\mathbf{x}) = \sum_k \delta\mathbf{v}_k(\mathbf{x}),
    \qquad
    \Delta v_\mathrm{rw}(\mathbf{x}) = \left[\sum_k |\delta\mathbf{v}_k(\mathbf{x})|^2\right]^{1/2}.
\end{equation}

We evaluate these quantities at three fixed locations spanning galactocentric distances of 12, 18, and 26 kpc, and along the integrated backward trajectories of the 16 progenitor orbits at both the orbit position and at $\pm100$ pc offsets perpendicular to the velocity. As an independent cross-check, we build a natural cubic spline through the snapshot forces $\mathbf{F}_k$ at each fixed probe and integrate the residual $\mathbf{F}_\mathrm{lin}(t) -\mathbf{F}_\mathrm{cub}(t)$ across each interval by Simpson's rule. The cubic and Lagrange estimators of $|\delta\mathbf{v}_k|$ agree to within $1\%$ across the three probe radii, confirming that Eq.~\ref{eq:dvk} is a faithful estimate of the linearity-assumption residual independent of the finite-difference scheme used for $\ddot{\mathbf{F}}_k$.

Accumulated over the full 5 Gyr backward integration along each of the 16 progenitor orbits, the on-track velocity bias $|\Delta \mathbf{v}_\mathrm{coh}|/v_\mathrm{orb}$ (where $v_\mathrm{orb}$ is the orbital velocity at that point.) remains below 0.1\% in all cases, and the across-stream differential bias at $\pm 100$ pc offsets is roughly an order of magnitude smaller. The largest residuals occur on orbits with the smallest pericenters ($r_\mathrm{peri} \lesssim 12$ kpc), where the inner halo evolves on the shortest timescales and $|\ddot{\mathbf{F}}|$ is largest.

\subsection{Comparison with nearest-snapshot interpolation}\label{app:step}
 
A complementary check is to replace linear interpolation with nearest-snapshot ``step'' interpolation (\texttt{interpLinear=false}), in which $\mathbf{F}_\mathrm{step}(t)$ takes the value at whichever snapshot is closer to $t$. Within an interval, the residual is no longer pinned at both endpoints,
\begin{equation}\label{eq:F_step}
    \mathbf{F}_\mathrm{step}(t) - \mathbf{F}_\mathrm{true}(t)
    = -\dot{\mathbf{F}}_k \tau - \tfrac{1}{2} \ddot{\mathbf{F}}_k \tau^2 + \mathcal{O}(\tau^3)
    \qquad (\tau < \Delta t_k / 2),
\end{equation}
with an analogous expression around $t_{k+1}$ for $\tau \geq \Delta t_k / 2$. The pointwise residual is therefore $\mathcal{O}(\dot{\mathbf{F}}_k \Delta t_k)$, larger than linear's $\mathcal{O}(\ddot{\mathbf{F}}_k \Delta t_k^2)$ by a factor $T_\mathrm{evol}/\Delta t_k \gg 1$, where $T_\mathrm{evol}$ is the local timescale of halo evolution. The integral over the full interval, however, benefits from antisymmetric cancellation across the handoff at $t_k + \Delta t_k / 2$,

\begin{equation}\label{eq:dvk_step}
    \delta\mathbf{v}_k^\mathrm{step} = \tfrac{1}{8} \ddot{\mathbf{F}}_k \Delta t_k^{\,3} + \mathcal{O}(\Delta t_k^{\,4}),
\end{equation}
recovering the same scaling in $\Delta t_k$ as Eq.~\ref{eq:dvk} with a prefactor 1.5 times larger. This is analogous to comparing midpoint and trapezoidal quadrature, both of which converge as $h^3$ per interval despite differing pointwise behavior.
 
We generate particle-spray streams from 16 progenitors in the step-interpolated potential. The resulting streams show no visual differences in morphology or orbital structure relative to the linearly-interpolated runs, and none that would alter the classification metrics employed in this work. We adopt the linear-interpolation scheme for all production runs. The residuals bounded above are orders of magnitude smaller than the morphological features reported in Section~\ref{sec:results}, confirming that the stream complexity we measure is physical rather than a numerical artifact of potential reconstruction.

\subsection{Time cadence and interpolation order}\label{app:cadence_validation}

To inform the choice of time cadence and interpolation scheme for the runs, we performed a controlled comparison in an idealized setting: an $N = 4 \times 10^6$ isolated halo undergoing a radial-orbit instability into a triaxial state, evolved with a tree-based Poisson solver. We store BFE-reconstructed potential at $\sim 3500$ timesteps over $\sim 10$ orbital periods, then reintegrate test orbits in the BFE-reconstructed potential subsampled at different cadence and compare to the tree-code trajectories. This setup directly mirrors our production pipeline, where the FIRE-2 simulations provide the underlying ``truth'' and the BFE serves as the reconstructed potential for stream integration. The comparison yields several relevant takeaways.

First, the orbit-reconstruction error decreases with finer time sampling but saturates once the difference between the tree-code and BFE-reconstructed force fields dominates. The cadence at which this saturation is reached depends on the orbital timescale at the relevant radii. For our FIRE-2 runs, the empirical residuals measured in Section~\ref{app:curvature} provide the relevant constraint: the $\sim$25 Myr snapshot cadence yields velocity biases below $0.1\%$ of $v_{\rm orb}$ for typical orbits, with the largest residuals on the innermost orbits ($r_{\rm peri} \lesssim 10$ kpc) where the potential evolves on the shortest timescales. Second, linear interpolation between snapshots reduces the orbit-reconstruction error at coarse cadences (comparable to or longer than the characteristic time for large-scale potential variations) compared to nearest-snapshot ``step'' interpolation, consistent with the analysis in Section~\ref{app:step}, and also mitigates the integrator-stuttering effect in which adaptive Runge--Kutta schemes discard and repeat timesteps at force discontinuities. Higher-order time interpolation (quadratic, cubic) yields negligible additional improvement at a greater computational cost. Third, the integrator tolerance contributes negligibly to the cumulative error budget at our production setting (\texttt{Agama}'s default relative tolerance is $10^{-8}$, but the error stays the same even when the tolerance is set to $10^{-4}$, which speeds up the integration by a factor of a few). The dominant uncertainty in the integrated trajectories is the BFE projection itself, calibrated in Appendix~\ref{app:bfe_shot_noise}. Finally, the alternative DPRKN8 integration method available in \texttt{Agama} is more computationally efficient than DOP853 and less prone to performance degradation for nearest-snapshot potentials.

\section{Validating the Host Potential and Stream Formation Models}\label{app:force_err_Nbody}
\begin{figure*}[ht]
    \centering
    \includegraphics[width=\linewidth]{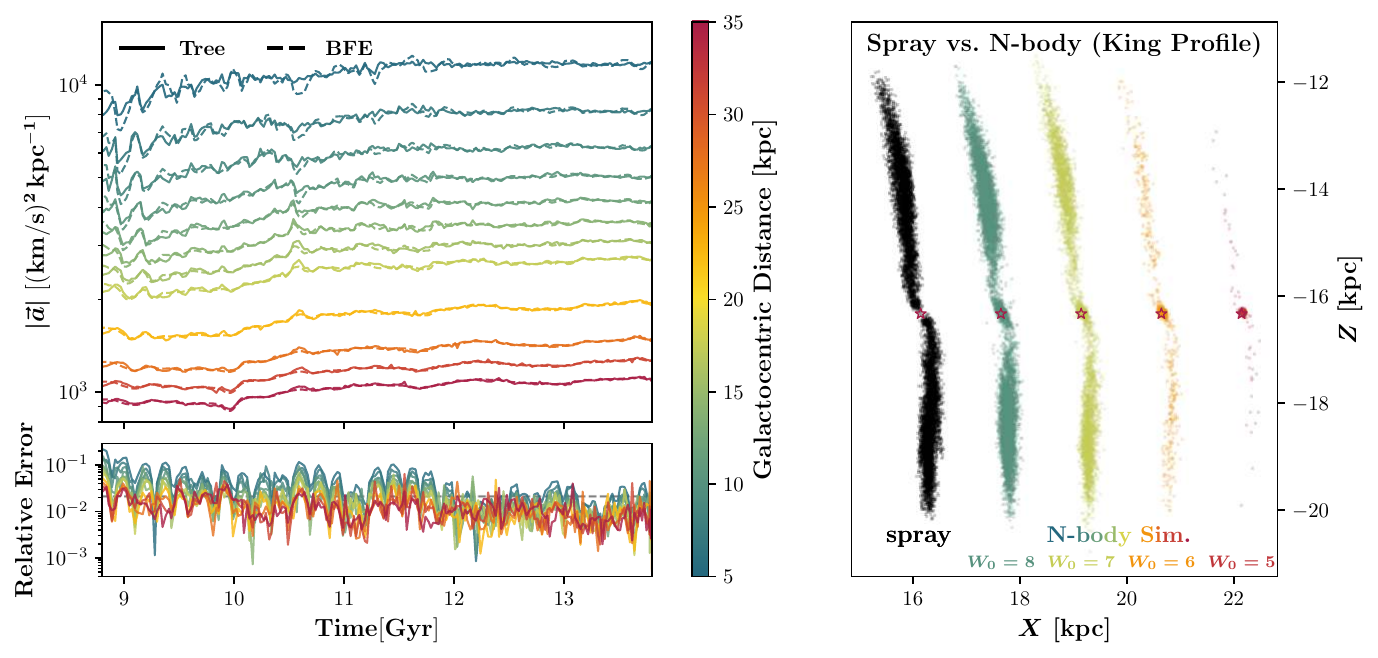}
    \caption{\textit{Left:} Accelerations at 12 fixed galactocentric locations as a function of time, computed from a Barnes--Hut tree evaluation of the FIRE-2 particle distribution (solid) and from our basis function expansion (BFE) representation (dashed). Colors indicate galactocentric distance. The bottom panel shows the relative error $|\vec{a}_{\rm BFE} -\vec{a}_{\rm tree}| / |\vec{a}_{\rm tree}|$, which stays below $\sim$2\% outside of $\sim$7 kpc. \textit{Right:} Example particle-spray stream (\citealt{chen2025improved}; black) compared to a collisionless N-body simulations (colored) for the same progenitor orbit and King density profile ($M = 2.7 \times 10^4\,M_\odot$, $r_s = 4$ pc) in the \mi{} host potential. The \citet{chen2025improved} spray model is calibrated for $W_0 = 8$. The spray stream probes the same phase-space region as the N-body run, though a progenitor with lower $W_0$ releases fewer unbound particles and shows less structure as expected.} \label{fig:force_err_streams}
\end{figure*}

In this section, we validate two methodological choices that underpin our stream catalog: (i) that our BFE representation of each time-evolving host potential accurately reproduces the underlying tidal field from the FIRE-2 particle distribution, and (ii) that our particle-spray stream formation model samples the same phase-space region as a collisionless N-body simulation for the same progenitor.

\subsection{BFE force accuracy}

We compute accelerations at 12 fixed spatial locations sampling galactocentric distances from $\sim$5 to $\sim$35 kpc over the last 5 Gyr. At each location, we evaluate the acceleration directly from the FIRE-2 particle distribution using a GPU-based Barnes--Hut tree\footnote{Available at \url{https://github.com/appy2806/Nbody_streams/}. \label{foot:Nbody_streams}} with opening angle $\theta = 0.5$ and Plummer-equivalent softenings matching those used in FIRE-2 ($\epsilon_{\rm DM} = 28$ pc, and $\epsilon_{\rm baryon} = 4$ pc). We compare these direct evaluations to the acceleration computed from our BFE representation at $l_{\rm max} = 4$.

Figure~\ref{fig:force_err_streams} (left) shows the two acceleration estimates as a function of time at each location, with the relative error in the bottom panel. The BFE reproduces the tree-based acceleration to better than $\sim$2\% at all galactocentric distances outside of $\sim$7 kpc, with the largest discrepancies confined to the innermost location. These results are consistent with \citet{arora2024efficient}, which found that the same BFE framework recovers the total acceleration field to within $\sim$1\% outside $\sim$10 kpc and within $\sim$10\% in the innermost regions at the present-day snapshot. The time evolution of the two estimates tracks each other closely, confirming that our interpolation between snapshots preserves the tidal field on the relevant orbital timescales.

\subsection{Particle-spray versus collisionless N-body}

We validate the particle-spray stream formation model against direct collisionless N-body simulations of GC progenitors evolving in the same \mi{} potential. We sample progenitor initial conditions from a King distribution function using \texttt{Agama} for a fixed progenitor mass ($M = 2.7 \times 10^4\,M_\odot$) and scale radius ($r_s = 4$ pc), varying the central potential depth $W_0$. Each progenitor is integrated forward for 5 Gyr with 25{,}000 particles in a direct $\mathcal{O}(N^2)$ N-body solver with Plummer-equivalent softening $\epsilon = 0.3$ pc and a time step of $10^{-5}$ Gyr using a GPU-based collisionless N-body code\footref{foot:Nbody_streams}. We then run the particle-spray model of \citet{chen2025improved} for the same progenitor orbit and density profile within our time-evolving \mi{} BFE potential.

Figure~\ref{fig:force_err_streams} (right) compares the resulting streams. The \citet{chen2025improved} spray prescription is calibrated against N-body experiments at $W_0 = 8$, and at that central potential depth the spray stream probes the same phase-space region as the collisionless N-body run. At other values of $W_0$, the balance between bound and unbound particles shifts as the progenitor's central concentration changes, but the large-scale phase-space structure of the resulting stream remains consistent. Across the range of $W_0$ explored, the spray model reproduces the bulk phase-space structure of the N-body streams to within the statistical uncertainty of a single realization, validating its use across the progenitor population in our catalog.

\section{Calibration of the Disturbance Metric} \label{app:wass_dis_Gauss}
\begin{figure*}[ht]
    \centering
    \includegraphics[width=0.5\linewidth]{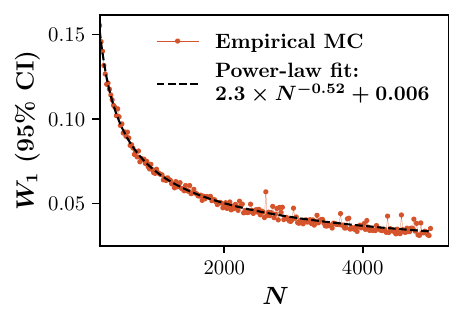}
    \caption{Monte Carlo calibration of the \(W_1\) null distribution as a function of sample size \(N\). Orange markers: 95\% percentile \(W_1\) values from 2,000 trials of a standard normal sample of size \(N\) each compared to a fixed reference normal with $N_{\rm ref}=20{,}000$. The black dashed line is the best-fit curve \(T(N)=aN^{p}+b\) (fit parameters annotated). The curve follows the expected \(\sim N^{-1/2}\) scaling.}
    \label{fig:wasster_metric_samples}
\end{figure*}

We quantify local departures from Gaussianity using the 1D Wasserstein--1 distance (Earth-Mover's Distance) between the empirical cross-track distribution in a bin and a reference standard normal ($z_{\rm std}$). For one-dimensional cumulative distribution functions \(F(x)\) and \(Z_{\rm ref}(x)\) the \(W_1\) distance has the simple form
\begin{equation}
  W_1(F,Z_{\rm ref}) \;=\; \int_{-\infty}^{\infty} \left| F(x) - Z_{\rm ref}(x) \right| \, dx,
\end{equation}

which corresponds to the minimum ``work'' required to transform one distribution into the other (mass moved \(\times\) distance moved). We take $Z_{\rm ref}$ to be a reference standard normal sampled with \(N_{\rm ref}=20{,}000\) draws (generated with \texttt{numpy.random.randn}) and compute \(W_1\) using \texttt{scipy.stats.wasserstein\_distance}.

To convert the raw \(W_1\) into a significance-normalized metric that accounts for finite sample size, we calibrate the null distribution of \(W_1\) for samples drawn from a true standard normal. Let \(Z_N\) denote an empirical sample of size \(N\) from a standard normal with CDF \(Z_N(x)\). Then
\begin{equation}
  W_1(Z_N,Z_{\rm ref}) \;=\; \int_{-\infty}^{\infty} \left| Z_N(x) - Z_{\rm ref}(x) \right| \, dx.
\end{equation}

By Donsker's theorem \citep{donsker1951invariance}, the empirical process \(\sqrt{N}\,\Big(Z_N(x)-Z_{\rm ref}(x)\Big)\) converges in distribution to a Brownian bridge \(G(x)\), so

\begin{equation}
  \sqrt{N}\,W_1(Z_N,Z_{\rm ref}) \xrightarrow{d} \int |G(x)|\,dx,
\end{equation}

and the typical scale of \(W_1(Z_N,Z_{\rm ref})\) follows \( \propto N^{-1/2}\).

We calibrate the 95\% significance threshold \(T(N)\) for finite \(N\) by Monte Carlo. For each sample size \(N\) we draw 2,000 independent samples from a standard normal, compute \(W_1(Z_N,Z_{\rm ref})\) for each trial, and record the 95th percentile value of the resulting distribution. We then fit the empirical dependence with
\begin{equation}
  T(N) \approx a\,N^{p} + b,
  \label{eq:fitting_WD}
\end{equation}
where theory predicts \(p\approx-0.5\) and \(b\approx0\). Figure~\ref{fig:wasster_metric_samples} shows the calibration results. Orange markers are the 95-th percentile $W_1$ values obtained from 2,000 trials at each $N$. The best-fit parameters from Equation~\ref{eq:fitting_WD} are
\[
  a = 2.3,\qquad p = -0.52,\qquad b = 0.006,
\]

consistent with the theoretical scaling and providing a smooth 95\% confidence threshold across the sample-size range used in our analysis.

We define the normalized disturbance metric $\xi$ for an empirical bin distribution \(F\) with sample size \(N\) as
\begin{equation}
  \xi \;\equiv\; \frac{W_1(F,Z_{\rm ref})}{T(N)}.\label{eq:nonGauss_metric}
\end{equation}

\begin{nolinenumbers}
By construction, \(\xi \le 1\) indicates that the empirical distribution is consistent with a standard normal at the 5\% significance level. Larger values quantify increasingly strong departures from Gaussianity, with $\xi = 2$ corresponding to a Wasserstein distance twice the 95th-percentile threshold.

\noindent\textit{Implementation notes.}
The reference sample \(Z_{\rm ref}\) is drawn once and held fixed for all trials to avoid additional noise in the calibration. In practice we compute \(W_1\) using raw sample vectors (\texttt{SciPy} implementation uses the empirical CDFs internally), and evaluate \(\xi\) bin-by-bin using each bin's sample size \(N\). This provides a straightforward, robust test for 1D Gaussianity that performs well for small and moderate \(N\) and is symmetric by construction, unlike asymmetric divergence measure such as Kullback--Leibler divergence. 
\end{nolinenumbers}

\section{Supplementary Figures}\label{app:supp}
Here, we present supplementary figures that provide further context to results from the main text.

\begin{figure*}[ht]
    \centering
    \includegraphics[width=\linewidth]{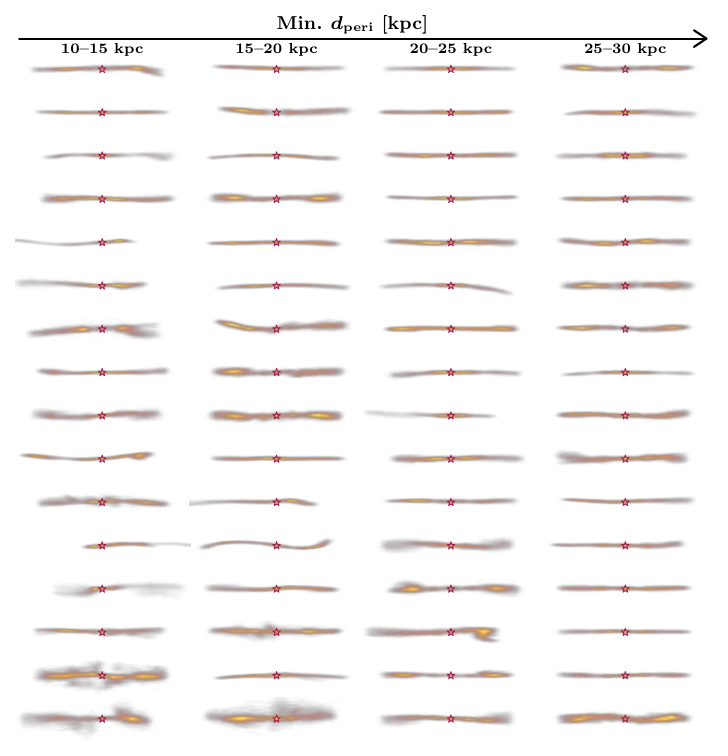}
    \caption{A random selection of 16 streams per pericenter bin (columns, $d_{\rm peri} = 10$--$30$ kpc in intervals of 5 kpc, annotated), sorted by increasing global disturbance score ($\Dglobal{}$) from top to bottom within each column. Inner-region streams show pronounced off-track features including localized spurs and broad cross-track heating and, outer-region streams are generally smoother, though localized features remain present.}
\label{fig:streams_grid_peri_2d}
\end{figure*}

\begin{figure*}[ht]
    \centering
    \includegraphics[width=\linewidth]{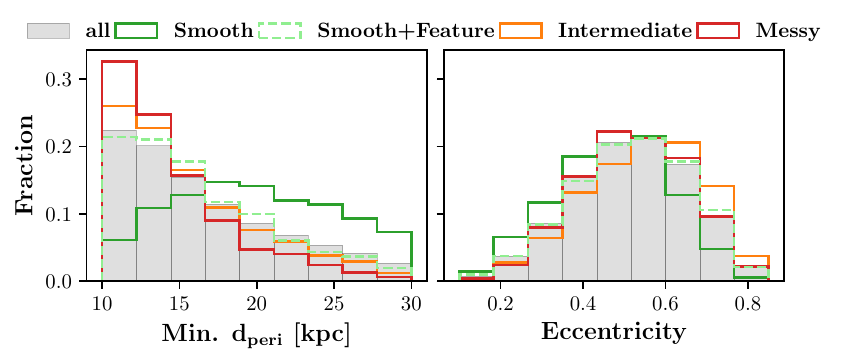}
    \caption{Distributions of progenitor minimum pericentric distance (left) and orbital eccentricity (right) for streams split by the four morphological tiers (colored-lines) in Table~\ref{tab:xi_classification}. Histograms are normalized within each tier to show the fraction of streams. Smooth streams (green) are concentrated at larger pericenters, with the highest fraction of outer-halo orbits. Smooth + Feature streams (light green, dashed) follow a similar but broader distribution. Intermediate and Messy streams peak at smaller pericenters. Eccentricity distributions are more similar across tiers, though Smooth streams favor slightly more circular orbits.}\label{fig:streams_peri_eccen_tier4}

\end{figure*}

\begin{figure*}[ht]
    \centering
    \includegraphics[width=\linewidth]{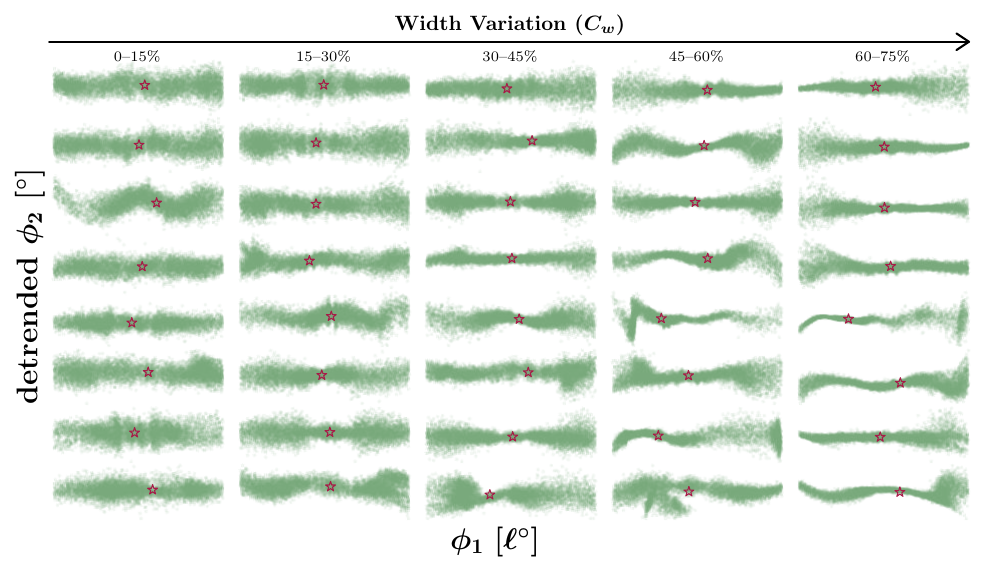}
    \caption{Random selection of smooth streams (green) arranged in bins of $C_w$ (columns, $0$--$75\%$ in steps of $15\%$), sorted by increasing peak disturbance score ($\Dpeak{}$) from top to bottom within each column. Axes have unequal aspect to highlight along-track width variation. Streams with higher $C_w$ tend to show more pronounced localized features, reflected in their higher $\Dpeak{}$ values.}
    \label{fig:streams_grid_cw_2d}
\end{figure*}

\setlength{\bibsep}{0pt} % Reduces space between entries
\renewcommand{\baselinestretch}{1.0} % Remove paragraph spacing
\bibliography{ref}{}

\begin{thebibliography}{}
\expandafter\ifx\csname natexlab\endcsname\relax\def\natexlab#1{#1}\fi
\providecommand{\url}[1]{\href{#1}{#1}}
\providecommand{\dodoi}[1]{doi:~\href{http://doi.org/#1}{\nolinkurl{#1}}}
\providecommand{\doeprint}[1]{\href{http://ascl.net/#1}{\nolinkurl{http://ascl.net/#1}}}
\providecommand{\doarXiv}[1]{\href{https://arxiv.org/abs/#1}{\nolinkurl{https://arxiv.org/abs/#1}}}

\bibitem[{N.~C. Amorisco {et~al.}(2016)Amorisco, G{\'o}mez, Vegetti, \& White}]{amorisco2016gaps}
Amorisco, N.~C., G{\'o}mez, F.~A., Vegetti, S., \& White, S.~D. 2016, \bibinfo{title}{Gaps in globular cluster streams: giant molecular clouds can cause them too,} Monthly Notices of the Royal Astronomical Society: Letters, 463, L17

\bibitem[{S. Ansar {et~al.}(2025)Ansar, Pearson, Sanderson, Arora, Hopkins, Wetzel, Cunningham, \& Quinn}]{ansar2025bar}
Ansar, S., Pearson, S., Sanderson, R.~E., {et~al.} 2025, \bibinfo{title}{Bar formation and destruction in the FIRE-2 simulations,} The Astrophysical Journal, 978, 37

\bibitem[{A. Arora {et~al.}(2024{\natexlab{a}})Arora, Garavito-Camargo, Sanderson, Cunningham, Wetzel, Panithanpaisal, \& Barry}]{arora2024lmc}
Arora, A., Garavito-Camargo, N., Sanderson, R.~E., {et~al.} 2024{\natexlab{a}}, \bibinfo{title}{LMC-driven anisotropic boosts in stream--subhalo interactions,} The Astrophysical Journal, 974, 286

\bibitem[{A. Arora {et~al.}(2022)Arora, Sanderson, Panithanpaisal, Cunningham, Wetzel, \& Garavito-Camargo}]{arora2022stability}
Arora, A., Sanderson, R.~E., Panithanpaisal, N., {et~al.} 2022, \bibinfo{title}{On the stability of tidal streams in action space,} The Astrophysical Journal, 939, 2

\bibitem[{A. Arora {et~al.}(2024{\natexlab{b}})Arora, Sanderson, Regan, Garavito-Camargo, Bregou, Panithanpaisal, Wetzel, Cunningham, Loebman, Dropulic, {et~al.}}]{arora2024efficient}
Arora, A., Sanderson, R., Regan, C., {et~al.} 2024{\natexlab{b}}, \bibinfo{title}{Efficient and accurate force replay in cosmological-baryonic simulations,} The Astrophysical Journal, 977, 23

\bibitem[{A. Arora {et~al.}(2025)Arora, Garavito-Camargo, Sanderson, Weinberg, Petersen, Varela-Lavin, G{\'o}mez, Johnston, Laporte, Shipp, {et~al.}}]{arora2025shaping}
Arora, A., Garavito-Camargo, N., Sanderson, R.~E., {et~al.} 2025, \bibinfo{title}{Shaping the Milky Way: The interplay of mergers and cosmic filaments,} The Astrophysical Journal, 988, 190

\bibitem[{E. Balbinot \& M. Gieles(2018)Balbinot \& Gieles}]{balbinot2018devil}
Balbinot, E., \& Gieles, M. 2018, \bibinfo{title}{The devil is in the tails: the role of globular cluster mass evolution on stream properties,} Monthly Notices of the Royal Astronomical Society, 474, 2479

\bibitem[{N. Banik \& J. Bovy(2019)Banik \& Bovy}]{banik2019effects}
Banik, N., \& Bovy, J. 2019, \bibinfo{title}{Effects of baryonic and dark matter substructure on the Pal 5 stream,} Monthly Notices of the Royal Astronomical Society, 484, 2009

\bibitem[{N. Banik \& J. Bovy(2021)Banik \& Bovy}]{banik2021n}
Banik, N., \& Bovy, J. 2021, \bibinfo{title}{On N-body simulations of globular cluster streams,} Monthly Notices of the Royal Astronomical Society, 504, 648

\bibitem[{N. Banik {et~al.}(2021)Banik, Bovy, Bertone, Erkal, \& de~Boer}]{banik2021novel}
Banik, N., Bovy, J., Bertone, G., Erkal, D., \& de~Boer, T. 2021, \bibinfo{title}{Novel constraints on the particle nature of dark matter from stellar streams,} Journal of Cosmology and Astroparticle Physics, 2021, 043

\bibitem[{M. Barry {et~al.}(2023)Barry, Wetzel, Chapman, Samuel, Sanderson, \& Arora}]{barry2023dark}
Barry, M., Wetzel, A., Chapman, S., {et~al.} 2023, \bibinfo{title}{The dark side of FIRE: predicting the population of dark matter subhaloes around Milky Way-mass galaxies,} Monthly Notices of the Royal Astronomical Society, 523, 428

\bibitem[{K. {Bechtol} {et~al.}(2026){Bechtol}, {Sevilla-Noarbe}, {Drlica-Wagner}, {Yanny}, {Gruendl}, {Sheldon}, {Rykoff}, {De Vicente}, {Adamow}, {Anbajagane}, {Becker}, {Bernstein}, {Carnero Rosell}, {Gschwend}, {Gorsuch}, {Hartley}, {Jarvis}, {Jeltema}, {Kron}, {Manning}, {O'Donnell}, {Pieres}, {Rodr{\'\i}guez-Monroy}, {Sanchez Cid}, {Tabbutt}, {Tan}, {Toribio San Cipriano}, {Tucker}, {Weaverdyck}, {Yamamoto}, {Abbott}, {Aguena}, {Alarc{\'o}n}, {Allam}, {Amon}, {Andrade-Oliveira}, {Avila}, {Bernardinelli}, {Bertin}, {Blazek}, {Brooks}, {Burke}, {Carretero}, {Castander}, {Cawthon}, {Chang}, {Choi}, {Conselice}, {Costanzi}, {Crocce}, {da Costa}, {Davis}, {Desai}, {Diehl}, {Dodelson}, {Doel}, {Doux}, {Fert{\'e}}, {Flaugher}, {Fosalba}, {Frieman}, {Garc{\'\i}a-Bellido}, {Gatti}, {Gaztanaga}, {Giannini}, {Gruen}, {Gutierrez}, {Herner}, {Hinton}, {Hollowood}, {Honscheid}, {Huterer}, {Jeffrey}, {Krause}, {Kuehn}, {Lahav}, {Lee}, {Lidman}, {Lima}, {Lin}, {Marshall}, {Mena-Fern{\'a}ndez}, {Miquel}, {Mohr}, {Muir},
  {Myles}, {Ogando}, {Palmese}, {Plazas Malag{\'o}n}, {Porredon}, {Prat}, {Raveri}, {Romer}, {Roodman}, {Samuroff}, {Sanchez}, {Scarpine}, {Smith}, {Soares-Santos}, {Suchyta}, {Tarle}, {Troxel}, {Vikram}, {Walker}, {Weller}, {Wiseman}, {Zhang}, \& {DES Collaboration}}]{Bechtol:2026}
{Bechtol}, K., {Sevilla-Noarbe}, I., {Drlica-Wagner}, A., {et~al.} 2026, \bibinfo{title}{{Dark Energy Survey Year 6 Results: Photometric Dataset for Cosmology},} \apjs, 282, 62, \dodoi{10.3847/1538-4365/ae18d3}

\bibitem[{V. Belokurov {et~al.}(2006)Belokurov, Zucker, Evans, Gilmore, Vidrih, Bramich, Newberg, Wyse, Irwin, Fellhauer, {et~al.}}]{belokurov2006field}
Belokurov, V., Zucker, D., Evans, N., {et~al.} 2006, \bibinfo{title}{The field of streams: Sagittarius and its siblings,} The Astrophysical Journal Letters, 642, L137

\bibitem[{A. Bonaca {et~al.}(2019{\natexlab{a}})Bonaca, Conroy, Price-Whelan, \& Hogg}]{bonaca2019multiple}
Bonaca, A., Conroy, C., Price-Whelan, A.~M., \& Hogg, D.~W. 2019{\natexlab{a}}, \bibinfo{title}{Multiple components of the Jhelum stellar stream,} The Astrophysical Journal Letters, 881, L37

\bibitem[{A. Bonaca {et~al.}(2012)Bonaca, Geha, \& Kallivayalil}]{bonaca2012cold}
Bonaca, A., Geha, M., \& Kallivayalil, N. 2012, \bibinfo{title}{A cold Milky Way stellar stream in the direction of triangulum,} The Astrophysical Journal Letters, 760, L6

\bibitem[{A. Bonaca {et~al.}(2019{\natexlab{b}})Bonaca, Hogg, Price-Whelan, \& Conroy}]{bonaca2019spur}
Bonaca, A., Hogg, D.~W., Price-Whelan, A.~M., \& Conroy, C. 2019{\natexlab{b}}, \bibinfo{title}{The Spur and the Gap in GD-1: Dynamical evidence for a dark substructure in the Milky Way halo,} The Astrophysical Journal, 880, 38

\bibitem[{A. Bonaca \& A.~M. Price-Whelan(2025)Bonaca \& Price-Whelan}]{bonaca2025stellar}
Bonaca, A., \& Price-Whelan, A.~M. 2025, \bibinfo{title}{Stellar streams in the Gaia era,} New Astronomy Reviews, 100, 101713

\bibitem[{A. Bonaca {et~al.}(2020)Bonaca, Conroy, Hogg, Cargile, Caldwell, Naidu, Price-Whelan, Speagle, \& Johnson}]{bonaca2020high}
Bonaca, A., Conroy, C., Hogg, D.~W., {et~al.} 2020, \bibinfo{title}{High-resolution spectroscopy of the GD-1 stellar stream localizes the perturber near the orbital plane of Sagittarius,} The Astrophysical Journal Letters, 892, L37

\bibitem[{J. Bovy {et~al.}(2016)Bovy, Bahmanyar, Fritz, \& Kallivayalil}]{bovy2016shape}
Bovy, J., Bahmanyar, A., Fritz, T.~K., \& Kallivayalil, N. 2016, \bibinfo{title}{The shape of the inner milky way halo from observations of the Pal 5 and GD--1 Stellar Streams,} The Astrophysical Journal, 833, 31

\bibitem[{J. Bovy {et~al.}(2017)Bovy, Erkal, \& Sanders}]{bovy2017linear}
Bovy, J., Erkal, D., \& Sanders, J.~L. 2017, \bibinfo{title}{Linear perturbation theory for tidal streams and the small-scale CDM power spectrum,} Monthly Notices of the Royal Astronomical Society, 466, 628

\bibitem[{R.~A.~N. {Brooks} {et~al.}(2025){Brooks}, {Garavito-Camargo}, {Johnston}, {Price-Whelan}, {Sanders}, \& {Lilleengen}}]{brooks2025lmccalls}
{Brooks}, R. A.~N., {Garavito-Camargo}, N., {Johnston}, K.~V., {et~al.} 2025, \bibinfo{title}{{LMC Calls, Milky Way Halo Answers: Disentangling the Effects of the MW{\textendash}LMC Interaction on Stellar Stream Populations},} \apj, 978, 79, \dodoi{10.3847/1538-4357/ad93a7}

\bibitem[{R.~G. Carlberg(2012)Carlberg}]{carlberg2012dark}
Carlberg, R.~G. 2012, \bibinfo{title}{Dark matter sub-halo counts via star stream crossings,} The Astrophysical Journal, 748, 20

\bibitem[{R.~G. Carlberg \& C.~J. Grillmair(2013)Carlberg \& Grillmair}]{carlberg2013gaps}
Carlberg, R.~G., \& Grillmair, C.~J. 2013, \bibinfo{title}{Gaps in the GD-1 star stream,} The Astrophysical Journal, 768, 171

\bibitem[{Y. Chen \& O.~Y. Gnedin(2024)Chen \& Gnedin}]{chen2024catalogue}
Chen, Y., \& Gnedin, O.~Y. 2024, \bibinfo{title}{Catalogue of model star clusters in the Milky Way and M31 galaxies,} Monthly Notices of the Royal Astronomical Society, 527, 3692

\bibitem[{Y. Chen {et~al.}(2025)Chen, Valluri, Gnedin, \& Ash}]{chen2025improved}
Chen, Y., Valluri, M., Gnedin, O.~Y., \& Ash, N. 2025, \bibinfo{title}{Improved particle spray algorithm for modeling globular cluster streams,} The Astrophysical Journal Supplement Series, 276, 32

\bibitem[{R. Chiba {et~al.}(2021)Chiba, Friske, \& Sch{\"o}nrich}]{chiba2021resonance}
Chiba, R., Friske, J.~K., \& Sch{\"o}nrich, R. 2021, \bibinfo{title}{Resonance sweeping by a decelerating Galactic bar,} Monthly Notices of the Royal Astronomical Society, 500, 4710

\bibitem[{C. Conroy {et~al.}(2009)Conroy, Gunn, \& White}]{conroy2009propagation}
Conroy, C., Gunn, J.~E., \& White, M. 2009, \bibinfo{title}{The propagation of uncertainties in stellar population synthesis modeling. I. The relevance of uncertain aspects of stellar evolution and the initial mass function to the derived physical properties of galaxies,} The Astrophysical Journal, 699, 486

\bibitem[{ {Dark Energy Survey Collaboration} {et~al.}(2016){Dark Energy Survey Collaboration}, {Abbott}, {Abdalla}, {Aleksi{\'c}}, {Allam}, {Amara}, {Bacon}, {Balbinot}, {Banerji}, {Bechtol}, {Benoit-L{\'e}vy}, {Bernstein}, {Bertin}, {Blazek}, {Bonnett}, {Bridle}, {Brooks}, {Brunner}, {Buckley-Geer}, {Burke}, {Caminha}, {Capozzi}, {Carlsen}, {Carnero-Rosell}, {Carollo}, {Carrasco-Kind}, {Carretero}, {Castander}, {Clerkin}, {Collett}, {Conselice}, {Crocce}, {Cunha}, {D'Andrea}, {da Costa}, {Davis}, {Desai}, {Diehl}, {Dietrich}, {Dodelson}, {Doel}, {Drlica-Wagner}, {Estrada}, {Etherington}, {Evrard}, {Fabbri}, {Finley}, {Flaugher}, {Foley}, {Fosalba}, {Frieman}, {Garc{\'\i}a-Bellido}, {Gaztanaga}, {Gerdes}, {Giannantonio}, {Goldstein}, {Gruen}, {Gruendl}, {Guarnieri}, {Gutierrez}, {Hartley}, {Honscheid}, {Jain}, {James}, {Jeltema}, {Jouvel}, {Kessler}, {King}, {Kirk}, {Kron}, {Kuehn}, {Kuropatkin}, {Lahav}, {Li}, {Lima}, {Lin}, {Maia}, {Makler}, {Manera}, {Maraston}, {Marshall}, {Martini}, {McMahon},
  {Melchior}, {Merson}, {Miller}, {Miquel}, {Mohr}, {Morice-Atkinson}, {Naidoo}, {Neilsen}, {Nichol}, {Nord}, {Ogando}, {Ostrovski}, {Palmese}, {Papadopoulos}, {Peiris}, {Peoples}, {Percival}, {Plazas}, {Reed}, {Refregier}, {Romer}, {Roodman}, {Ross}, {Rozo}, {Rykoff}, {Sadeh}, {Sako}, {S{\'a}nchez}, {Sanchez}, {Santiago}, {Scarpine}, {Schubnell}, {Sevilla-Noarbe}, {Sheldon}, {Smith}, {Smith}, {Soares-Santos}, {Sobreira}, {Soumagnac}, {Suchyta}, {Sullivan}, {Swanson}, {Tarle}, {Thaler}, {Thomas}, {Thomas}, {Tucker}, {Vieira}, {Vikram}, {Walker}, {Wechsler}, {Weller}, {Wester}, {Whiteway}, {Wilcox}, {Yanny}, {Zhang}, \& {Zuntz}}]{des2016dark}
{Dark Energy Survey Collaboration}, {Abbott}, T., {Abdalla}, F.~B., {et~al.} 2016, \bibinfo{title}{{The Dark Energy Survey: more than dark energy - an overview},} \mnras, 460, 1270, \dodoi{10.1093/mnras/stw641}

\bibitem[{E. Darragh-Ford {et~al.}(2023)Darragh-Ford, Hunt, Price-Whelan, \& Johnston}]{darragh2023escargot}
Darragh-Ford, E., Hunt, J.~A., Price-Whelan, A.~M., \& Johnston, K.~V. 2023, \bibinfo{title}{ESCARGOT: Mapping Vertical Phase Spiral Characteristics Throughout the Real and Simulated Milky Way,} The Astrophysical Journal, 955, 74

\bibitem[{E. Darragh-Ford {et~al.}(2025)Darragh-Ford, Garavito-Camargo, Arora, Wechsler, Mansfield, Besla, Petersen, Weinberg, Varela-Lavin, Buch, {et~al.}}]{darragh2025shaping}
Darragh-Ford, E., Garavito-Camargo, N., Arora, A., {et~al.} 2025, \bibinfo{title}{Shaping the Milky Way. II. The dark matter halo's response to the LMC's passage in a cosmological context,} arXiv preprint arXiv:2511.02031

\bibitem[{W. Dehnen(1993)Dehnen}]{dehnen1993family}
Dehnen, W. 1993, \bibinfo{title}{A family of potential--density pairs for spherical galaxies and bulges,} Monthly Notices of the Royal Astronomical Society, 265, 250

\bibitem[{W. Dehnen(2000)Dehnen}]{dehnen2000effect}
Dehnen, W. 2000, \bibinfo{title}{The Effect of the Outer Lindblad Resonance of the Galactic Baron the Local Stellar Velocity Distribution,} The Astronomical Journal, 119, 800

\bibitem[{A.~M. Dillamore {et~al.}(2022)Dillamore, Belokurov, Evans, \& Price-Whelan}]{dillamore2022impact}
Dillamore, A.~M., Belokurov, V., Evans, N.~W., \& Price-Whelan, A.~M. 2022, \bibinfo{title}{The impact of a massive Sagittarius dSph on GD-1-like streams,} Monthly Notices of the Royal Astronomical Society, 516, 1685

\bibitem[{A.~M. Dillamore \& J.~L. Sanders(2026)Dillamore \& Sanders}]{dillamore2026geometry}
Dillamore, A.~M., \& Sanders, J.~L. 2026, \bibinfo{title}{Geometry of the Milky Way’s dark matter from dynamical models of the tilted stellar halo,} Monthly Notices of the Royal Astronomical Society, stag226

\bibitem[{M.~D. Donsker(1951)Donsker}]{donsker1951invariance}
Donsker, M.~D. 1951, An invariance principle for certain probability limit theorems, Memoirs of the American Mathematical Society No.~6 (American Mathematical Society), 1--12

\bibitem[{D. Erkal \& V. Belokurov(2015)Erkal \& Belokurov}]{erkal2015forensics}
Erkal, D., \& Belokurov, V. 2015, \bibinfo{title}{Forensics of subhalo--stream encounters: the three phases of gap growth,} Monthly Notices of the Royal Astronomical Society, 450, 1136

\bibitem[{D. Erkal {et~al.}(2016)Erkal, Belokurov, Bovy, \& Sanders}]{erkal2016number}
Erkal, D., Belokurov, V., Bovy, J., \& Sanders, J.~L. 2016, \bibinfo{title}{The number and size of subhalo-induced gaps in stellar streams,} Monthly Notices of the Royal Astronomical Society, 463, 102

\bibitem[{D. Erkal {et~al.}(2017)Erkal, Koposov, \& Belokurov}]{erkal2017sharper}
Erkal, D., Koposov, S.~E., \& Belokurov, V. 2017, \bibinfo{title}{A sharper view of Pal 5's tails: discovery of stream perturbations with a novel non-parametric technique,} Monthly Notices of the Royal Astronomical Society, 470, 60

\bibitem[{D. Erkal {et~al.}(2018)Erkal, Li, Koposov, Belokurov, Balbinot, Bechtol, Buncher, Drlica-Wagner, Kuehn, Marshall, {et~al.}}]{erkal2018modelling}
Erkal, D., Li, T., Koposov, S.~E., {et~al.} 2018, \bibinfo{title}{Modelling the Tucana III stream--a close passage with the LMC,} Monthly Notices of the Royal Astronomical Society, 481, 3148

\bibitem[{D. Erkal {et~al.}(2019)Erkal, Belokurov, Laporte, Koposov, Li, Grillmair, Kallivayalil, Price-Whelan, Evans, Hawkins, {et~al.}}]{erkal2019total}
Erkal, D., Belokurov, V., Laporte, C., {et~al.} 2019, \bibinfo{title}{The total mass of the Large Magellanic Cloud from its perturbation on the Orphan stream,} Monthly Notices of the Royal Astronomical Society, 487, 2685

\bibitem[{A. Eyre \& J. Binney(2011)Eyre \& Binney}]{eyre2011mechanics}
Eyre, A., \& Binney, J. 2011, \bibinfo{title}{The mechanics of tidal streams,} Monthly Notices of the Royal Astronomical Society, 413, 1852

\bibitem[{M.~A. Fardal {et~al.}(2015)Fardal, Huang, \& Weinberg}]{fardal2015generation}
Fardal, M.~A., Huang, S., \& Weinberg, M.~D. 2015, \bibinfo{title}{Generation of mock tidal streams,} Monthly Notices of the Royal Astronomical Society, 452, 301

\bibitem[{P. Ferguson {et~al.}(2021)Ferguson, Shipp, Drlica-Wagner, Li, Cerny, Tavangar, Pace, Marshall, Riley, Adamow, {et~al.}}]{ferguson2021delve}
Ferguson, P., Shipp, N., Drlica-Wagner, A., {et~al.} 2021, \bibinfo{title}{DELVE-ing into the Jet: A Thin Stellar Stream on a Retrograde Orbit at 30 kpc,} The Astronomical Journal, 163, 18

\bibitem[{ {Gaia Collaboration} {et~al.}(2018){Gaia Collaboration}, Brown, Vallenari, Prusti, De~Bruijne, Babusiaux, Juh{\'a}sz, Marschalk{\'o}, Marton, Moln{\'a}r, {et~al.}}]{gaia2018gaia}
{Gaia Collaboration}, Brown, A., Vallenari, A., {et~al.} 2018, \bibinfo{title}{Gaia data release 2 summary of the contents and survey properties,} Astronomy \& Astrophysics, 616

\bibitem[{N. {Garavito-Camargo} {et~al.}(2021){Garavito-Camargo}, {Besla}, {Laporte}, {Price-Whelan}, {Cunningham}, {Johnston}, {Weinberg}, \& {G{\'o}mez}}]{garavito2021quantifying}
{Garavito-Camargo}, N., {Besla}, G., {Laporte}, C. F.~P., {et~al.} 2021, \bibinfo{title}{{Quantifying the Impact of the Large Magellanic Cloud on the Structure of the Milky Way's Dark Matter Halo Using Basis Function Expansions},} The Astrophysical Journal, 919, 109, \dodoi{10.3847/1538-4357/ac0b44}

\bibitem[{N. Garavito-Camargo {et~al.}(2024)Garavito-Camargo, Price-Whelan, Samuel, Cunningham, Patel, Wetzel, Johnston, Arora, Sanderson, Garrison, {et~al.}}]{garavito2024corotation}
Garavito-Camargo, N., Price-Whelan, A.~M., Samuel, J., {et~al.} 2024, \bibinfo{title}{On the Corotation of Milky Way Satellites: LMC-mass Satellites Induce Apparent Motions in Outer Halo Tracers,} The Astrophysical Journal, 975, 100

\bibitem[{S. Garrison-Kimmel {et~al.}(2018)Garrison-Kimmel, Hopkins, Wetzel, El-Badry, Sanderson, Bullock, Ma, van~de Voort, Hafen, Faucher-Gigu{\`e}re, {et~al.}}]{garrison2018origin}
Garrison-Kimmel, S., Hopkins, P.~F., Wetzel, A., {et~al.} 2018, \bibinfo{title}{The origin of the diverse morphologies and kinematics of Milky Way-mass galaxies in the FIRE-2 simulations,} Monthly Notices of the Royal Astronomical Society, 481, 4133

\bibitem[{M. Gieles {et~al.}(2021)Gieles, Erkal, Antonini, Balbinot, \& Penarrubia}]{gieles2021supra}
Gieles, M., Erkal, D., Antonini, F., Balbinot, E., \& Penarrubia, J. 2021, \bibinfo{title}{A supra-massive population of stellar-mass black holes in the globular cluster Palomar 5,} Nature Astronomy, 5, 957

\bibitem[{O.~Y. {Gnedin} {et~al.}(2014){Gnedin}, {Ostriker}, \& {Tremaine}}]{gnedin2014globular}
{Gnedin}, O.~Y., {Ostriker}, J.~P., \& {Tremaine}, S. 2014, \bibinfo{title}{{Co-evolution of Galactic Nuclei and Globular Cluster Systems},} \apj, 785, 71, \dodoi{10.1088/0004-637X/785/1/71}

\bibitem[{{\'A}. Gonz{\'a}lez(2010)Gonz{\'a}lez}]{gonzalez2010measurement}
Gonz{\'a}lez, {\'A}. 2010, \bibinfo{title}{Measurement of areas on a sphere using Fibonacci and latitude--longitude lattices,} Mathematical geosciences, 42, 49

\bibitem[{C.~J. Grillmair(2006)Grillmair}]{grillmair2006detection}
Grillmair, C.~J. 2006, \bibinfo{title}{Detection of a 60°-long dwarf galaxy debris stream,} The Astrophysical Journal Letters, 645, L37

\bibitem[{M.~Y. {Grudi{\'c}} {et~al.}(2023){Grudi{\'c}}, {Hafen}, {Rodriguez}, {Guszejnov}, {Lamberts}, {Wetzel}, {Boylan-Kolchin}, \& {Faucher-Gigu{\`e}re}}]{grudic2023greatballs}
{Grudi{\'c}}, M.~Y., {Hafen}, Z., {Rodriguez}, C.~L., {et~al.} 2023, \bibinfo{title}{{Great balls of FIRE - I. The formation of star clusters across cosmic time in a Milky Way-mass galaxy},} \mnras, 519, 1366, \dodoi{10.1093/mnras/stac3573}

\bibitem[{L.~S.~M. Guerra {et~al.}(2026)Guerra, O'Neil, Lisanti, Roy, Sanderson, Kundu, Arora, Necib, Shipp, \& Shen}]{guerra2026probing}
Guerra, L. S.~M., O'Neil, S., Lisanti, M., {et~al.} 2026, \bibinfo{title}{Probing Atomic Dark Matter with Stellar Streams in Milky Way-Mass Galaxies,} arXiv preprint arXiv:2603.20367

\bibitem[{J.~J. Han {et~al.}(2022)Han, Naidu, Conroy, Bonaca, Zaritsky, Caldwell, Cargile, Johnson, Chandra, Speagle, {et~al.}}]{han2022tilt}
Han, J.~J., Naidu, R.~P., Conroy, C., {et~al.} 2022, \bibinfo{title}{A tilt in the dark matter halo of the galaxy,} The Astrophysical Journal, 934, 14

\bibitem[{C.~R. Harris {et~al.}(2020)Harris, Millman, van~der Walt, Gommers, Virtanen, Cournapeau, Wieser, Taylor, Berg, Smith, Kern, Picus, Hoyer, van Kerkwijk, Brett, Haldane, del R{\'{i}}o, Wiebe, Peterson, G{\'{e}}rard-Marchant, Sheppard, Reddy, Weckesser, Abbasi, Gohlke, \& Oliphant}]{numpy}
Harris, C.~R., Millman, K.~J., van~der Walt, S.~J., {et~al.} 2020, \bibinfo{title}{Array programming with {NumPy},} Nature, 585, 357, \dodoi{10.1038/s41586-020-2649-2}

\bibitem[{K. Hattori {et~al.}(2016)Hattori, Erkal, \& Sanders}]{hattori2016shepherding}
Hattori, K., Erkal, D., \& Sanders, J.~L. 2016, \bibinfo{title}{Shepherding tidal debris with the Galactic bar: the Ophiuchus stream,} Monthly Notices of the Royal Astronomical Society, 460, 497

\bibitem[{A. Helmi \& S.~D. White(1999)Helmi \& White}]{helmi1999building}
Helmi, A., \& White, S.~D. 1999, \bibinfo{title}{Building up the stellar halo of the Galaxy,} Monthly Notices of the Royal Astronomical Society, 307, 495

\bibitem[{C. Holm-Hansen {et~al.}(2025)Holm-Hansen, Chen, \& Gnedin}]{holm2025catalog}
Holm-Hansen, C., Chen, Y., \& Gnedin, O.~Y. 2025, \bibinfo{title}{Catalog of Mock Stellar Streams in Milky Way-Like Galaxies,} arXiv preprint arXiv:2510.09604

\bibitem[{P.~F. Hopkins(2015)Hopkins}]{hopkins2015code}
Hopkins, P.~F. 2015, \bibinfo{title}{A new class of accurate, mesh-free hydrodynamic simulation methods,} Monthly Notices of the Royal Astronomical Society, 450, 53

\bibitem[{P.~F. Hopkins {et~al.}(2018)Hopkins, Wetzel, Kere{\v{s}}, Faucher-Gigu{\`e}re, Quataert, Boylan-Kolchin, Murray, Hayward, Garrison-Kimmel, Hummels, {et~al.}}]{hopkins2018fire}
Hopkins, P.~F., Wetzel, A., Kere{\v{s}}, D., {et~al.} 2018, \bibinfo{title}{FIRE-2 simulations: physics versus numerics in galaxy formation,} Monthly Notices of the Royal Astronomical Society, 480, 800

\bibitem[{J.~A. Hunt \& E. Vasiliev(2025)Hunt \& Vasiliev}]{hunt2025milky}
Hunt, J.~A., \& Vasiliev, E. 2025, \bibinfo{title}{Milky Way dynamics in light of Gaia,} New Astronomy Reviews, 101721

\bibitem[{J.~A. Hunt {et~al.}(2025)Hunt, Petersen, Weinberg, Johnston, Bernet, Daniel, Hyman, Price-Whelan, Arora, Collaboration, {et~al.}}]{hunt2025dark}
Hunt, J.~A., Petersen, M.~S., Weinberg, M.~D., {et~al.} 2025, \bibinfo{title}{The dark matter wake of a galactic bar revealed by multichannel Singular Spectral Analysis,} arXiv preprint arXiv:2510.09751

\bibitem[{J.~D. {Hunter}(2007){Hunter}}]{matplotlib}
{Hunter}, J.~D. 2007, \bibinfo{title}{{Matplotlib: A 2D Graphics Environment},} Computing in Science and Engineering, 9, 90, \dodoi{10.1109/MCSE.2007.55}

\bibitem[{R. Ibata {et~al.}(2002)Ibata, Lewis, Irwin, \& Quinn}]{ibata2002uncovering}
Ibata, R., Lewis, G., Irwin, M., \& Quinn, T. 2002, \bibinfo{title}{Uncovering cold dark matter halo substructure with tidal streams,} Monthly Notices of the Royal Astronomical Society, 332, 915

\bibitem[{R. Ibata {et~al.}(2020)Ibata, Thomas, Famaey, Malhan, Martin, \& Monari}]{ibata2020detection}
Ibata, R., Thomas, G., Famaey, B., {et~al.} 2020, \bibinfo{title}{Detection of strong epicyclic density spikes in the GD-1 stellar stream: an absence of evidence for the influence of dark matter subhalos?} The Astrophysical Journal, 891, 161

\bibitem[{R. Ibata {et~al.}(2024)Ibata, Malhan, Tenachi, Ardern-Arentsen, Bellazzini, Bianchini, Bonifacio, Caffau, Diakogiannis, Errani, {et~al.}}]{ibata2024charting}
Ibata, R., Malhan, K., Tenachi, W., {et~al.} 2024, \bibinfo{title}{Charting the galactic acceleration field. II. A Global Mass Model of the Milky Way from the STREAMFINDER Atlas of Stellar Streams detected in Gaia DR3,} The Astrophysical Journal, 967, 89

\bibitem[{R.~A. Ibata {et~al.}(2019)Ibata, Malhan, \& Martin}]{ibata2019streams}
Ibata, R.~A., Malhan, K., \& Martin, N.~F. 2019, \bibinfo{title}{The streams of the gaping abyss: a population of entangled stellar streams surrounding the inner galaxy,} The Astrophysical Journal, 872, 152

\bibitem[{{\v{Z}}. Ivezi{\'c} {et~al.}(2019)Ivezi{\'c}, Kahn, Tyson, Abel, Acosta, Allsman, Alonso, AlSayyad, Anderson, Andrew, {et~al.}}]{ivezic2019lsst}
Ivezi{\'c}, {\v{Z}}., Kahn, S.~M., Tyson, J.~A., {et~al.} 2019, \bibinfo{title}{LSST: from science drivers to reference design and anticipated data products,} The Astrophysical Journal, 873, 111

\bibitem[{K.~V. Johnston {et~al.}(2002)Johnston, Spergel, \& Haydn}]{johnston2002lumpy}
Johnston, K.~V., Spergel, D.~N., \& Haydn, C. 2002, \bibinfo{title}{How lumpy is the Milky Way’s dark matter halo?} The Astrophysical Journal, 570, 656

\bibitem[{K.~V. Johnston {et~al.}(1999)Johnston, Zhao, Spergel, \& Hernquist}]{johnston1999tidal}
Johnston, K.~V., Zhao, H., Spergel, D.~N., \& Hernquist, L. 1999, \bibinfo{title}{Tidal streams as probes of the Galactic potential,} The Astrophysical Journal, 512, L109

\bibitem[{T. Kluyver {et~al.}(2016)Kluyver, Ragan-Kelley, P{\'e}rez, Granger, Bussonnier, Frederic, Kelley, Hamrick, Grout, Corlay, Ivanov, Avila, Abdalla, \& Willing}]{kluyver2016jupyter}
Kluyver, T., Ragan-Kelley, B., P{\'e}rez, F., {et~al.} 2016, in Positioning and Power in Academic Publishing: Players, Agents and Agendas, ed. F.~Loizides \& B.~Schmidt, IOS Press, 87 -- 90

\bibitem[{S.~E. Koposov {et~al.}(2010)Koposov, Rix, \& Hogg}]{koposov2010constraining}
Koposov, S.~E., Rix, H.-W., \& Hogg, D.~W. 2010, \bibinfo{title}{Constraining the milky way potential with a six-dimensional phase-space map of the GD-1 stellar stream,} The Astrophysical Journal, 712, 260

\bibitem[{S.~E. Koposov {et~al.}(2023)Koposov, Erkal, Li, Da~Costa, Cullinane, Ji, Kuehn, Lewis, Pace, Shipp, {et~al.}}]{koposov2023s}
Koposov, S.~E., Erkal, D., Li, T.~S., {et~al.} 2023, \bibinfo{title}{S 5: Probing the Milky Way and Magellanic Clouds potentials with the 6D map of the Orphan--Chenab stream,} Monthly Notices of the Royal Astronomical Society, 521, 4936

\bibitem[{A.~H. K{\"u}pper {et~al.}(2012)K{\"u}pper, Lane, \& Heggie}]{kupper2012more}
K{\"u}pper, A.~H., Lane, R.~R., \& Heggie, D.~C. 2012, \bibinfo{title}{More on the structure of tidal tails,} Monthly Notices of the Royal Astronomical Society, 420, 2700

\bibitem[{H. Li {et~al.}(2017)Li, Gnedin, Gnedin, Meng, Semenov, \& Kravtsov}]{li2017star}
Li, H., Gnedin, O.~Y., Gnedin, N.~Y., {et~al.} 2017, \bibinfo{title}{Star cluster formation in cosmological simulations. I. Properties of young clusters,} The Astrophysical Journal, 834, 69

\bibitem[{T.~S. Li {et~al.}(2021)Li, Koposov, Erkal, Ji, Shipp, Pace, Hilmi, Kuehn, Lewis, Mackey, {et~al.}}]{li2021broken}
Li, T.~S., Koposov, S.~E., Erkal, D., {et~al.} 2021, \bibinfo{title}{Broken into pieces: ATLAS and Aliqa Uma as one single stream,} The Astrophysical Journal, 911, 149

\bibitem[{S. Lilleengen {et~al.}(2023)Lilleengen, Petersen, Erkal, Pe{\~n}arrubia, Koposov, Li, Cullinane, Ji, Kuehn, Lewis, {et~al.}}]{lilleengen2023effect}
Lilleengen, S., Petersen, M.~S., Erkal, D., {et~al.} 2023, \bibinfo{title}{The effect of the deforming dark matter haloes of the Milky Way and the Large Magellanic Cloud on the Orphan--Chenab stream,} Monthly Notices of the Royal Astronomical Society, 518, 774

\bibitem[{G. Limberg {et~al.}(2025)Limberg, Ji, Li, Erkal, Koposov, Pace, Li, Awad, Senkevich, Bland-Hawthorn, {et~al.}}]{limberg2025s}
Limberg, G., Ji, A.~P., Li, T.~S., {et~al.} 2025, \bibinfo{title}{$ S\^{} 5$: Tidal Disruption in Crater 2 and Formation of Diffuse Dwarf Galaxies in the Local Group,} arXiv preprint arXiv:2512.02177

\bibitem[{B. Lowing {et~al.}(2011)Lowing, Jenkins, Eke, \& Frenk}]{lowing2011halo}
Lowing, B., Jenkins, A., Eke, V., \& Frenk, C. 2011, \bibinfo{title}{A halo expansion technique for approximating simulated dark matter haloes,} Monthly Notices of the Royal Astronomical Society, 416, 2697

\bibitem[{K. Malhan \& R.~A. Ibata(2019)Malhan \& Ibata}]{malhan2019constraining}
Malhan, K., \& Ibata, R.~A. 2019, \bibinfo{title}{Constraining the Milky Way halo potential with the GD-1 stellar stream,} Monthly Notices of the Royal Astronomical Society, 486, 2995

\bibitem[{K. Malhan {et~al.}(2019)Malhan, Ibata, Carlberg, Valluri, \& Freese}]{malhan2019butterfly}
Malhan, K., Ibata, R.~A., Carlberg, R.~G., Valluri, M., \& Freese, K. 2019, \bibinfo{title}{Butterfly in a cocoon, understanding the origin and morphology of globular cluster streams: the case of GD-1,} The Astrophysical Journal, 881, 106

\bibitem[{K. Malhan {et~al.}(2018)Malhan, Ibata, \& Martin}]{malhan2018ghostly}
Malhan, K., Ibata, R.~A., \& Martin, N.~F. 2018, \bibinfo{title}{Ghostly tributaries to the Milky Way: charting the halo’s stellar streams with the Gaia DR2 catalogue,} Monthly Notices of the Royal Astronomical Society, 481, 3442

\bibitem[{C. Mateu(2023)Mateu}]{mateu2023galstreams}
Mateu, C. 2023, \bibinfo{title}{galstreams: A library of Milky Way stellar stream footprints and tracks,} Monthly Notices of the Royal Astronomical Society, 520, 5225

\bibitem[{F. McCluskey {et~al.}(2024)McCluskey, Wetzel, Loebman, Moreno, Faucher-Gigu{\`e}re, \& Hopkins}]{mccluskey2024disc}
McCluskey, F., Wetzel, A., Loebman, S.~R., {et~al.} 2024, \bibinfo{title}{Disc settling and dynamical heating: histories of Milky Way-mass stellar discs across cosmic time in the FIRE simulations,} Monthly Notices of the Royal Astronomical Society, 527, 6926

\bibitem[{W. McKinney {et~al.}(2011)McKinney {et~al.}}]{mckinney2011pandas}
McKinney, W., {et~al.} 2011, \bibinfo{title}{pandas: a foundational Python library for data analysis and statistics,} Python for high performance and scientific computing, 14, 1

\bibitem[{Y. Mellier {et~al.}(2025)Mellier, Abdurro{\'u}f, Barroso, Ach{\'u}carro, Adamek, Adam, Addison, Aghanim, Aguena, Ajani, {et~al.}}]{mellier2025euclid}
Mellier, Y., Abdurro{\'u}f, A., Barroso, J.~A., {et~al.} 2025, \bibinfo{title}{Euclid. I. Overview of the Euclid mission,} Astronomy \& Astrophysics

\bibitem[{G. Monari {et~al.}(2019)Monari, Famaey, Siebert, Wegg, \& Gerhard}]{monari2019signatures}
Monari, G., Famaey, B., Siebert, A., Wegg, C., \& Gerhard, O. 2019, \bibinfo{title}{Signatures of the resonances of a large Galactic bar in local velocity space,} Astronomy \& Astrophysics, 626, A41

\bibitem[{E.~O. Nadler {et~al.}(2021)Nadler, Banerjee, Adhikari, Mao, \& Wechsler}]{nadler2021effects}
Nadler, E.~O., Banerjee, A., Adhikari, S., Mao, Y.-Y., \& Wechsler, R.~H. 2021, \bibinfo{title}{The effects of dark matter and baryonic physics on the Milky Way subhalo population in the presence of the large magellanic cloud,} The Astrophysical Journal Letters, 920, L11

\bibitem[{T. Nguyen {et~al.}(2025)Nguyen, Pei, Li, Shipp, Dodelson, Erkal, Ferguson, Starkenburg, Rau, Riley, {et~al.}}]{nguyen2025forecasting}
Nguyen, T., Pei, R., Li, Z., {et~al.} 2025, \bibinfo{title}{Forecasting Dark Matter Subhalo Constraints from Stellar Streams using Implicit Likelihood Inference,} arXiv preprint arXiv:2512.07960

\bibitem[{J. Nibauer \& A. Bonaca(2025)Nibauer \& Bonaca}]{nibauer2025galactic}
Nibauer, J., \& Bonaca, A. 2025, \bibinfo{title}{Galactic Accelerations from the GD-1 Stream Suggest a Tilted Dark Matter Halo,} The Astrophysical Journal Letters, 985, L22

\bibitem[{J. Nibauer {et~al.}(2024)Nibauer, Bonaca, Lisanti, Erkal, \& Hastings}]{nibauer2024slant}
Nibauer, J., Bonaca, A., Lisanti, M., Erkal, D., \& Hastings, Z. 2024, \bibinfo{title}{Slant, fan, and narrow: the response of stellar streams to a tilting galactic disk,} The Astrophysical Journal, 969, 55

\bibitem[{J. Nibauer {et~al.}(2025)Nibauer, Bonaca, Price-Whelan, Spergel, \& Greene}]{nibauer2025measurement}
Nibauer, J., Bonaca, A., Price-Whelan, A.~M., Spergel, D.~N., \& Greene, J.~E. 2025, \bibinfo{title}{Measurement of Dark Matter Substructure from the Kinematics of the GD-1 Stellar Stream,} arXiv preprint arXiv:2510.02247

\bibitem[{J. {Nibauer} {et~al.}(2025){Nibauer}, {Bonaca}, {Spergel}, {Price-Whelan}, {Greene}, {Starkman}, \& {Johnston}}]{nibauer2025streamsculptor}
{Nibauer}, J., {Bonaca}, A., {Spergel}, D.~N., {et~al.} 2025, \bibinfo{title}{{StreamSculptor: Hamiltonian Perturbation Theory for Stellar Streams in Flexible Potentials with Differentiable Simulations},} \apj, 983, 68, \dodoi{10.3847/1538-4357/adb8e8}

\bibitem[{N. Panithanpaisal {et~al.}(2021)Panithanpaisal, Sanderson, Wetzel, Cunningham, Bailin, \& Faucher-Gigu{\`e}re}]{panithanpaisal2021galaxy}
Panithanpaisal, N., Sanderson, R.~E., Wetzel, A., {et~al.} 2021, \bibinfo{title}{The Galaxy Progenitors of Stellar Streams around Milky Way--mass Galaxies in the FIRE Cosmological Simulations,} The Astrophysical Journal, 920, 10

\bibitem[{N. Panithanpaisal {et~al.}(2026)Panithanpaisal, Sanderson, Rodriguez, Starkenburg, Pearson, Bonaca, Hopkins, Cook, Arora, \& Weatherford}]{panithanpaisal2025breaking}
Panithanpaisal, N., Sanderson, R.~E., Rodriguez, C.~L., {et~al.} 2026, \bibinfo{title}{Breaking Down the CosmoGEMS: Toward Modeling and Understanding Globular Cluster Stellar Streams in a Fully Cosmological Context,} The Astrophysical Journal, 997, 182

\bibitem[{S. Pearson {et~al.}(2024)Pearson, Bonaca, Chen, \& Gnedin}]{pearson2024forecasting}
Pearson, S., Bonaca, A., Chen, Y., \& Gnedin, O.~Y. 2024, \bibinfo{title}{Forecasting the Population of Globular Cluster Streams in Milky Way--type Galaxies,} The Astrophysical Journal, 976, 54

\bibitem[{S. Pearson {et~al.}(2017)Pearson, Price-Whelan, \& Johnston}]{pearson2017gaps}
Pearson, S., Price-Whelan, A.~M., \& Johnston, K.~V. 2017, \bibinfo{title}{Gaps and length asymmetry in the stellar stream Palomar 5 as effects of Galactic bar rotation,} Nature Astronomy, 1, 633

\bibitem[{S. Pearson {et~al.}(2019)Pearson, Starkenburg, Johnston, Williams, Ibata, \& Khan}]{pearson2019detecting}
Pearson, S., Starkenburg, T.~K., Johnston, K.~V., {et~al.} 2019, \bibinfo{title}{Detecting Thin Stellar Streams in External Galaxies: Resolved Stars and Integrated Light,} The Astrophysical Journal, 883, 87

\bibitem[{M.~S. Petersen {et~al.}(2021)Petersen, Weinberg, \& Katz}]{petersen2021using}
Petersen, M.~S., Weinberg, M.~D., \& Katz, N. 2021, \bibinfo{title}{Using commensurabilities and orbit structure to understand barred galaxy evolution,} Monthly Notices of the Royal Astronomical Society, 500, 838

\bibitem[{J. Pfeffer {et~al.}(2018)Pfeffer, Kruijssen, Crain, \& Bastian}]{pfeffer2018mosaics}
Pfeffer, J., Kruijssen, J.~D., Crain, R.~A., \& Bastian, N. 2018, \bibinfo{title}{The E-MOSAICS project: simulating the formation and co-evolution of galaxies and their star cluster populations,} Monthly Notices of the Royal Astronomical Society, 475, 4309

\bibitem[{A. {Phillips} {et~al.}(2026){Phillips}, {Conroy}, {Nibauer}, {Wang}, {Chandra}, {Bonaca}, {Strader}, \& {MacLeod}}]{Phillips_DirectNbody}
{Phillips}, A., {Conroy}, C., {Nibauer}, J., {et~al.} 2026, \bibinfo{title}{{The Binary Populations of Stellar Streams are Set by Cluster Dynamics},} arXiv e-prints, arXiv:2603.06790, \dodoi{10.48550/arXiv.2603.06790}

\bibitem[{ {Planck Collaboration} {et~al.}(2016){Planck Collaboration}, {Ade}, {Aghanim}, {Arnaud}, {Ashdown}, {Aumont}, {Baccigalupi}, {Banday}, {Barreiro}, {Bartlett}, {Bartolo}, {Battaner}, {Battye}, {Benabed}, {Beno{\^\i}t}, {Benoit-L{\'e}vy}, {Bernard}, {Bersanelli}, {Bielewicz}, {Bock}, {Bonaldi}, {Bonavera}, {Bond}, {Borrill}, {Bouchet}, {Boulanger}, {Bucher}, {Burigana}, {Butler}, {Calabrese}, {Cardoso}, {Catalano}, {Challinor}, {Chamballu}, {Chary}, {Chiang}, {Chluba}, {Christensen}, {Church}, {Clements}, {Colombi}, {Colombo}, {Combet}, {Coulais}, {Crill}, {Curto}, {Cuttaia}, {Danese}, {Davies}, {Davis}, {de Bernardis}, {de Rosa}, {de Zotti}, {Delabrouille}, {D{\'e}sert}, {Di Valentino}, {Dickinson}, {Diego}, {Dolag}, {Dole}, {Donzelli}, {Dor{\'e}}, {Douspis}, {Ducout}, {Dunkley}, {Dupac}, {Efstathiou}, {Elsner}, {En{\ss}lin}, {Eriksen}, {Farhang}, {Fergusson}, {Finelli}, {Forni}, {Frailis}, {Fraisse}, {Franceschi}, {Frejsel}, {Galeotta}, {Galli}, {Ganga}, {Gauthier}, {Gerbino}, {Ghosh}, {Giard},
  {Giraud-H{\'e}raud}, {Giusarma}, {Gjerl{\o}w}, {Gonz{\'a}lez-Nuevo}, {G{\'o}rski}, {Gratton}, {Gregorio}, {Gruppuso}, {Gudmundsson}, {Hamann}, {Hansen}, {Hanson}, {Harrison}, {Helou}, {Henrot-Versill{\'e}}, {Hern{\'a}ndez-Monteagudo}, {Herranz}, {Hildebrandt}, {Hivon}, {Hobson}, {Holmes}, {Hornstrup}, {Hovest}, {Huang}, {Huffenberger}, {Hurier}, {Jaffe}, {Jaffe}, {Jones}, {Juvela}, {Keih{\"a}nen}, {Keskitalo}, {Kisner}, {Kneissl}, {Knoche}, {Knox}, {Kunz}, {Kurki-Suonio}, {Lagache}, {L{\"a}hteenm{\"a}ki}, {Lamarre}, {Lasenby}, {Lattanzi}, {Lawrence}, {Leahy}, {Leonardi}, {Lesgourgues}, {Levrier}, {Lewis}, {Liguori}, {Lilje}, {Linden-V{\o}rnle}, {L{\'o}pez-Caniego}, {Lubin}, {Mac{\'\i}as-P{\'e}rez}, {Maggio}, {Maino}, {Mandolesi}, {Mangilli}, {Marchini}, {Maris}, {Martin}, {Martinelli}, {Mart{\'\i}nez-Gonz{\'a}lez}, {Masi}, {Matarrese}, {McGehee}, {Meinhold}, {Melchiorri}, {Melin}, {Mendes}, {Mennella}, {Migliaccio}, {Millea}, {Mitra}, {Miville-Desch{\^e}nes}, {Moneti}, {Montier}, {Morgante}, {Mortlock},
  {Moss}, {Munshi}, {Murphy}, {Naselsky}, {Nati}, {Natoli}, {Netterfield}, {N{\o}rgaard-Nielsen}, {Noviello}, {Novikov}, {Novikov}, {Oxborrow}, {Paci}, {Pagano}, {Pajot}, {Paladini}, {Paoletti}, {Partridge}, {Pasian}, {Patanchon}, {Pearson}, {Perdereau}, {Perotto}, {Perrotta}, {Pettorino}, {Piacentini}, {Piat}, {Pierpaoli}, {Pietrobon}, {Plaszczynski}, {Pointecouteau}, {Polenta}, {Popa}, {Pratt}, \& {Pr{\'e}zeau}}]{collaboration2015planck}
{Planck Collaboration}, {Ade}, P.~A.~R., {Aghanim}, N., {et~al.} 2016, \bibinfo{title}{{Planck 2015 results. XIII. Cosmological parameters},} \aap, 594, A13, \dodoi{10.1051/0004-6361/201525830}

\bibitem[{C. Power {et~al.}(2003)Power, Navarro, Jenkins, Frenk, White, Springel, Stadel, \& Quinn}]{power2003inner}
Power, C., Navarro, J.~F., Jenkins, A., {et~al.} 2003, \bibinfo{title}{The inner structure of $\Lambda$CDM haloes—I. A numerical convergence study,} Monthly Notices of the Royal Astronomical Society, 338, 14

\bibitem[{A.~M. Price-Whelan(2017)Price-Whelan}]{price2017gala}
Price-Whelan, A.~M. 2017, \bibinfo{title}{Gala: A Python package for galactic dynamics,} Journal of Open Source Software, 2, 388

\bibitem[{A.~M. Price-Whelan \& A. Bonaca(2018)Price-Whelan \& Bonaca}]{price2018off}
Price-Whelan, A.~M., \& Bonaca, A. 2018, \bibinfo{title}{Off the beaten path: Gaia reveals GD-1 stars outside of the main stream,} The Astrophysical Journal Letters, 863, L20

\bibitem[{A.~M. Price-Whelan {et~al.}(2016{\natexlab{a}})Price-Whelan, Johnston, Valluri, Pearson, K{\"u}pper, \& Hogg}]{price2016chaotic}
Price-Whelan, A.~M., Johnston, K.~V., Valluri, M., {et~al.} 2016{\natexlab{a}}, \bibinfo{title}{Chaotic dispersal of tidal debris,} Monthly Notices of the Royal Astronomical Society, 455, 1079

\bibitem[{A.~M. Price-Whelan {et~al.}(2016{\natexlab{b}})Price-Whelan, Sesar, Johnston, \& Rix}]{price2016spending}
Price-Whelan, A.~M., Sesar, B., Johnston, K.~V., \& Rix, H.-W. 2016{\natexlab{b}}, \bibinfo{title}{Spending too much time at the Galactic bar: chaotic fanning of the Ophiuchus stream,} The Astrophysical Journal, 824, 104

\bibitem[{I. Puerari \& H. Dottori(1992)Puerari \& Dottori}]{puerari1992fourier}
Puerari, I., \& Dottori, H. 1992, \bibinfo{title}{Fourier analysis of structure in spiral galaxies,} Astronomy and Astrophysics Supplement Series (ISSN 0365-0138), vol. 93, no. 3, June 1992, p. 469-493. Research supported by CNPq and FINEP., 93, 469

\bibitem[{Y. Qian {et~al.}(2022)Qian, Arshad, \& Bovy}]{qian2022structure}
Qian, Y., Arshad, Y., \& Bovy, J. 2022, \bibinfo{title}{The structure of accreted stellar streams,} Monthly Notices of the Royal Astronomical Society, 511, 2339

\bibitem[{A.~H. Riley {et~al.}(2025)Riley, Shipp, Simpson, Bieri, Fattahi, Brown, Oman, Fragkoudi, G{\'o}mez, Grand, {et~al.}}]{riley2025auriga}
Riley, A.~H., Shipp, N., Simpson, C.~M., {et~al.} 2025, \bibinfo{title}{Auriga Streams--I: disrupting satellites surrounding Milky Way-mass haloes at multiple resolutions,} Monthly Notices of the Royal Astronomical Society, 542, 2443

\bibitem[{L.~V. Sales {et~al.}(2008)Sales, Helmi, Starkenburg, Morrison, Engle, Harding, Mateo, Olszewski, \& Sivarani}]{sales2008genealogy}
Sales, L.~V., Helmi, A., Starkenburg, E., {et~al.} 2008, \bibinfo{title}{On the genealogy of the Orphan Stream,} Monthly Notices of the Royal Astronomical Society, 389, 1391

\bibitem[{J.~L. Sanders {et~al.}(2020)Sanders, Lilley, Vasiliev, Evans, \& Erkal}]{sanders2020models}
Sanders, J.~L., Lilley, E.~J., Vasiliev, E., Evans, N.~W., \& Erkal, D. 2020, \bibinfo{title}{Models of distorted and evolving dark matter haloes,} Monthly Notices of the Royal Astronomical Society, 499, 4793

\bibitem[{R.~E. Sanderson {et~al.}(2017)Sanderson, Hartke, \& Helmi}]{sanderson2017modeling}
Sanderson, R.~E., Hartke, J., \& Helmi, A. 2017, \bibinfo{title}{Modeling the gravitational potential of a cosmological dark matter halo with stellar streams,} The Astrophysical Journal, 836, 234

\bibitem[{R.~E. Sanderson {et~al.}(2020)Sanderson, Wetzel, Loebman, Sharma, Hopkins, Garrison-Kimmel, Faucher-Gigu{\`e}re, Kere{\v{s}}, \& Quataert}]{sanderson2020synthetic}
Sanderson, R.~E., Wetzel, A., Loebman, S., {et~al.} 2020, \bibinfo{title}{Synthetic gaia surveys from the fire cosmological simulations of milky way-mass galaxies,} The Astrophysical Journal Supplement Series, 246, 6

\bibitem[{S. Shao {et~al.}(2021)Shao, Cautun, Deason, \& Frenk}]{shao2021twisted}
Shao, S., Cautun, M., Deason, A., \& Frenk, C.~S. 2021, \bibinfo{title}{The twisted dark matter halo of the Milky Way,} Monthly Notices of the Royal Astronomical Society, 504, 6033

\bibitem[{N. Shipp {et~al.}(2018)Shipp, Drlica-Wagner, Balbinot, Ferguson, Erkal, Li, Bechtol, Belokurov, Buncher, Carollo, {et~al.}}]{shipp2018stellar}
Shipp, N., Drlica-Wagner, A., Balbinot, E., {et~al.} 2018, \bibinfo{title}{Stellar streams discovered in the dark energy survey,} The Astrophysical Journal, 862, 114

\bibitem[{N. Shipp {et~al.}(2019)Shipp, Li, Pace, Erkal, Drlica-Wagner, Yanny, Belokurov, Wester, Koposov, Kuehn, {et~al.}}]{shipp2019proper}
Shipp, N., Li, T., Pace, A., {et~al.} 2019, \bibinfo{title}{Proper motions of stellar streams discovered in the Dark Energy Survey,} The Astrophysical Journal, 885, 3

\bibitem[{N. Shipp {et~al.}(2021)Shipp, Erkal, Drlica-Wagner, Li, Pace, Koposov, Cullinane, Da~Costa, Ji, Kuehn, {et~al.}}]{shipp2021measuring}
Shipp, N., Erkal, D., Drlica-Wagner, A., {et~al.} 2021, \bibinfo{title}{Measuring the Mass of the Large Magellanic Cloud with Stellar Streams Observed by S 5,} The Astrophysical Journal, 923, 149

\bibitem[{N. Shipp {et~al.}(2025)Shipp, Riley, Simpson, Bieri, Necib, Arora, Fragkoudi, G{\'o}mez, Grand, \& Marinacci}]{shipp2025auriga}
Shipp, N., Riley, A.~H., Simpson, C.~M., {et~al.} 2025, \bibinfo{title}{Auriga Streams II: orbital properties of tidally disrupting satellites of Milky Way-mass galaxies,} Monthly Notices of the Royal Astronomical Society, 542, 1109

\bibitem[{D. Spergel {et~al.}(2015)Spergel, Gehrels, Baltay, Bennett, Breckinridge, Donahue, Dressler, Gaudi, Greene, Guyon, {et~al.}}]{spergel2015wide}
Spergel, D., Gehrels, N., Baltay, C., {et~al.} 2015, \bibinfo{title}{Wide-field infrarred survey telescope-astrophysics focused telescope assets WFIRST-AFTA 2015 report,} arXiv preprint arXiv:1503.03757

\bibitem[{N. Starkman {et~al.}(2023)Starkman, Bovy, Webb, Calvetti, \& Somersalo}]{starkman2023fast}
Starkman, N., Bovy, J., Webb, J.~J., Calvetti, D., \& Somersalo, E. 2023, \bibinfo{title}{On the fast track: Rapid construction of stellar stream paths,} Monthly Notices of the Royal Astronomical Society, 522, 5022

\bibitem[{N. Starkman {et~al.}(2025)Starkman, Nibauer, Bovy, Webb, Tavangar, Price-Whelan, \& Bonaca}]{starkman2025stream}
Starkman, N., Nibauer, J., Bovy, J., {et~al.} 2025, \bibinfo{title}{Stream members only: Data-driven characterization of stellar streams with mixture density networks,} The Astrophysical Journal, 980, 253

\bibitem[{K. Tavangar {et~al.}(2022)Tavangar, Ferguson, Shipp, Drlica-Wagner, Koposov, Erkal, Balbinot, Garc{\'\i}a-Bellido, Kuehn, Lewis, {et~al.}}]{tavangar2022fire}
Tavangar, K., Ferguson, P., Shipp, N., {et~al.} 2022, \bibinfo{title}{From the fire: a deeper look at the Phoenix stream,} The Astrophysical Journal, 925, 118

\bibitem[{M. Valluri {et~al.}(2025)Valluri, Fagrelius, Koposov, Li, Gnedin, Bell, Carlberg, Cooper, Aguilar, Ahlen, {et~al.}}]{valluri2025gd}
Valluri, M., Fagrelius, P., Koposov, S.~E., {et~al.} 2025, \bibinfo{title}{GD-1 Stellar Stream and Cocoon in the DESI Early Data Release,} The Astrophysical Journal, 980, 71

\bibitem[{E. {van der Velden}(2020){van der Velden}}]{cmasher}
{van der Velden}, E. 2020, \bibinfo{title}{{CMasher: Scientific colormaps for making accessible, informative and 'cmashing' plots},} The Journal of Open Source Software, 5, 2004, \dodoi{10.21105/joss.02004}

\bibitem[{G. Van~Rossum \& F.~L. Drake(2009)Van~Rossum \& Drake}]{python}
Van~Rossum, G., \& Drake, F.~L. 2009, Python 3 Reference Manual (Scotts Valley, CA: CreateSpace)

\bibitem[{E. Vasiliev(2019)Vasiliev}]{vasiliev2019agama}
Vasiliev, E. 2019, \bibinfo{title}{AGAMA: action-based galaxy modelling architecture,} Monthly Notices of the Royal Astronomical Society, 482, 1525

\bibitem[{E. Vasiliev(2024)Vasiliev}]{vasiliev2024dear}
Vasiliev, E. 2024, \bibinfo{title}{Dear Magellanic Clouds, welcome back!,} Monthly Notices of the Royal Astronomical Society, 527, 437

\bibitem[{E. Vasiliev {et~al.}(2021)Vasiliev, Belokurov, \& Erkal}]{vasiliev2021tango}
Vasiliev, E., Belokurov, V., \& Erkal, D. 2021, \bibinfo{title}{Tango for three: Sagittarius, LMC, and the Milky Way,} Monthly Notices of the Royal Astronomical Society, 501, 2279

\bibitem[{C. Vera-Ciro \& A. Helmi(2013)Vera-Ciro \& Helmi}]{vera2013constraints}
Vera-Ciro, C., \& Helmi, A. 2013, \bibinfo{title}{Constraints on the shape of the Milky Way dark matter halo from the Sagittarius stream,} The Astrophysical Journal Letters, 773, L4

\bibitem[{P. Virtanen {et~al.}(2020)Virtanen, Gommers, Oliphant, Haberland, Reddy, Cournapeau, Burovski, Peterson, Weckesser, Bright, {van der Walt}, Brett, Wilson, Millman, Mayorov, Nelson, Jones, Kern, Larson, Carey, Polat, Feng, Moore, {VanderPlas}, Laxalde, Perktold, Cimrman, Henriksen, Quintero, Harris, Archibald, Ribeiro, Pedregosa, {van Mulbregt}, \& {SciPy 1.0 Contributors}}]{scipy}
Virtanen, P., Gommers, R., Oliphant, T.~E., {et~al.} 2020, \bibinfo{title}{{{SciPy} 1.0: Fundamental Algorithms for Scientific Computing in Python},} Nature Methods, 17, 261, \dodoi{10.1038/s41592-019-0686-2}

\bibitem[{Y. Wang {et~al.}(2020)Wang, Athanassoula, \& Mao}]{wang2020basis}
Wang, Y., Athanassoula, E., \& Mao, S. 2020, \bibinfo{title}{Basis function expansions for galactic dynamics: Spherical versus cylindrical coordinates,} Astronomy \& Astrophysics, 639, A38

\bibitem[{N.~C. Weatherford {et~al.}(2024)Weatherford, Rasio, Chatterjee, Fragione, K{\i}ro{\u{g}}lu, \& Kremer}]{weatherford2024stellar}
Weatherford, N.~C., Rasio, F.~A., Chatterjee, S., {et~al.} 2024, \bibinfo{title}{Stellar Escape from Globular Clusters. II. Clusters May Eat Their Own Tails,} The Astrophysical Journal, 967, 42

\bibitem[{J.~J. Webb \& J. Bovy(2019)Webb \& Bovy}]{webb2019searching}
Webb, J.~J., \& Bovy, J. 2019, \bibinfo{title}{Searching for the GD-1 stream progenitor in Gaia DR2 with direct N-body simulations,} Monthly Notices of the Royal Astronomical Society, 485, 5929

\bibitem[{S. Weerasooriya {et~al.}(2025)Weerasooriya, Starkenburg, Cunningham, \& Johnston}]{weerasooriya2025dancing}
Weerasooriya, S., Starkenburg, T., Cunningham, E.~C., \& Johnston, K.~V. 2025, \bibinfo{title}{Dancing Streams In Merging Halos: Stellar Streams in a MW--LMC-like merger,} arXiv preprint arXiv:2505.14792

\bibitem[{M.~D. Weinberg(1992)Weinberg}]{weinberg1992detection}
Weinberg, M.~D. 1992, \bibinfo{title}{Detection of a large-scale stellar bar in the Milky Way,} Astrophysical Journal, Part 1 (ISSN 0004-637X), vol. 384, Jan. 1, 1992, p. 81-94., 384, 81

\bibitem[{M.~D. Weinberg(2001)Weinberg}]{weinberg2001noise}
Weinberg, M.~D. 2001, \bibinfo{title}{Noise-driven evolution in stellar systems--I. Theory,} Monthly Notices of the Royal Astronomical Society, 328, 311

\bibitem[{M.~D. Weinberg(2023)Weinberg}]{weinberg2023new}
Weinberg, M.~D. 2023, \bibinfo{title}{New dipole instabilities in spherical stellar systems,} Monthly Notices of the Royal Astronomical Society, 525, 4962

\bibitem[{M.~D. Weinberg \& M.~S. Petersen(2021)Weinberg \& Petersen}]{weinberg2021using}
Weinberg, M.~D., \& Petersen, M.~S. 2021, \bibinfo{title}{Using multichannel singular spectrum analysis to study galaxy dynamics,} Monthly Notices of the Royal Astronomical Society, 501, 5408

\bibitem[{P.~D. Welch(1967)Welch}]{welch1967use}
Welch, P.~D. 1967, \bibinfo{title}{The use of fast Fourier transform for the estimation of power spectra: A method based on time averaging over short, modified periodograms,} IEEE Transactions on audio and electroacoustics, 15, 70

\bibitem[{A. Wetzel {et~al.}(2023)Wetzel, Hayward, Sanderson, Ma, Angl{\'e}s-Alc{\'a}zar, Feldmann, Chan, El-Badry, Wheeler, Garrison-Kimmel, {et~al.}}]{wetzel2023public}
Wetzel, A., Hayward, C.~C., Sanderson, R.~E., {et~al.} 2023, \bibinfo{title}{Public data release of the FIRE-2 cosmological zoom-in simulations of galaxy formation,} The Astrophysical Journal Supplement Series, 265, 44

\bibitem[{A. Wetzel {et~al.}(2025)Wetzel, Samuel, Gandhi, Ponnada, Su, Arora, Angles-Alcazar, Hayward, Sanderson, Feldmann, {et~al.}}]{wetzel2025second}
Wetzel, A., Samuel, J., Gandhi, P.~J., {et~al.} 2025, \bibinfo{title}{Second public data release of the FIRE-2 cosmological zoom-in simulations of galaxy formation,} arXiv preprint arXiv:2508.06608

\bibitem[{A.~R. Wetzel {et~al.}(2016)Wetzel, Hopkins, Kim, Faucher-Gigu{\`e}re, Kere{\v{s}}, \& Quataert}]{wetzel2016reconciling}
Wetzel, A.~R., Hopkins, P.~F., Kim, J.-h., {et~al.} 2016, \bibinfo{title}{Reconciling dwarf galaxies with $\Lambda$CDM cosmology: simulating a realistic population of satellites around a Milky Way--mass galaxy,} The Astrophysical Journal Letters, 827, L23

\bibitem[{T.~D. Yavetz {et~al.}(2023)Yavetz, Johnston, Pearson, Price-Whelan, \& Hamilton}]{yavetz2023stream}
Yavetz, T.~D., Johnston, K.~V., Pearson, S., Price-Whelan, A.~M., \& Hamilton, C. 2023, \bibinfo{title}{Stream Fanning and Bifurcations: Observable Signatures of Resonances in Stellar Stream Morphology,} The Astrophysical Journal, 954, 215

\bibitem[{T.~D. Yavetz {et~al.}(2021)Yavetz, Johnston, Pearson, Price-Whelan, \& Weinberg}]{yavetz2021separatrix}
Yavetz, T.~D., Johnston, K.~V., Pearson, S., Price-Whelan, A.~M., \& Weinberg, M.~D. 2021, \bibinfo{title}{Separatrix divergence of stellar streams in galactic potentials,} Monthly Notices of the Royal Astronomical Society, 501, 1791

\bibitem[{A. Zonca {et~al.}(2019)Zonca, Singer, Lenz, Reinecke, Rosset, Hivon, \& Gorski}]{healpy2019}
Zonca, A., Singer, L.~P., Lenz, D., {et~al.} 2019, \bibinfo{title}{healpy: equal area pixelization and spherical harmonics transforms for data on the sphere in Python,} Journal of Open Source Software, 4, 1298

\end{thebibliography}
\bibliographystyle{aasjournalv7}

\end{document}